\documentclass[10pt,journal,compsoc]{IEEEtran}

\ifCLASSOPTIONcompsoc
\usepackage[nocompress]{cite}
\else
\usepackage{cite}
\fi
\ifCLASSINFOpdf
\else
\fi

\usepackage{xcolor,colortbl}

\definecolor{Gray}{gray}{0.85}
\definecolor{LightCyan}{rgb}{0.88,1,1}
\newcolumntype{a}{>{\columncolor{Gray}}c}
\newcolumntype{b}{>{\columncolor{white}}c}

\usepackage{amssymb}
\usepackage{hyperref}
\usepackage{epstopdf}

\usepackage{graphicx}
\usepackage[caption=false]{subfig}

\usepackage{enumerate}
\usepackage{slashbox}
\usepackage{amsmath}
\usepackage{verbatim}

\usepackage{microtype}
\DisableLigatures{encoding = *, family = * }

\usepackage[nolist]{acronym}
\begin{acronym}
	\acro{AAA}{Authentication, Authorization and Accounting}
	\acro{ACL}{Access Control List}
	\acro{AKI}{Accountable Key Infrastructure}
	\acro{API}{Application Programming Interface}
    \acro{BSM}{Basic Safety Message}
    \acro{BYOD}{Bring Your Own Device}
    \acro{BF}{Bloom Filter}
	\acro{C2C-CC}{Car2Car Communication Consortium}
	\acro{C2I}{Car-to-Infrastructure}
	\acro{C$^2$RL}{Compressed \ac{CRL}}
	\acro{CA}{Certification Authority}
	\acro{CN}{Common Name}
	\acro{CAM}{Cooperative Awareness Message}
	\acro{CAMP VSC3}{Crash Avoidance Metrics Partnership Vehicle Safety Consortium}
	\acro{CDF}{Cumulative Distribution Function}
	\acro{CIA}{Confidentiality, Integrity and Availability}
	\acro{CRL}{Certificate Revocation List}
	\acro{CDN}{Content Delivery Network}
	\acro{COCA}{Cornell OnLine Certification Authority}
	\acro{CSR}{Certificate Signing Request}
	\acro{DAA}{Direct Anonymous Attestation}
	\acro{DDoS}{Distributed DoS}
	\acro{DDH}{Decisional Diffie-Helman}
	\acro{DENM}{Decentralized Environmental Notification Message}
	\acro{DHT}{Distributed Hash Table}
	\acro{DL/ECIES}{Discrete Logarithm and Elliptic Curve Integrated Encryption Scheme}
	\acro{DoS}{Denial of Service}
	\acro{DoT}{Department of Transportation}
	\acro{DPA}{Data Protection Agency}
	\acro{DSRC}{Dedicated Short Range Communication}
	\acro{DSS}{Digital Signature Standard}
	\acro{DTLS}{Datagram \ac{TLS}}
	\acro{ECU}{Electronic Control Unit}
	\acro{EDR}{Event Data Recorder}
	\acro{ETSI}{European Telecommunications Standards Institute}
	\acro{ECDSA}{Elliptic Curve Digital Signature Algorithm}
	\acro{ECC}{Elliptic Curve Cryptography}
	\acro{EVITA}{E-safety Vehicle Intrusion protected Applications}
	\acro{FOT}{Field Operational Testing}
	\acro{FPGA}{Field-Programmable Gate Array}
	\acro{GPA}{Global Passive Adversary}
	\acro{GN}{GeoNetworking}
	\acro{GS-VLR}{Group Signatures with Verifier Local Revocation}
	\acro{GS}{Group Signatures}
	\acro{GM}{Group Manager}
	\acro{GBA}{Generic Bootstrapping Architecture}
	\acro{GPS}{Global Positioning System}
	\acro{GUI}{Graphic User Interface}
	\acro{HSM}{Hardware Security Module}
	\acro{HTTP}{Hypertext Transfer Protocol}
	\acro{IEEE}{Institute of Electrical and Electronics Engineers}
	\acro{IETF}{Internet Engineering Task Force}
	\acro{IoT}{Internet of Things}
	\acro{ITS}{Intelligent Transport System}
	\acro{IT}{Information Technologies}
	\acro{IMSI}{International Mobile Subscriber Identity}
	\acro{IMEI}{International Mobile Station Equipment Identity}
	\acro{IdP}{Identity Provider}
	\acro{IDS}{Intrusion Detection System}
	\acro{ISP}{Internet Service Provider}
	\acro{LEA}{Law Enforcement Agency}
	\acro{LCPP}{Lightweight Conditional Privacy Preservation}
	\acro{LTC}{Long Term Certificate}
	\acro{LTCA}{Long Term CA}
	\acro{H-LTCA}{Home-LTCA}
	\acro{F-LTCA}{Foreign-LTCA}
	\acro{LDAP}{Lightweight Directory Access Protocol}
	\acro{LBS}{Location Based Service}
	\acro{LTE}{Long Term Evolution}
	\acro{LuST}{Luxembourg SUMO Traffic}
	\acro{MAC}{Message Authentication Code}
	\acro{MCA}{Message \ac{CA}}
	\acro{MEA}{Misbehavior Evaluation Authority}
	\acro{MANET}{Mobile Ad-hoc Network}
	\acro{MPB}{Most Pieces Broadcast}
	\acro{NoW}{Network on Wheel}
	\acro{OBU}{On-Board Unit}
	\acro{OEM}{Original Equipment Manufacturer}
	\acro{OCSP}{Online Certificate Status Protocol}
	\acro{PCA}{Pseudonym CA}
	\acro{PDP}{Policy Decision Point}
	\acro{PEP}{Policy Enforcement Point}
	\acro{PIR}{Private Information Retrieval}
	\acro{PKC}{Public Key Cryptography}
	\acro{PKCS}{Public Key Cryptosystem}
	\acro{PKI}{Public-Key Infrastructure}
	\acro{PRECIOSA}{Privacy Enabled Capability in Co-operative Systems and Safety Applications}
	\acro{PRESERVE}{Preparing Secure Vehicle-to-X Communication Systems}
	\acro{P2P}{peer-to-peer}
	\acro{PS}{Participatory Sensing}
	\acro{RA}{Resolution Authority}
	\acro{REST}{Representational State Transfer}
	\acro{RBAC}{Role Based Access Control}
	\acro{RCA}{Root \acs{CA}}
	\acro{RSU}{Roadside Unit}
	\acro{RTC}{Real Time Clock}
	\acro{SAML}{Security Assertion Markup Language}
	\acro{SAS}{Sample Aggregation Service}
	\acro{SCMS}{Security Credential Management System}
	\acro{SCORE@F}{Système COopératif Routier Expérimental Français}
	\acro{SDSI}{Simple Distributed Security Infrastructure}
	\acro{SRAAC}{Secure Revocable Anonymous Authenticated Inter-Vehicle Communication}
	\acro{SeVeCom}{Secure Vehicle Communication}
	\acro{SIT}{Sichere Informationstechnologie}
	\acro{SLC}{Short-Lived Certificate}
	\acro{SoA}{Service-oriented-Approach}
	\acro{SIFS}{Short Inter Frame Space}
	\acro{SSO}{Single-Sign-On}
	\acro{SSL}{Secure Sockets Layer}
	\acro{SOAP}{Simple Object Access Protocol}
	\acro{TACK}{Temporary Anonymous Certified Key}
	\acro{TS}{Task Service}
	\acro{TLS}{Transport Layer Security}
	\acro{TPM}{Trusted Platform Module}
	\acro{TTP}{Trusted Third Party}
	\acro{TVR}{Ticket Validation Repository}
	\acro{URI}{Uniform Resource Identifier}
	\acro{VANET}{Vehicular Ad-hoc Network}
	\acro{V2I}{Vehicle-to-Infrastructure}
	\acro{V2V}{Vehicle-to-Vehicle}
	\acro{V2X}{\ac{V2V}/\ac{V2I}}
	\acro{VC}{Vehicular Communication}
	\acro{VM}{Virtual Machine}
	\acro{VSS}{\ac{VC} Security Subsystem}
	\acro{WAVE}{Wireless Access in Vehicular Environments}
	\acro{WSDL}{Web Services Discovery Language}
	\acro{W3C}{World Wide Web Consortium}
	\acro{V}{Vehicle}
	\acro{VANET}{Vehicular Ad-hoc Network}
	\acro{VLR}{Verifier-Local Revocation}
	\acro{VPKI}{Vehicular Public-Key Infrastructure}
	\acro{VM}{Virtual Machine}
	\acro{WS}{Web Service}
	\acro{WoT}{Web of Trust}
	\acro{WSACA}{\ac{WAVE} Service Advertisement \ac{CA}}
	\acro{XML}{Extensible Markup Language}
	\acro{XACML}{eXtensible Access Control Markup Language}
	\acro{3G}{3rd Generation}
\end{acronym}

\usepackage{makecell}

\usepackage{arydshln}

\usepackage{xcolor}

\usepackage{algpseudocode}

\usepackage{setspace}

\errorcontextlines\maxdimen

\makeatletter
\newcommand*{\algrule}[1][\algorithmicindent]{\makebox[#1][l]{\hspace*{.5em}\thealgruleextra\vrule height \thealgruleheight depth \thealgruledepth}}%
\newcommand*{\thealgruleextra}{}
\newcommand*{\thealgruleheight}{.75\baselineskip}
\newcommand*{\thealgruledepth}{.25\baselineskip}

\newcount\ALG@printindent@tempcnta
\def\ALG@printindent{%
	\ifnum \theALG@nested>0
	\ifx\ALG@text\ALG@x@notext
	\else
	\unskip
	\addvspace{-1pt}
	\ALG@printindent@tempcnta=1
	\loop
	\algrule[\csname ALG@ind@\the\ALG@printindent@tempcnta\endcsname]%
	\advance \ALG@printindent@tempcnta 1
	\ifnum \ALG@printindent@tempcnta<\numexpr\theALG@nested+1\relax
	\repeat
	\fi
	\fi
}%
\usepackage{etoolbox}
\patchcmd{\ALG@doentity}{\noindent\hskip\ALG@tlm}{\ALG@printindent}{}{\errmessage{failed to patch}}
\makeatother

\newbox\statebox
\newcommand{\myState}[1]{%
	\setbox\statebox=\vbox{#1}%
	\edef\thealgruleheight{\dimexpr \the\ht\statebox+1pt\relax}%
	\edef\thealgruledepth{\dimexpr \the\dp\statebox+1pt\relax}%
	\ifdim\thealgruleheight<.75\baselineskip
	\def\thealgruleheight{\dimexpr .75\baselineskip+1pt\relax}%
	\fi
	\ifdim\thealgruledepth<.25\baselineskip
	\def\thealgruledepth{\dimexpr .25\baselineskip+1pt\relax}%
	\fi
	\State #1%
	\def\thealgruleheight{\dimexpr .75\baselineskip+1pt\relax}%
	\def\thealgruledepth{\dimexpr .25\baselineskip+1pt\relax}%
}

\usepackage{footnote}

\usepackage{rotating}

\usepackage{soul}

\usepackage{algorithm}

\usepackage{tikz}
\usetikzlibrary{shadows,positioning,calc}
\tikzset{multiple/.style = {double copy shadow={shadow xshift=1ex,shadow
			yshift=-1.5ex,draw=black!30},fill=white,draw=black,thick,minimum height = 1cm,minimum
		width=2cm},
	ordinary/.style = {rectangle,draw,thick,minimum height = 1cm,minimum width=2cm}}

\usetikzlibrary{decorations.pathreplacing}

\usepackage{booktabs}
\usepackage{multirow}
\usepackage{siunitx}

\usepackage{algpseudocode}

\makeatletter
\renewcommand{\ALG@beginalgorithmic}{\footnotesize} 
\makeatother

\usepackage{afterpage}
\newlength{\oldtextfloatsep}

\usepackage[utf8]{inputenc}

\usepackage[normalem]{ulem}

\usepackage{booktabs} 

\usepackage{blindtext}

\usepackage{xcolor,colortbl}

\begin{document}
%
\title{Scalable \& Resilient Vehicle-Centric Certificate Revocation List Distribution in \acl{VC} Systems} 


%
%
%
%

\author{Mohammad~Khodaei,~\IEEEmembership{Member,~IEEE,}
	and~Panos~Papadimitratos,~\IEEEmembership{Fellow,~IEEE}}

\markboth{} 
{}
\IEEEtitleabstractindextext{%
\begin{abstract}

\acresetall
In spite of progress in securing \ac{VC} systems, there is no consensus on how to distribute \acp{CRL}. The main challenges lie exactly in (i) crafting an efficient and timely distribution of \acp{CRL} for numerous anonymous credentials, \emph{pseudonyms}, (ii) maintaining strong privacy for vehicles prior to revocation events, even with \emph{honest-but-curious} system entities, (iii) and catering to computation and communication constraints of on-board units with intermittent connectivity to the infrastructure. Relying on peers to distribute the \acp{CRL} is a double-edged sword: \emph{abusive peers} could ``pollute'' the process, thus degrading the timely \acp{CRL} distribution. In this paper, we propose a \emph{vehicle-centric} solution that addresses all these challenges and thus closes a gap in the literature. Our scheme radically reduces \ac{CRL} distribution overhead: each vehicle receives \acp{CRL} corresponding only to its region of operation and its actual trip duration. Moreover, a ``fingerprint'' of \ac{CRL} `pieces' is attached to a subset of (verifiable) pseudonyms for fast \ac{CRL} `piece' validation (while mitigating resource depletion attacks abusing the \ac{CRL} distribution). Our experimental evaluation shows that our scheme is efficient, scalable, dependable, and practical: with no more than 25 KB/s of traffic load, the latest \ac{CRL} can be delivered to 95\% of the vehicles in a region (15$\times$15 KM) within 15s, i.e., more than 40 times faster than the state-of-the-art. Overall, our scheme is a comprehensive solution that complements standards and can catalyze the deployment of secure and privacy-protecting \ac{VC} systems. 

\end{abstract}

}

\maketitle

\IEEEdisplaynontitleabstractindextext

%
\IEEEpeerreviewmaketitle

\section{Introduction}
\label{sec:crl-dis-introduction}

\acresetall

\ac{V2V} and \ac{V2I} communications seek to enhance transportation safety and efficiency with a gamut of applications, ranging from collision avoidance alerts to traffic conditions updates; moreover, they can integrate and enrich \acp{LBS}~\cite{ETSI-102-638, papadimitratos2009vehicular} and vehicular social networks~\cite{jin2016security}, and provide infotainment services. It has been well-understood that \ac{VC} systems are vulnerable to attacks and that the privacy of their users is at stake. As a result, security and privacy solutions have been developed by standardization bodies (IEEE 1609.2 WG~\cite{1609-2016} and \acs{ETSI}~\cite{ETSI-102-638, cits_platform_annex1, cits_platform_annex2}), harmonization efforts (\ac{C2C-CC}~\cite{c2c}), and projects (\acs{SeVeCom}~\cite{papadimitratos2007architecture}, \acs{PRESERVE}~\cite{preserve-url}, and CAMP~\cite{whyte2013security}). A consensus towards using \ac{PKC} to protect \ac{V2X} communication is reached: a set of Certification Authorities (CAs) constitutes the \ac{VPKI}, providing multiple anonymous credentials, termed \emph{pseudonyms}, to legitimate vehicles. Vehicles switch from one pseudonym to a non-previously used one towards unlinkability of digitally signed messages, and improved sender privacy for \ac{V2V}/\ac{V2I} messages. Pseudonymity is conditional in the sense that the corresponding long-term vehicle identity (\ac{LTC}) can be retrieved by the \ac{VPKI} entities if deviating from system policies.

In fact, vehicles can be compromised or faulty and disseminate erroneous information across the \ac{V2X} network~\cite{Papadi:C:08, raya2006certificaterevocation}. They should be held \emph{accountable} for such actions and credentials (their \acp{LTC} and their pseudonyms) can be revoked. To efficiently revoke a set of pseudonyms, one can disclose a single entry for all (revoked) pseudonyms of the vehicle~\cite{fischer2006secure, stumpf2007trust, laberteaux2008security, haas2009design}. However, upon a revocation event, all non-revoked (but expired) pseudonyms belonging to the ``misbehaving'' vehicle would also be linked. Linking pseudonyms with lifetimes prior to a revocation event implies that all the corresponding digitally signed messages will be trivially linked. Even if revocation is justified, this does not imply that a user \emph{``deserves''} to abolish privacy prior to the revocation event. Avoiding such a situation, i.e., achieving what is termed in the literature as \emph{perfect-forward-privacy}~\cite{schaub2009privacy}, can be guaranteed if the \ac{VPKI} entities are \emph{fully-trustworthy}~\cite{haas2011efficient}. However, we need to guarantee strong user privacy even in the presence of \emph{honest-but-curious} \ac{VPKI} entity; recent revelations of mass surveillance show that assuming service providers are fully-trustworthy is no longer a viable approach.

A main concern, relevant to all proposals in the literature~\cite{papadimitratos2008certificate, haas2011efficient, laberteaux2008security, haas2009design, nowatkowski2010certificate, nowatkowski2009cooperative} is efficiency and scalability, essentially low communication and computation overhead even as system dimension grows. Consider first typical operational constraints: the average daily commute time is less than an hour (on average 29.2 miles and 46 minutes per day)~\cite{whyte2013security, acs-survey, newsroom-how-much-motorists-drive} while the latencies for the dissemination of a full \ac{CRL} can exceed the actual trip duration~\cite{DOTHS812014}. One can compress \ac{CRL} using a \acf{BF}~\cite{raya2006certificaterevocation, raya2007eviction, rigazzi2017optimized}; however, the size of a \ac{CRL} grows linearly with the number of revoked pseudonyms, thus necessitates larger \acp{BF}. More so, a sizable portion of the \ac{CRL} information is irrelevant to a receiving vehicle and can be left unused. This, at the system level, constitutes waste of computation, communication (bandwidth), and storage resources. In turn, it leads to higher latency for all vehicles to reconstruct the \ac{CRL}, i.e., a degradation of timely distribution.

Alternatively, vehicles can only validate revocation status of (their neighbors') pseudonyms via an \ac{OCSP}. Even if a \ac{VPKI} can comfortably handle such a demanding load~\cite{khodaei2014ScalableRobustVPKI}, \ac{OCSP} cannot be used as a standalone solution in \ac{VC} systems: it requires continuous connectivity and significant bandwidth dedicated to revocation traffic, thus impractical due to the network volatility and scale~\cite{raya2006certificaterevocation}. Moreover, what would be the course of action if the \ac{VPKI} were not reachable for other reasons, e.g., during a \ac{DoS} attack? So, the challenge is \emph{how can one distribute the most relevant revocation information to a given vehicle, per trip, and ensure timely revocation even without uninterrupted connectivity to the \ac{VPKI}?}

The computation overhead for the verification of the \ac{CRL} could interfere with safety- and time-critical operations especially if one considers typical \ac{VC} rates of 10 safety beacons per second, and thus processing of possibly hundreds of messages from neighboring vehicles per second. Simply put, with existing computation and communication overhead and the time-critical nature of safety applications, minimizing the overhead for \ac{CRL} verification and distribution is paramount.

From a different viewpoint, we need to allocate as little bandwidth as possible for the \ac{CRL} distribution in order not to interfere with safety-critical operations or enable an attacker to broadcast a fake \ac{CRL} at a high rate. However, this should be hand in hand with timely \ac{CRL} distribution. This can be achieved with the use of \acp{RSU}~\cite{papadimitratos2008certificate}; however, dense deployment of \acp{RSU} in a large-scale environment is costly. If the deployment is sparse, a significant delay could be introduced. Alternatively, the \ac{CRL} can be distributed in a peer-to-peer, epidemic manner~\cite{laberteaux2008security, haas2011efficient, haas2009design}. This is a double-edged sword: \emph{abusive peers}, seeking to compromise the trustworthiness of the system, could pollute the \ac{CRL} distribution and mount a clogging \ac{DoS} attack. Such an attack relates other content dissemination, e.g., vehicular social networks~\cite{jin2016security}, yet it is critical to mitigate it for \ac{CRL} distribution: delaying or preventing legitimate users from obtaining the most up-to-date \ac{CRL} pieces would result in prolonging the operation of a malicious compromised vehicle in the system.

Moreover, a vehicle could be wrongly identified as a misbehaving entity, e.g., a false positive in an event-based misbehavior detection~\cite{raya2007eviction, CAMP-VSC3-Phase2}, or a vehicle sensor malfunction, e.g., a communication breakdown or a problem in a safety-critical system (e.g., steering or braking~\cite{google-self-driving-cars-failure}). Thus, the \ac{LTC} as well as all pseudonyms of the ``misbehaving'' entity would be revoked, with the revocation information to be distributed/made available to all vehicles in the network. This raises two challenges: misbehavior could be falsely identified, or the sensor malfunctioning could be transient, thus distributing ``imprecise'' revocation information, likely in high volumes. The effect would be waste of communication (bandwidth) and higher convergence time to reconstruct the \ac{CRL}, i.e., degradation of timely \ac{CRL} distribution. Repairing a faulty sensor or patching the vehicle software or resolving the erroneous eviction implies the vehicle can rejoin the system as legitimate participant. This means the vehicle needs to obtain a fresh batch of pseudonyms, e.g., for another three years~\cite{whyte2013security}, i.e., imposing unnecessary workload on the \ac{VPKI} entities. Therefore, a flexible design with easily reversible revocation status that allows to temporarily evict a misbehaving or malfunctioning vehicle from the system until the issue is resolved would address this challenge. By the same token, we need to ensure that such temporal eviction of a vehicle does not harm user privacy; in other words, exactly because such a flexible design allows resolution and temporal eviction, a single \ac{VPKI} entity should not be able to link the corresponding (successive) pseudonyms of the same vehicle, upon rejoining the system.

Furthermore, the \ac{VPKI} system must narrow down the vulnerability window, i.e., the interval between a revocation event/incident until all vehicles successfully obtain the latest revocation information. Thus, if a new revocation event happens when the base-\ac{CRL} has already been distributed, the \ac{VPKI} entities should release $\Delta\textnormal{-}$\acp{CRL} in order to close down the vulnerability window. However, frequent distribution of such revocation information could affect the performance of \ac{VC} systems. More important, abusive peers could mount a clogging \ac{DoS} attack on the $\Delta\textnormal{-}$\acp{CRL} distribution. A lightweight, i.e., incurring low computation and communication overhead, and resilient mechanism for distributing $\Delta\textnormal{-}$\acp{CRL} would address this challenge.

Despite the plethora of research efforts, none addresses all challenges at hand. In this paper, we show how to \emph{efficiently revoke a very large volume of pseudonyms while providing strong user privacy protection, even in the presence of honest-but-curious \ac{VPKI} entities. Our system effectively, resiliently, and in a timely manner disseminate the authentic \ac{CRL} throughout a large-scale (multi-domain) \ac{VC} system.} Moreover, \emph{we ensure that the \ac{CRL} distribution incurs low overhead and prevents abuse of the distribution mechanism. Furthermore, our flexible design allows to temporarily evict a vehicle from the system without compromising user privacy. At the same time, it facilitates rejoining the system as a legitimate participant upon resolving the issue without imposing unnecessary workload on the \ac{VPKI} entities, by frequently refilling pseudonyms pool, and, most important, shields the system from clogging \ac{DoS} attacks leveraging the \ac{CRL} and $\Delta\textnormal{-}$\ac{CRL} distribution.}

\emph{Contributions:} Our comprehensive security and privacy-preserving solution systematically addresses all key aspects of \ac{CRL}-based revocation, i.e., security, privacy, and efficiency. This is based on few simple yet powerful, as it turns out, ideas. We propose making the \ac{CRL} acquisition process \emph{vehicle-centric}: each vehicle only receives the pieces of \acp{CRL} corresponding to its targeted region and its actual trip duration, i.e., obtaining only region- and time-relevant revocation information. Moreover, randomly chosen pseudonyms issued by the \ac{VPKI} are selected to piggyback a notification about new \ac{CRL}-update events and an authenticator for efficiently validating pieces of the latest \ac{CRL}; in other words, validation of the \ac{CRL} pieces \emph{almost for free}. These novel features dramatically reduce the \ac{CRL} size and \ac{CRL} validation overhead, while they significantly increase its resiliency against resource depletion attacks. Moreover, we propose a secure, efficient, and resilient distribution mechanism for $\Delta\textnormal{-}$\ac{CRL} updates to narrow down the vulnerability window. Furthermore, our scheme facilitates eviction of a ``misbehaving'' vehicle temporarily, i.e., for an interval until the issue is resolved.

In the rest of the paper, we critically survey the literature (Sec.~\ref{sec:crl-dis-related-work}) and explain the system model (Sec.~\ref{sec:crl-dis-model-requirements}). We present system design (Sec.~\ref{sec:crl-dis-design}), followed by qualitative and quantitative analysis (Sec.~\ref{sec:crl-dis-scheme-analysis-evaluation}). We then conclude the paper (Sec.~\ref{sec:crl-dis-conclusions}).


\section{Related Work}
\label{sec:crl-dis-related-work}

The need to evict misbehaving or compromised~\cite{Papadi:C:08} vehicles from a \ac{VC} system is commonly accepted, because such vehicles can threaten the safety of vehicles and users and degrade transportation efficiency. \ac{CRL} distribution is of central importance and it is the final and definitive line of defense~\cite{1609-2016, ETSI-102-638, gerlach2007security, papadimitratos2007architecture, papadimitratos2007architecture, raya2007eviction}: only the \acs{VPKI} can \emph{``ultimately''} revoke a vehicle by including its unexpired certificates' serial numbers in a \ac{CRL}.

The literature proposes distribution of the \ac{CRL} via \acp{RSU}~\cite{papadimitratos2008certificate} and car-to-car epidemic communication~\cite{laberteaux2008security, haas2009design, haas2011efficient}, with enhancements on the distribution of pieces~\cite{nowatkowski2010certificate, nowatkowski2009cooperative} evaluated in~\cite{nowatkowski2010scalable}. A na\"ive solution would be to digitally sign the entire \ac{CRL} and broadcast it; however, it imposes difficulties in downloading a large \ac{CRL} file and exchanging it over short contact period (with an \ac{RSU} or a peer). Splitting the digitally signed \ac{CRL} into multiple pieces is vulnerable to \emph{pollution} attacks: in the absence of fine-grained authentication, per \ac{CRL} piece, an adversary can delay or even prevent reception by injecting fake pieces. Thus, the straightforward solution is to have the \ac{VPKI} prepare the \ac{CRL}, split it into multiple pieces, sign each piece, and distribute all of them across the \ac{VC} system. \acp{RSU} can broadcast \ac{CRL} pieces randomly or in a round-robin fashion~\cite{papadimitratos2008certificate}, and vehicles can relay pieces until all vehicles receive all pieces necessary to reconstruct the \ac{CRL}~\cite{laberteaux2008security}. Erasure codes can be used to enhance the fault-tolerance of the \ac{CRL} piece distribution in the highly volatile \ac{VC} environment~\cite{papadimitratos2008certificate, ardelean2009CRLImplementation}.

Signing each \ac{CRL} piece so that it is self-verifiable, incurs significant computation overhead, which grows linearly with the number of \ac{CRL} pieces, both for the \ac{VPKI} and for the receiving vehicles. Furthermore, an attacker could aggressively forge \ac{CRL} pieces for a \ac{DDoS} attack leveraging signature verification delays~\cite{hsiao2011flooding} that can prevent vehicles from obtaining the genuine \ac{CRL} pieces. A \emph{``precode-and-hash''} scheme~\cite{nguyen2016secure} proposes to calculate a hash value of each pre-coded piece, sign it, and disseminate it with higher priority. Each relaying node can apply a different precode to the original \ac{CRL} and act as a secondary source. However, by applying different encodings to the original \ac{CRL} file, another receiver cannot reconstruct the entire \ac{CRL} from the pieces, encoded differently by various relaying nodes. To mitigate pollution and \ac{DoS} attacks, we propose to piggyback a fingerprint (a \ac{BF}~\cite{bloom1970space, mitzenmacher2002compressed}) for \ac{CRL} pieces into a subset of pseudonyms to validating \ac{CRL} pieces ``for free''.

To efficiently revoke an ensemble of pseudonyms, one can enable revocation of multiple pseudonyms with a single \ac{CRL} entry, to reduce the \ac{CRL} size, e.g.,~\cite{laberteaux2008security, haas2009design, rabieh2018srpv}. Despite a huge reduction in size, such schemes do not provide \emph{perfect-forward-privacy}~\cite{schaub2009privacy}: upon a revocation event and \ac{CRL} release, all the ``non-revoked'' but previously expired pseudonyms belonging to the evicted entity would be linked as well. Although forward-privacy can be achieved by leveraging a hash chain~\cite{haas2011efficient}, the pseudonyms' issuer can trivially link all pseudonyms belonging to a vehicle, and thus the pseudonymously authenticated messages~\cite{khodaei2018PrivacyUniformity, khodaei2018NowhereToHide, khodaei2018MixzoneEverywhere}, towards tracking it for the entire duration of its presence in the system~\cite{fischer2006secure, stumpf2007trust, laberteaux2008security, haas2009design, haas2011efficient}. More precisely, the \acs{CA} specifies a \emph{``time interval''} so that each vehicle receives $\mathbb{D}$ pseudonyms during the pseudonym acquisition process~\cite{haas2011efficient}. As a result, for each batch of revoked pseudonyms, a single key is disclosed. But, upon a revocation event, all pseudonyms within an interval are linked, because one can decrypt all pseudonym serial numbers; thus, no \emph{perfect-forward-privacy} is achieved for that period. On the contrary, in our scheme, upon a revocation event and \ac{CRL} release, it is infeasible to link the previously non-revoked (but expired) pseudonyms belonging to a misbehaving vehicle. This is so due to the utilization of a hash chain during the pseudonym issuance process, thus achieving perfect-forward-privacy~\cite{khodaei2018VehicleCentric}.

Compressing \acp{CRL} using a \ac{BF} was proposed for compact storage of revocation entries~\cite{raya2007eviction}, or to efficiently distribute them across the network~\cite{raya2006certificaterevocation, raya2007eviction, rigazzi2017optimized}. However, the challenge is twofold: scalability and efficiency. Their \ac{CRL} size still grows linearly with the number of revoked pseudonyms, while a substantial portion of the compressed \ac{CRL} can be irrelevant to a receiving vehicle and be left unused. Moreover, compressing \acp{CRL} using a \ac{BF} does not necessarily reduce the size of a \ac{CRL} as vehicles can be provided with possibly hundreds of pseudonyms~\cite{khodaei2018VehicleCentric, 1609-2016, simplicio2018acpc}. Unlike such schemes~\cite{raya2006certificaterevocation, raya2007eviction, rigazzi2017optimized}, we do not compress the \ac{CRL}: our scheme disseminates only trip-relevant revocation information to vehicles and it utilizes a \ac{BF} to provide a condensed authenticator for the \ac{CRL} pieces. Our scheme leverages and \emph{enhances} the functionality of the state-of-the-art \ac{VPKI} system~\cite{khodaei2018Secmace} towards efficiently revoking a batch of pseudonyms without compromising user privacy backwards: upon a revocation event, all pseudonyms prior to the revocation event remain unlinkable.

Alternatively, vehicles could validate pseudonym status information through \ac{OCSP}. But, due to intermittent network connectivity, significant usage of the bandwidth by time- and safety-critical operations, and substantial overhead for the \ac{VPKI} (if it is reachable), \ac{OCSP} cannot be used as a standalone solution~\cite{raya2006certificaterevocation}. A hybrid solution could rely on distributing certificate status information to other mobile nodes~\cite{marias2005adopt, ganan2013coach}; however, the system would be subject to the reachability (of sufficiently many cooperative) and the trustworthiness of such nodes. In our scheme, we ensure that the latest \ac{CRL} is efficiently, effectively, and timely distributed among all vehicles without any assumption on persistent reachability and trustworthiness of specific mobile nodes.

Temporal eviction of a misbehaving or malfunctioning vehicle from the \ac{VC} system has received limited attention. There are several situations that a vehicle should be temporarily evicted from the system until the issue is resolved and the vehicle can rejoin the system, e.g., in case a malfunctioning sensor disseminating false information. \ac{SCMS}~\cite{whyte2013security, kumar2017binary} supports only permanent eviction of a misbehaving vehicle by including a linkage seed into a \ac{CRL}. Towards temporal revocation of credentials,~\cite{junior2018privacy, simplicioprivacy} propose a \emph{linkage hook} between any \emph{linkage seed} and the corresponding \emph{pre-linkage values} in the original \ac{SCMS} design. Thus, in order to temporarily revoke the credentials, the linkage hook is disclosed (instead of the linkage seed, used for permanently revoking the credentials). Temporal eviction of a subset of the certificates requires additional layers to be added to the tree. However, the disclosure of linkage hooks would trivially link all pseudonyms inside a given subtree. Our scheme facilitates eviction of a misbehaving vehicle temporarily, i.e., for a fine-grained interval, until the issue is resolved without compromising user privacy (prior to the revocation event).

\ac{SCMS}~\cite{whyte2013security, kumar2017binary} issues pseudonyms with the help of two Linkage Authorities (LAs): a batch of 20-40 pseudonyms, valid for a week, are issued for each vehicle without having a single \ac{VPKI} entity able to link them. In case of revocation, a single entry is disclosed to invalidate the batch of revoked pseudonyms. As a result of this binding, the size of a \ac{CRL} linearly grows with the number of compromised vehicles (and not with the number of revoked pseudonyms). However, due to large number of pseudonyms carried by each vehicle for a long period, e.g., 3 years~\cite{whyte2013security} or 25 years~\cite{kumar2017binary}, the size of a \ac{CRL} could be huge~\cite{simplicio2018acpc}. The \ac{VPKI}~\cite{whyte2013security, kumar2017binary} could provision vehicles for a long period, e.g., 25 years worth of pseudonyms, with a decryption key for, e.g., a weekly batch of pseudonyms, delivered periodically~\cite{verheul2016activate, kumar2017binary, simplicio2018acpc}. This would eliminate the need for frequently recurring bidirectional connectivity to the \ac{VPKI} to obtain pseudonyms. To evict a vehicle, the \ac{VPKI} can stop delivering the corresponding decryption key to the vehicle \ac{HSM}. Still, it is imperative to distribute the \ac{CRL} and cover the (week long) period and the corresponding revoked pseudonyms. Furthermore, having released a \ac{CRL} towards the end of a week, signed messages with the private keys corresponding to the recently revoked pseudonyms (included in the \ac{CRL}) can be linked, i.e., achieving no \emph{perfect-forward-privacy} for that period~\cite{DOTHS812014}.


\section{Model and Requirements}
\label{sec:crl-dis-model-requirements}

\subsection{System Model and Assumptions}
\label{subsec:crl-dis-system-model-assumptions}

A \ac{VPKI} consists of a set of Certification Authorities (CAs) with distinct roles: the \ac{RCA}, the highest-level authority, certifies other lower-level authorities; the \ac{LTCA} is responsible for the vehicle registration and the \acf{LTC} issuance, and the \ac{PCA} issues pseudonyms for the registered vehicles. Pseudonyms have a lifetime (a validity period), typically ranging from minutes to hours; in principle, the shorter the pseudonym lifetime is, the higher the unlinkability and thus the higher privacy protection can be achieved. We assume that each vehicle is registered only with its \emph{\ac{H-LTCA}}, the \emph{policy decision and enforcement point}, reachable by the registered vehicles. Without loss of generality, a \emph{domain} can be defined as a set of vehicles in a region, registered with the \ac{H-LTCA}, subject to the same administrative regulations and policies~\cite{khodaei2015VTMagazine}. There can be several \acp{PCA}, each active in one or more domains. Trust between two domains can be established with the help of the \ac{RCA}, or through cross certification.

All vehicles (\acp{OBU}) registered in the system are provided with \acp{HSM}, ensuring that private keys never leave the \ac{HSM}. Moreover, we assume that there is a misbehavior detection system, e.g.,~\cite{bissmeyer2014misbehavior}, that triggers the revocation. The \acf{RA} can initiate a process to resolve and revoke all pseudonyms of a misbehaving vehicle: it interacts with the corresponding \acp{PCA} and \ac{LTCA} (a detailed protocol description in~\cite{khodaei2012secure, khodaei2014ScalableRobustVPKI, khodaei2018Secmace}) to resolve and revoke all credentials issued for a misbehaving vehicle. Consequently, the misbehaving vehicle can no longer obtain credentials from the \ac{VPKI}. The \ac{VPKI} is responsible for distributing the \acp{CRL} and notifying all legitimate entities about the revocation; this implies a new \ac{CRL}-update event.

\subsection{Adversarial Model and Requirements}
\label{subsec:crl-dis-adversarial-model}

We extend the general adversary model in secure vehicular communications~\cite{papadimitratos2006securing} to include \ac{VPKI} entities that are \emph{honest-but-curious}, i.e., entities complying with security protocols and policies, but motivated to profile users. In a \ac{VC} environment, internal adversaries, i.e., malicious, compromised, or non-cooperative clients, and external adversaries, i.e., unauthorized entities, raise four challenges. More specifically in the context of this work, adversaries can try to (i) exclude revoked pseudonym serial numbers from a \ac{CRL}, (ii) add valid pseudonyms by forging a fake \ac{CRL} (piece), or (iii) prevent legitimate entities from obtaining genuine and the most up-to-date \ac{CRL} (pieces), or delay the \ac{CRL} distribution by replaying old, spreading fake \ac{CRL} (pieces), or performing a \ac{DoS} attack. This allows wrong-doers to remain operational in the \ac{VC} system using their current revoked pseudonym sets. Moreover, they might be simply non-cooperative or malicious, tempted to prevent other vehicles from receiving a notification on a new \ac{CRL}-update event, thus preventing them from requesting to download the \acp{CRL}. Lastly, (iv) \ac{VPKI} entities (in collusion with vehicle communication observers) could potentially link messages signed under (non-revoked but expired) pseudonyms prior to the revocation events, e.g., inferring sensitive information from the \acp{CRL} towards linking pseudonyms, and thus tracking vehicles backwards. The \acp{PCA} operating in a domain (or across domains) could also collude, i.e., share information that each of them individually has, to harm user privacy.

Security and privacy requirements for \ac{V2X} communications have been specified in the literature~\cite{papadimitratos2006securing}, and additional requirements for \ac{VPKI} entities in~\cite{khodaei2018Secmace}. The security and privacy requirements for the \ac{CRL} distribution are: \emph{fine-grained authentication, integrity, and non-repudiation}, \emph{unlinkability (perfect-forward-privacy)}, \emph{availability}, \emph{efficiency}, and \emph{explicit and/or implicit notification on revocation events}~\cite{khodaei2018VehicleCentric}. Beyond these requirements, the revocation mechanism should be flexible and provide the option for easily reversible certificate status. More precisely, the system should facilitate temporal eviction of a misbehaving or malfunctioning vehicle from the system. By the same token, the revocation mechanism should allow reinstating an entity previously revoked upon resolving a misbehaving or malfunctioning issue that led to the revocation; the temporarily-evicted entity could rejoin the system and continue operating in the \ac{VC} system. Even if a vehicle is preloaded with pseudonyms for a longer period, e.g., a year, it should be able to leverage the already obtained credentials upon rejoining the system.


\section{Design}
\label{sec:crl-dis-design}

\subsection{Motivation and Overview} 
\label{subsec:crl-dis-motivation-and-overview}

\textbf{Preliminary assumptions:} We leverage a state-of-the-art \ac{VPKI} system~\cite{khodaei2018Secmace, khodaei2018VPKIaaS, khodaei2019Scaling} that provides pseudonyms in an \emph{on-demand} fashion: each vehicle \emph{``decides''} when to trigger the pseudonym acquisition process based on various factors~\cite{khodaei2016evaluating, khodaei2017RHyTHM}. Such a scheme requires sparse connectivity to the \ac{VPKI}, but it facilitates an \ac{OBU} to be \emph{preloaded} with pseudonyms proactively, covering a longer period, e.g., a week or a month, should the connectivity be expected heavily intermittent. The efficiency, scalability and robustness of the \ac{VPKI} system is systematically investigated~\cite{khodaei2016evaluating, khodaei2018Secmace, khodaei2018VPKIaaS, khodaei2019Scaling} with the \ac{VPKI} handles a large workload. Moreover, it enhances user privacy, notably preventing linking pseudonyms based on \emph{timing information}~\cite{khodaei2018PrivacyUniformity} (the instance of issuance and the pseudonym lifetime) as well as offers strong user privacy protection even in the presence of \emph{honest-but-curious} \ac{VPKI} entities. More precisely, a universally fixed interval, $\Gamma$, is specified by the \ac{H-LTCA} and all pseudonyms in that domain are issued with the lifetime ($\tau_{P}$) aligned with the \ac{VPKI} clock. Vehicles obtain pseudonyms on-the-fly as they operate, and the number of pseudonyms in a request is $\frac{\Gamma}{\tau_{p}}$, i.e., no prior calculation needed. As a result of this policy, at any point in time, all vehicles pseudonyms are indistinguishable based on issuance time thanks to this time alignment, i.e., eliminating any distinction among pseudonym sets of different vehicles, thus enhancing user privacy. We leverage and \emph{enhance} the functionality of this \ac{VPKI} system; in particular, our solution necessitates two modifications during pseudonym acquisition process, notably (i) implicitly binding pseudonyms issued to a given requester per $\Gamma$, and (ii) integrating a fingerprint into a subset of the pseudonyms for efficient \ac{CRL} validation (detailed description in~\cite{khodaei2018VehicleCentric}).

\textbf{High-level overview:} The default policy is to distribute all revocation information to all vehicles. Nonetheless, this approach ignores the locality, the temporal nature of pseudonyms, and other constraints, e.g., the average daily commute time. Locality could be geographical, i.e., credentials relative to the corresponding region, and temporal, i.e., relevance to the lifetime of pseudonyms with respect to the trip duration of a vehicle. To efficiently, effectively, and timely distribute the \acp{CRL} across the \ac{V2X} network, we propose making the \ac{CRL} acquisition process \emph{vehicle-centric}, i.e., through \emph{a content-based and context-sensitive ``publish-subscribe''} scheme~\cite{eugster2003many, huang2004publish}.

By starting a new trip, each vehicle only subscribes to receive the pieces of \acp{CRL}, i.e., the content, corresponding to its actual trip duration and its targeted region, i.e., the context. To reap the benefits of the ephemeral nature pseudonyms and the timely-aligned pseudonym provisioning policy, towards an effective, efficient, and scalable \ac{CRL} distribution, a fixed interval, $\Gamma_{CRL}$, is predetermined by the \acp{PCA} in the domain. They publicize revoked pseudonyms whose lifetimes fall within $\Gamma_{CRL}$, i.e., distributing only the serial number of these pseudonyms rather than publishing the entire \ac{CRL}. Note that $\Gamma$, the universally fixed interval to obtain pseudonyms~\cite{khodaei2018Secmace}, and $\Gamma_{CRL}$ are not necessarily aligned due to the unpredictable nature of revocation events.

When a vehicle reliably connects to the \ac{VPKI}, it can obtain the \emph{``necessary''} \ac{CRL} pieces corresponding to its trip duration during the pseudonym acquisition phase. However, if reliable connectivity is not guaranteed, or if a vehicle obtained (possibly preloaded with enough) pseudonyms in advance, or a new revocation event happens, one can be notified about a new \ac{CRL}-update (revocation) event: a signed fingerprint (a \acf{BF}~\cite{bloom1970space, mitzenmacher2002compressed}) of \ac{CRL} pieces is broadcasted by \acp{RSU}; furthermore, it is appended in a subset of recently issued pseudonyms for a subset of vehicles (termed \emph{fingerprint-carrier} nodes); these pseudonyms are attached to as typically all \acp{CAM}. This essentially piggybacks a notification about the latest \ac{CRL}-update event and an authenticator for validating \ac{CRL} pieces. This provides \ac{CRL} validation for free in terms of computational overhead: pseudonyms are readily validated by the receiving vehicles since each vehicle verifies the signature on a pseudonym before validating the content of a \ac{CAM}, i.e., the verification of \ac{CRL} pieces does not incur extra computation overhead. This eliminates the need for signature verification, but a \ac{BF} membership test, for each \ac{CRL} piece as the fingerprint is signed by the \ac{PCA}.

\begin{figure} [!t]
	\begin{center}
		\centering
		\includegraphics[width=0.45\textwidth,height=0.45\textheight,keepaspectratio]{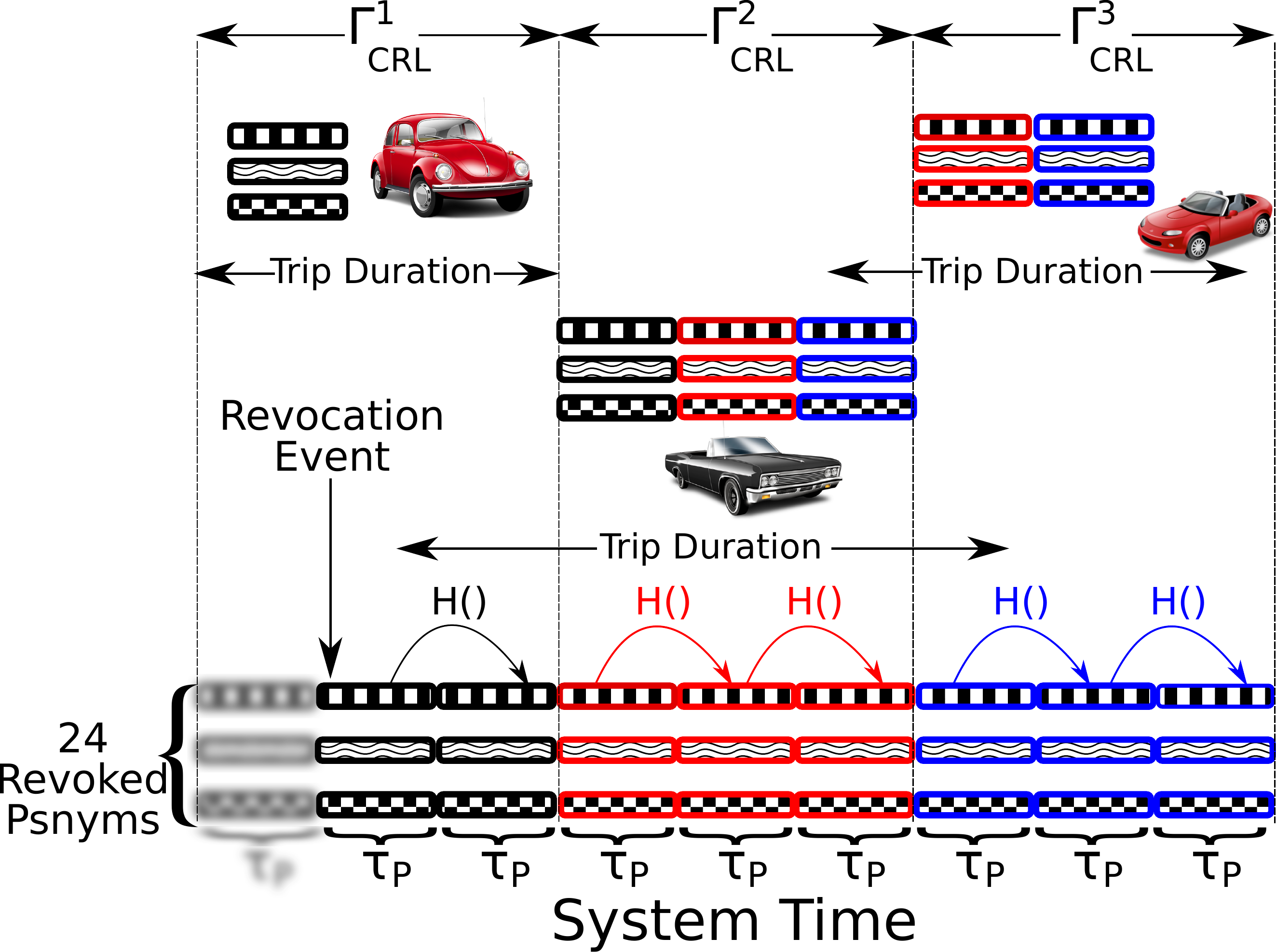}
		\vspace{-0.25em}
		\caption{A vehicle-centric approach: each vehicle only subscribes for pieces of \acp{CRL} corresponding to its trip duration.}
		\label{fig:crl-dis-vehicle-centric-overview}
	\end{center}
	\vspace{-1.75em}
\end{figure}

Our scheme does not require prior knowledge on trip duration to obtain \acp{CRL}, i.e., a vehicle can be oblivious to the trip duration. In fact, such information would not be relevant to the \ac{CRL} dissemination: due to the unpredictable nature of revocation events, the \acp{PCA} disseminate at each point revoked pseudonyms whose lifetimes fall within a $\Gamma_{CRL}$ interval. In other words, even if a vehicle knows the trip duration, it will not receive revocation information regarding the far future. In contrast, the revocation information is \emph{progressively} distributed among the vehicles. The reason is twofold: first, trivially, some revocation events are not yet scheduled; receiving \acp{CRL} within a near time horizon is more likely to include the \emph{latest} revocation information. Moreover, upon resolving the issue that led to the revocation of a malfunctioning or misbehaving entity before the next update of the \ac{CRL}, the corresponding credentials should not be included in the \ac{CRL}, i.e., achieving reversible revocation status. The percentage of the information that is relevant at a given point in time, and is included in the \ac{CRL}, is a function of pseudonym lifetime ($\tau_p$), and the size of the $\Gamma_{CRL}$ interval. More precisely, the shorter the pseudonym lifetime ($\tau_p$) and the longer the $\Gamma_{CRL}$ intervals are, the higher the number of revocation entries is, i.e., the larger the \ac{CRL} size.

Fig.~\ref{fig:crl-dis-vehicle-centric-overview} illustrates an example of 24 revoked pseudonyms to be distributed. A vehicle traveling within $\Gamma^{1}_{CRL}$ would possibly only face revoked pseudonyms with a lifetime falling in that interval, 6 pseudonyms, shown in black, instead of all 24 entries (the blurred pseudonyms are expired, thus not included in the \ac{CRL}). These 6 revoked pseudonyms within $\Gamma^{1}_{CRL}$ can be implicitly bound without compromising their unlinkability prior to the revocation event, in a way that one can simply derive subsequent pseudonyms from an anchor (the blurred pseudonyms are non-revoked but expired and they cannot be linked to the revoked ones). Thus, in this example, distributing 3 entries for that vehicle is sufficient. Another vehicle, however, traveling for a longer duration, from the middle of {\small $\Gamma^{1}_{CRL}$} till the beginning of {\small $\Gamma^{3}_{CRL}$}, would need to be provided with all 24 revocation entries, i.e., requiring 9 entries to derive all 24 revoked pseudonyms.

\begin{figure} [!t]
	\begin{center}
		\centering
		\subfloat[Revoked pseudonyms.] {
			\hspace{-1.5em}
			\includegraphics[trim=0cm 0cm 0cm 0cm, clip=true, width=0.2\textwidth,height=0.2\textheight,keepaspectratio]{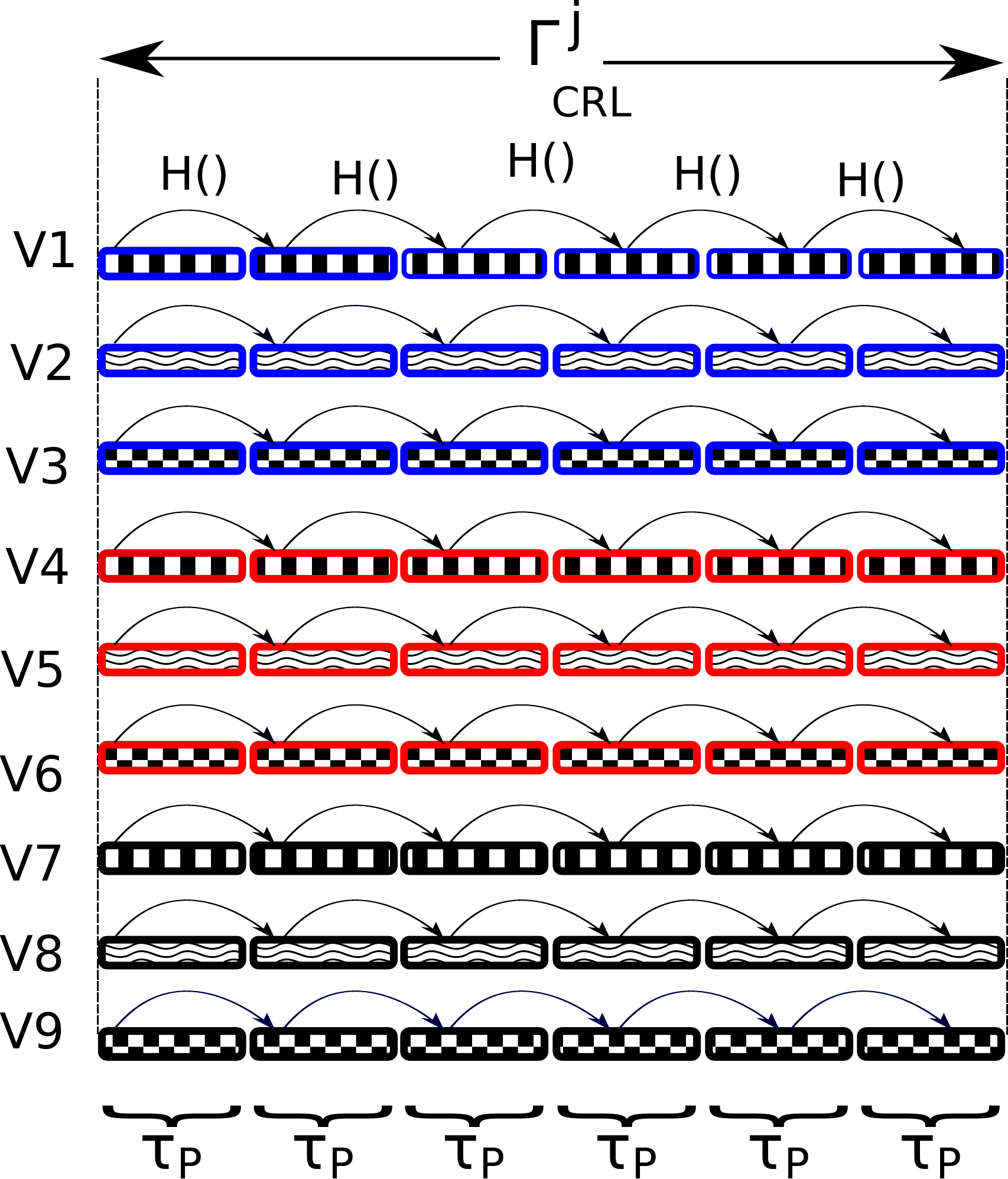}}
		\subfloat[A \ac{CRL} fingerprint construction.] { 
			\hspace{0.75em}
			\includegraphics[trim=0cm 0cm 0cm 0cm, clip=true, width=0.273\textwidth,height=0.273\textheight,keepaspectratio]{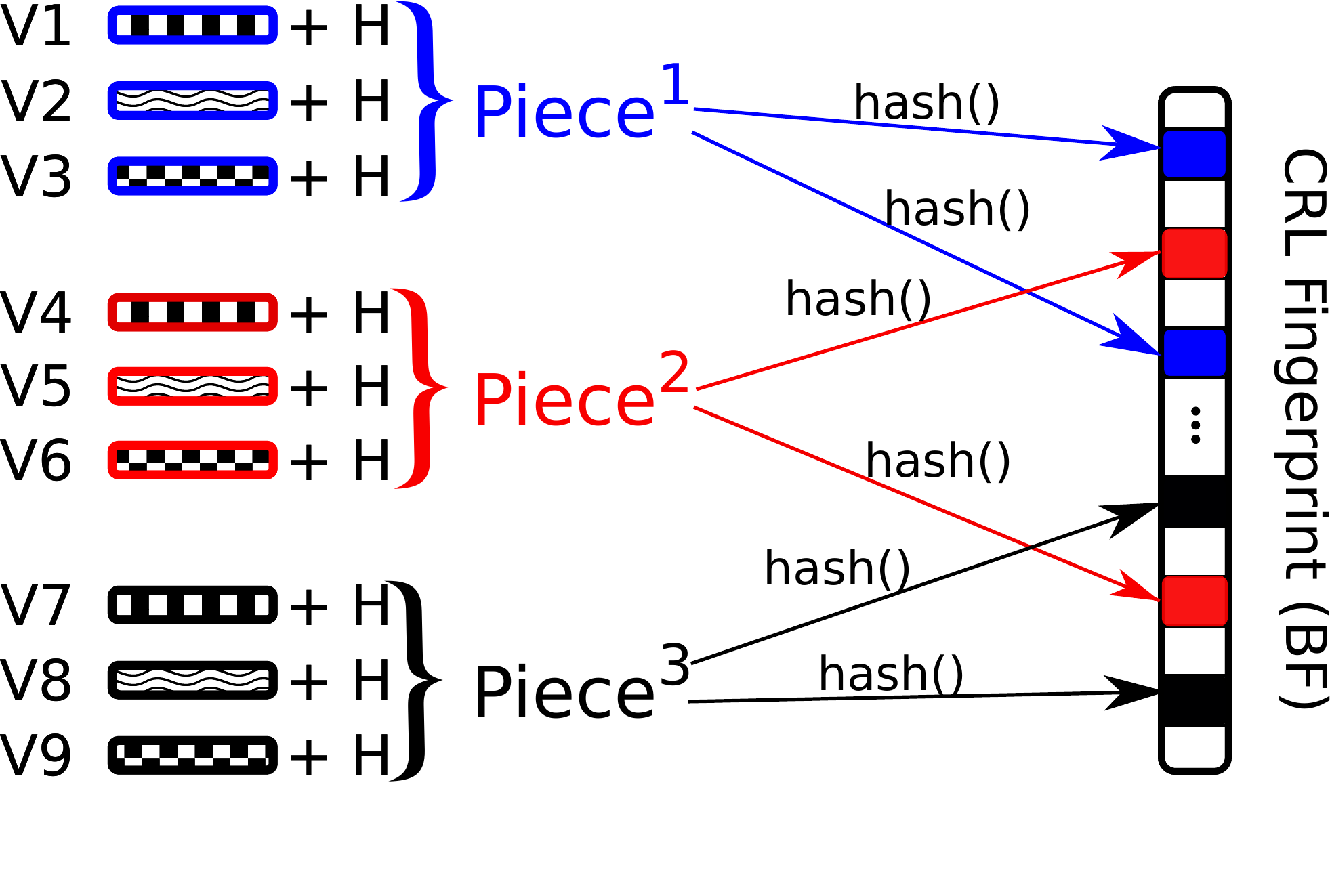}}
		\vspace{-0.25em}
		\caption{(a) Revoked pseudonyms in a $\Gamma^{j}_{CRL}$. (b) Fingerprint construction.} 
		\label{fig:crl-dis-bloomfilter-construction-by-pca}
	\end{center}
	\vspace{-2em}
\end{figure}

Fig.~\ref{fig:crl-dis-bloomfilter-construction-by-pca} shows how the \ac{PCA} condenses the revocation information corresponding to nine compromised vehicles, each having 6 pseudonyms within a $\Gamma^{j}_{CRL}$ interval. The \ac{PCA} utilizes the hash chain value for each set of revoked pseudonyms, generated during the pseudonym acquisition process~\cite{khodaei2018VehicleCentric}. Exactly because downloading a large \ac{CRL} file is challenging, with vehicles and \acp{RSU} disseminating or exchanging the \ac{CRL} over short contact (vehicle-to-vehicle or vehicle-to-\ac{RSU}) periods, the \ac{PCA} splits a large \ac{CRL} into several pieces constructed by including the serial number of the first revoked pseudonym along with the corresponding complementary\footnote{The complementary information is constructed by the \ac{PCA} during pseudonym acquisition process: the \ac{PCA} implicitly correlates a batch of pseudonyms belonging to each requester. This essentially enables efficient distribution of the \ac{CRL} because the \ac{PCA} only needs to include one entry per batch of pseudonyms without compromising their unlinkability. We refer interested readers to our earlier work~\cite{khodaei2018VehicleCentric} for the detailed protocol description.} information. In order to ensure the authenticity and integrity of each \ac{CRL} piece, the trivial solution is to sign each piece. However, this would incur significant computation and communication overhead, which grows linearly with the number of \ac{CRL} pieces, both for the \ac{VPKI} and for the receiving vehicles.

For efficient validation of \ac{CRL} pieces, the \ac{PCA} condenses authentication information for all the \ac{CRL} pieces by constructing a probabilistic data structure, i.e., a \acf{BF}. This condensed fingerprint is signed by the \ac{PCA} and periodically broadcasted by \acp{RSU}; moreover, it is also integrated into a subset of recently issued pseudonyms, which are broadcasted along with \acp{CAM}. This notifies other vehicles about a new revocation event (and thus an updated \ac{CRL}) as well as facilitates very fast validation of \ac{CRL} pieces. This authenticator (or notification) is a condensed authenticator for the \ac{CRL} pieces, and it can be broadcasted within a region very fast. Pseudonyms by default have to be authenticated by any vehicle. This implies validation of \ac{CRL} pieces almost for free. Note that the \ac{PCA} does not compress the \ac{CRL} using a \ac{BF}; rather, it utilizes a \ac{BF} for efficiently validating \ac{CRL} pieces. In fact, our vehicle-centric scheme does not need to compress \acp{CRL} because it effectively and progressively distributes revocation information corresponding to the actual vehicle trip duration.

\begin{figure} [!t]
	\vspace{-0em}
	\begin{center}
		\centering
		\includegraphics[trim=0cm 1.5cm 0cm 0cm, clip=true, width=0.45\textwidth,height=0.45\textheight,keepaspectratio]{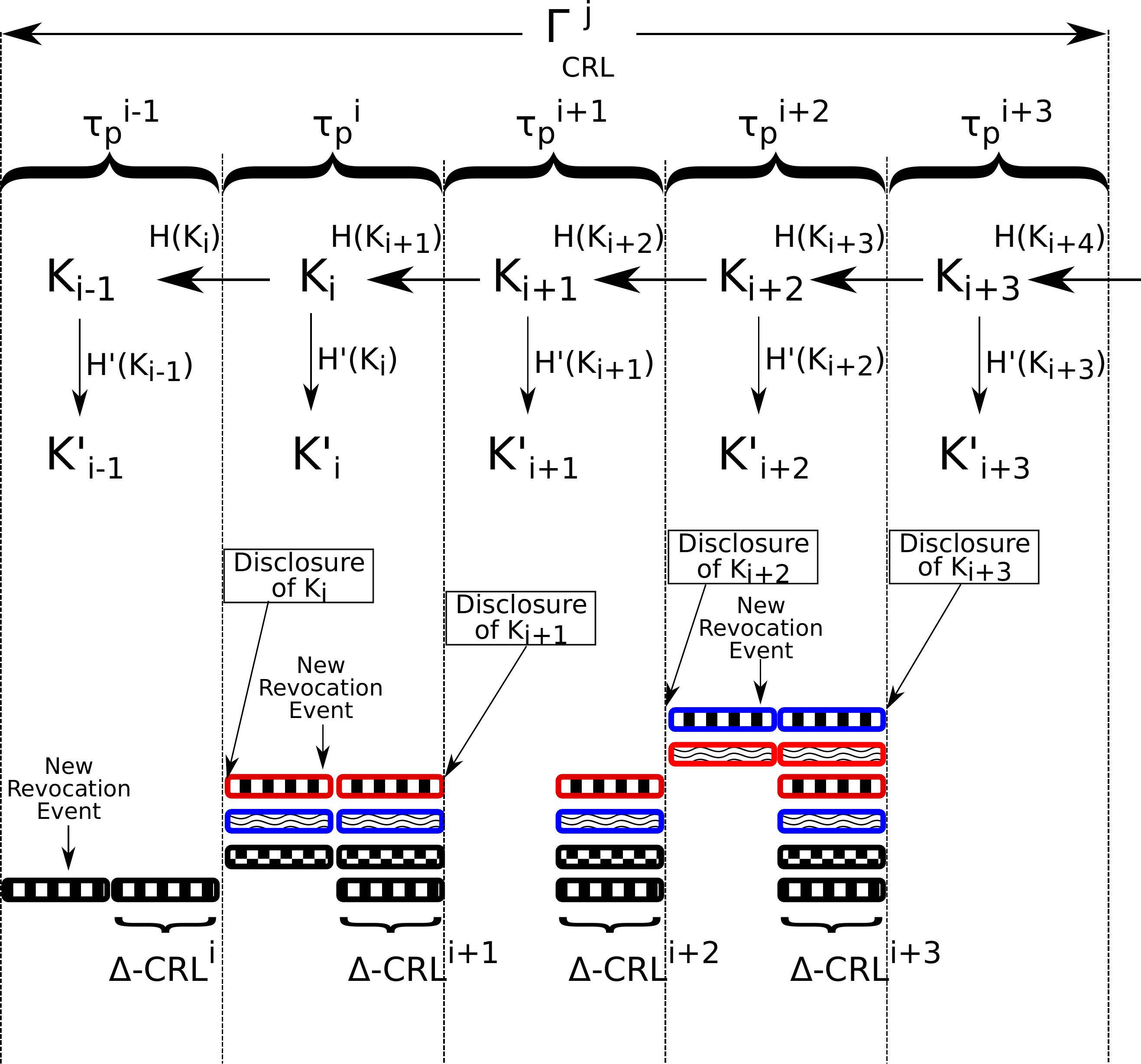}
		\vspace{-0.25em}
		\caption{Vehicle-centric $\Delta\textnormal{-}$\ac{CRL} distribution.}
		\label{fig:crl-dis-vehicle-centric-delta-crl}
	\end{center}
	\vspace{-2em}
\end{figure}

\textbf{$\Delta\textnormal{-}$\ac{CRL} updates:} Each $\Gamma_{CRL}$ interval could potentially have a base-\ac{CRL}, including non-expired revoked pseudonyms whose lifetimes fall within that $\Gamma_{CRL}$ interval. If new revocation event happen within a given $\Gamma_{CRL}$ interval, the \ac{PCA} releases $\Delta\textnormal{-}$\acp{CRL} to narrow down the vulnerability window, i.e., the interval between a revocation event until the time all vehicles successfully obtain the latest revocation information. $\Delta\textnormal{-}$\acp{CRL} disseminate the revoked but not-yet-expired pseudonyms that are not included in the base-\ac{CRL}. To ensure the authenticity of each $\Delta\textnormal{-}$\ac{CRL} (piece), a trivial solution would be for the \ac{PCA} to either sign each piece, i.e., the baseline scheme, or construct another fingerprint and sign it, i.e., similar to vehicle-centric \ac{CRL} distribution. However, due to unpredictable nature and potentially multiplicity of the revocation events, and more important, resilience considerations, i.e., performing a signature flooding attack~\cite{hsiao2011flooding} by broadcasting fake pieces, such solutions are not desirable.

Fig.~\ref{fig:crl-dis-vehicle-centric-delta-crl} shows how the latest revocation information is disseminated within a $\Gamma^{j}_{CRL}$ interval. The \ac{PCA} creates a hash chain, $N$ times, for each $\Gamma^{j}_{CRL}$, where $N$ is $\frac{\Gamma_{CRL}}{\tau_p}$ and it assigns the hash values sequentially to the time intervals, i.e., one key per $\tau_{P}$. Each hash value represents a cryptographic key to compute a \ac{MAC}, similar to~\cite{perrig2002tesla}, for all the $\Delta\textnormal{-}$\ac{CRL} pieces for that interval. The hash anchor is signed by the \ac{PCA} and is included in the base-\ac{CRL} pieces. Each $\Delta\textnormal{-}$\ac{CRL} piece contains the most recent revocation information, not included in the base-\ac{CRL}, and a \ac{MAC} of that piece. The \ac{PCA} then reveals these hash values in the reverse order of generation. At every pseudonym transition process\footnote{Pseudonym transition is the process of switching from one pseudonym (and the corresponding private key) to another one (ideally, non-previously used) to ensure message unlinkability.}, the \ac{PCA} discloses the most recent one-way hash value, i.e., the key to the \ac{MAC}, that it can disclose.

As Fig.~\ref{fig:crl-dis-vehicle-centric-delta-crl} shows, a new revocation event happened in interval $i-1$, e.g., a vehicle was compromised and it disseminates erroneous information across the network. Thus, all credentials corresponding to that vehicle should be revoked. The \ac{PCA} constructs a $\Delta\textnormal{-}$\ac{CRL} by including the serial number of the revoked pseudonym to be used during interval $i$. The \ac{PCA} distributes the $\Delta\textnormal{-}$\ac{CRL} pieces to be used during interval $i$ during the interval $i-1$. Upon a pseudonym transition, the \ac{PCA} discloses $K_i$, the cryptographic key to compute the \ac{MAC} of the $\Delta\textnormal{-}$\ac{CRL} pieces. In the next interval $i$, there is a new revocation event and three more vehicles are compromised. The \ac{PCA} constructs a $\Delta\textnormal{-}$\ac{CRL} piece, corresponding to interval $i+1$, and includes four entries (including the revoked one from the previous event). Again, upon a pseudonym transition, the \ac{PCA} reveals $K_{i+1}$ to authenticate the $\Delta\textnormal{-}$\ac{CRL} pieces corresponding to the interval $i+1$. Note that in order to facilitate a fine-grained and reversible pseudonym revocation status, the \ac{PCA} does not disclose the complementary information for the revoked pseudonyms in each $\Delta\textnormal{-}$\ac{CRL} piece (unlike the base-\acp{CRL}). Rather, it progressively and accumulatively distributes the revocation information in each pseudonym interval. This way, upon resolving the issues that led to the revocation of a misbehaving or malfunctioning vehicle, the \ac{PCA} can easily stop including the serial number of the pseudonyms, belonging to that vehicle, into the $\Delta\textnormal{-}$\acp{CRL}.

Note that within each time interval ($\tau^{i}_{P}$), the \ac{PCA} distributes $\Delta\textnormal{-}$\ac{CRL} pieces including the serial number of pseudonyms, whose lifetimes fall within the next time interval ($\tau^{i+1}_{P}$). However, vehicles cannot authenticate the $\Delta\textnormal{-}$\ac{CRL} pieces until the \ac{PCA} discloses the cryptographic key for that interval. Since vehicles know the schedule for disclosing the secret key (every $\tau_{P}$) and they are loosely synchronized with the \ac{VPKI} clock, they wait for the delayed disclosure of keys at every pseudonym transition\footnote{Note that we assume the \ac{VPKI} issues time-aligned pseudonyms, with non-overlapping interval, for all vehicles; thus, all vehicles know the exact time key disclosure by the \ac{PCA} and they change their pseudonyms at the same time~\cite{khodaei2018Secmace}.}. Upon receiving the secret key, vehicles can validate the buffered $\Delta\textnormal{-}$\ac{CRL} piece(s). However, after the disclosure of the secret key for an interval, vehicles do not accept any $\Delta\textnormal{-}$\ac{CRL} piece because such pieces could have been manipulated, i.e., forged by an adversary with the knowledge of the recently released key. Note that \acp{RSU} push $\Delta\textnormal{-}$\acp{CRL} as well as the corresponding secret keys throughout the entire \ac{VC} systems; as a result, the distribution of $\Delta\textnormal{-}$\acp{CRL} in this way does not affect (change) the distribution of base-\acp{CRL}.

This flexible design allows to temporarily evict an entity from the system: during each $\tau_{P}$, the \ac{PCA} only distributes serial numbers of revoked pseudonyms, whose validity intervals fall within the successive $\tau_{P}$. Upon resolving the misbehavior or malfunctioning, the \ac{PCA} is informed (by the \ac{RA}) and it does not include the serial number of the previously revoked entity into the \acp{CRL} or $\Delta\textnormal{-}$\acp{CRL}. Thus, our vehicle-centric scheme provides a flexible, reversible, and fine-grained revocation management towards evicting a malfunctioning or misbehaving vehicle for an interval until the issue is resolved. Moreover, this procedure does not harm user privacy: the \ac{PCA} cannot identify the actual identity of a misbehaving vehicle; also, upon resolving the issue, the \ac{PCA}, as the issuer, can only link successive pseudonyms until the end of $\Gamma$; thus, user privacy is not degraded. Further discussion on security and privacy in Sec.~\ref{sec:crl-dis-scheme-analysis-evaluation}.

\subsection{Security Protocols}
\label{subsec:crl-dis-security-protocols}

In a nutshell, the \acp{PCA} operating in a domain construct the \acp{CRL}~\cite{khodaei2018VehicleCentric} and $\Delta\textnormal{-}$\acp{CRL} by sorting the revoked pseudonyms based on their validity periods in a $\Gamma_{CRL}$ interval and push them to the \acp{RSU}. For ease of exposition, we assume there is one \ac{PCA}, even though the extension with multiple \acp{PCA} within a given domain is straightforward. \acp{RSU} and fingerprint-carrier peers publish the \ac{CRL}-update notification and the \ac{CRL} pieces (Sec.~\ref{subsubsec:crl-dis-operations-for-publishing-crl}). Moreover, \acp{RSU} push $\Delta\textnormal{-}$\acp{CRL} and the \ac{MAC} secret keys throughout the entire \ac{VC} systems (Protocol~\ref{protocol:crl-dis-delta-crl-construction-algo} in Sec.~\ref{subsubsec:crl-dis-operations-for-delta-crl}). Upon receiving a new revocation event, each vehicle broadcasts a query to its neighbors to fetch the (missing) pieces of the \ac{CRL}/$\Delta\textnormal{-}$\ac{CRL}, e.g., similarly to~\cite{das2004spawn}, corresponding to its actual trip duration (Sec.~\ref{subsubsec:crl-dis-operations-for-crl-subscription}). Finally, it parses recovered \ac{CRL} pieces and stores them locally (Protocol~\ref{crl-dis-algo-parsing-crl-pieces-by-vehicels} in Sec.~\ref{subsubsec:crl-dis-vehcile-operation-for-parsing-crl-pieces}). Due to space limitations, we refer readers to our prior work~\cite{khodaei2018VehicleCentric} for the pseudonym acquisition process and \ac{CRL} construction protocols. The notation is given in Table~\ref{table:crl-dis-protocols-notation}.

\begin{table}
	\centering
	\caption{Notation Used in the Protocols.}
	\vspace{-1.5em}
	\label{table:crl-dis-protocols-notation}
	\hspace{-0.95em}
	\resizebox{0.51\textwidth}{!}
	{
		\renewcommand{\arraystretch}{1.0001}
		\begin{tabular}{ | c | c || c | c | }
			\hline
			\textbf{Notation} & \textbf{Description} & \textbf{Notation} & \textbf{Description} \\\hline\hline
			$(P^{i}_{v})_{pca}$, $P^{i}_{v}$ & a valid psnym signed by the \acs{PCA} & $\gamma$ & frequency of distribution/broadcasting \\\hline 
			\shortstack{$(K^i_v, k^i_v)$} & \shortstack{psnym pub./priv. key pairs} & $p$ & \ac{BF} false positive rate \\\hline 
			$SN$, $SN_{P}$ & psnym serial number & m & \ac{BF} size \\\hline 
			\shortstack{$(msg)_{\sigma_{v}}$} & \shortstack{signed msg with vehicle's priv. key} & $k$ & \ac{BF} optimal hash functions \\\hline 
			$LTC$ & \acl{LTC} & \shortstack{$\Gamma$} & \shortstack{interval to issue time-aligned psnyms} \\\hline 
			$t_{now}, t_s, t_e$ & a fresh, starting, ending timestamp & $\Gamma_{CRL}$ & interval to release \acp{CRL} \\\hline 
			$Sign(Lk_{ca}, msg)$ & signing a msg with \acs{CA}'s priv. key & $\mathbb{B}$ & max. bandwidth for \ac{CRL} distribution \\\hline  
			$Verify(LTC_{ca}, msg)$ & verifying with the \acs{CA}'s pub. key & $\mathbb{R}$ & revocation rate \\\hline 
			$H()/H'()/H^{k}(), H$ & hash function ($k$ times), hash value & N & number of \ac{CRL} pieces in a $\Gamma_{CRL}$ interval \\\hline 
			$Append()$ & adding a revoked psnym to \acp{CRL} & n & number of remaining psnyms in each batch \\\hline
			$MAC(), MAC$ & Keyed hash function/value & $CRL_{v}$ & \ac{CRL} version \\\hline 
			$K_{i}$ & commitment key for interval $i$ & $\emptyset$ & Null or empty vector \\\hline 
			$K'_{i}$ & key to compute MAC for interval $i$ & $\tau_{P}$ & pseudonym lifetime \\\hline 
			BFTest() & \ac{BF} membership test & w, i, j, $\zeta$ & temporary variables \\\hline 
		\end{tabular} 
	}
	\vspace{-0em}	
\end{table}

\subsubsection{\textbf{Publishing the \ac{CRL}}}
\label{subsubsec:crl-dis-operations-for-publishing-crl}

Each \ac{RSU} continuously broadcasts the signed fingerprint of \ac{CRL} pieces, to notify vehicles in a region about any new revocation event. The transmission rate of the signed fingerprint corresponding to the current $\Gamma^{i}_{CRL}$ can gradually decrease towards the end of $\Gamma^{i}_{CRL}$; instead, the transmission rate of the signed fingerprint for $\Gamma^{i+1}_{CRL}$ can moderately increase. This ``ensures'' that all legitimate vehicles are notified about a new revocation event, thus being capable to request and efficiently validate \ac{CRL} pieces. Upon reception and validation of a query, an \ac{RSU} commences transmission across the wireless data link with a low-rate transmission (without any acknowledgment from peers). Upon receiving an authentic query for the missing \ac{CRL} pieces by a neighboring vehicle, a vehicle searches its local repository and randomly chooses one of the requested pieces and broadcasts it. A detailed protocol description in~\cite{khodaei2018VehicleCentric}.

\subsubsection{\textbf{$\Delta\textnormal{-}$\ac{CRL} Construction} (Protocol~\ref{protocol:crl-dis-delta-crl-construction-algo})}
\label{subsubsec:crl-dis-operations-for-delta-crl}

Upon a new revocation event within a $\Gamma_{CRL}$ interval, the \ac{PCA} distributes $\Delta\textnormal{-}$\ac{CRL} pieces to all vehicles. The number of revocation entries in a $\Delta\textnormal{-}$\ac{CRL} is proportional to the number of compromised vehicles and the number of revocation events. The \ac{PCA} constructs $\Delta\textnormal{-}$\ac{CRL} pieces by including all recently revoked pseudonyms whose lifetimes are valid for the next pseudonym interval ($\tau^{i}_{P}$) (steps~\ref{protocol:crl-dis-delta-crl-construction-algo}.2\textendash~\ref{protocol:crl-dis-delta-crl-construction-algo}.8). The \ac{PCA} then derives the corresponding \ac{MAC} keys\footnote{TESLA uses different hash functions to derive key $K_{i-1}$ and to compute \acp{MAC} to mitigate potential vulnerabilities of using the same key for different cryptographic operations~\cite{perrig2002tesla}.} (steps~\ref{protocol:crl-dis-delta-crl-construction-algo}.9\textendash~\ref{protocol:crl-dis-delta-crl-construction-algo}.10) and it splits the $\Delta\textnormal{-}$\ac{CRL} into multiple pieces according to the maximum allocated bandwidth, i.e., system parameter $\mathbb{B}$, for \ac{CRL} distribution (steps~\ref{protocol:crl-dis-delta-crl-construction-algo}.11\textendash~\ref{protocol:crl-dis-delta-crl-construction-algo}.13). It then appends an authenticator to each $\Delta\textnormal{-}$\ac{CRL} piece by calculating $MAC(K'_{i}, Piece^{\Delta^{w}_{i}}_{\Gamma_{CRL}^{j}})$ (step~\ref{protocol:crl-dis-delta-crl-construction-algo}.14). Note that $\Delta\textnormal{-}$\acp{CRL} for an interval $i$, i.e., $\tau^{i}_{P}$, are distributed within interval $i-1$, i.e., $\tau^{i-1}_{P}$, and the secret key $K'_{i}$ is distributed upon pseudonym transition (from $\tau^{i-1}_{P}$ to $\tau^{i}_{P}$).

\setlength{\textfloatsep}{0pt}
\begin{algorithm}[!t]
	\floatname{algorithm}{Protocol}
	\caption{$\Delta\textnormal{-}$\ac{CRL} Construction (by the \acs{PCA})}
	\label{protocol:crl-dis-delta-crl-construction-algo}
	\begin{algorithmic}[1]
		\Procedure{GenDeltaCRL}{$\Gamma_{CRL}^{j}, i, K_{i}, \mathbb{B}, t_{now}$}
		\myState{$Piece^{\Delta_{i}}_{\Gamma_{CRL}^{j}} \gets \emptyset$}
		\Repeat \Comment{{\scriptsize Fetching revoked pseudonym, not included in base-\ac{CRL}}}
		\vspace{-0.1em}
		\myState{$SN_{P} \gets fetchRevokedPsnyms(\Gamma_{CRL}^{j}, i, t_{now})$} 
		\If {$SN_{P} \not= Null$} 
		\myState{$Piece^{\Delta_{i}}_{\Gamma_{CRL}^{j}} \gets Append(SN_{P})$}
		\EndIf
		\Until{$SN_{P} == Null$} 
		\myState{$K_{i-1} \gets H(K_{i}) $} \Comment{{\scriptsize Calculating the key for interval $i-1$}} 
		\myState{$K'_{i} \gets H'(K_{i}) $} \Comment{{\scriptsize Calculating the key for interval $i$}} 
		\myState{$N \gets {\scriptsize {\Bigg \lceil} \dfrac{size(Piece^{\Delta_{i}}_{\Gamma_{CRL}^{j}})}{\mathbb{B}} {\Bigg \rceil}} $} \Comment{{\scriptsize Calculating number of pieces}}
		\For{$w\gets 0, N$} \Comment{{\scriptsize N: number of pieces}} 
		\myState{$\zeta \gets Split(Piece^{\Delta_{i}}_{\Gamma_{CRL}^{j}}, \mathbb{B}, N)$} 
		\myState{$Piece^{\Delta^{w}_{i}}_{\Gamma_{CRL}^{j}} \gets \{\zeta || MAC(K'_{i}, \zeta) || K_{i-1}\} $} 
		\EndFor
		\myState{\textbf{return} $\{(Piece^{\Delta^{1}_{i}}_{\Gamma_{CRL}^{j}}),\dots,(Piece^{\Delta^{N}_{i}}_{\Gamma_{CRL}^{j}})\}$}
		\EndProcedure
	\end{algorithmic}
\end{algorithm}


\subsubsection{\textbf{\ac{CRL} Subscription}} 
\label{subsubsec:crl-dis-operations-for-crl-subscription}

Each vehicle can receive necessary \ac{CRL} pieces corresponding to its actual trip duration from nearby \acp{RSU} or neighboring vehicles. A vehicle broadcasts a signed query to its neighbors, to receive the missing pieces of the revocation information of $\Gamma^{i}_{CRL}$ during which the vehicle wishes to travel. Having received a \ac{CRL} piece, it simply validates the piece by testing against the signed fingerprint (already obtained from \acp{RSU} or integrated in a subset of recently issued pseudonyms broadcasted in the network). If the \ac{BF} test is successful, it accepts that piece and keeps requesting until successfully receiving all remaining pieces. A detailed protocol description is available in~\cite{khodaei2018VehicleCentric}. In case of $\Delta\textnormal{-}$\ac{CRL}, each vehicle should buffer all received $\Delta\textnormal{-}$\ac{CRL} pieces with appended \acp{MAC} in order to validate them upon disclosure of the corresponding key. When the \ac{PCA} discloses the key at every $\tau_{P}$, each vehicle computes the \ac{MAC} of each piece using that key. If the two \acp{MAC} are the same, then the $\Delta\textnormal{-}$\ac{CRL} piece would be accepted; otherwise, dropped.


\subsubsection{\textbf{Parsing \ac{CRL} (Protocol~\ref{crl-dis-algo-parsing-crl-pieces-by-vehicels})}}
\label{subsubsec:crl-dis-vehcile-operation-for-parsing-crl-pieces}

Upon reception and validation of a \ac{CRL} piece, each vehicle derives the revoked pseudonym serial numbers from the obtained hash anchors, by calculating a hash value $n$ times: {\small $H(SN_{z} || H_{z}^{w}(Rnd_{z})$} (steps~\ref{crl-dis-algo-parsing-crl-pieces-by-vehicels}.2\textendash~\ref{crl-dis-algo-parsing-crl-pieces-by-vehicels}.10). Revocation entries can be stored in local storage, and searched with $O(log(n))$ time complexity. To enhance revocation status validation, a vehicle could generate a \ac{BF} locally~\cite{haas2011efficient} with constant computational cost ($O(1)$) for insertions and search operations but at a cost of a false positive rate. Note that the search operation is very efficient because revocation entries are sorted for the period they are valid for, i.e., in a $\tau_{P}$ interval.

\setlength{\textfloatsep}{0pt}
\begin{algorithm}[!t]
	\floatname{algorithm}{Protocol}
	\caption{Parsing a \ac{CRL} Piece (by the \acp{OBU})}
	\label{crl-dis-algo-parsing-crl-pieces-by-vehicels}
	\begin{algorithmic}[1]
		\Procedure{ParseCRL}{$Piece^{j}_{\Gamma_{CRL}^i}, N$}
		\myState {$\{SN_{z}, Rnd_{z}, n_{z}\}_{_{_{_{N}}}} \gets Piece^{j}_{\Gamma_{CRL}^i}$} 
		\myState {$CRL_{\Gamma_{CRL}^{i}} \gets \emptyset$} 
		\For{$z\gets 1, N$}  \Comment{{\scriptsize N: Number of entries in this piece}}
		\For{$w\gets 1, n_{z}$}  \Comment{{\scriptsize n: Number of remaining pseudonyms}}
		\myState{$CRL_{\Gamma_{CRL}^{i}} \gets Append(H(SN_{z} || H_{z}^{w}(Rnd_{z})))$} 
		\myState{$SN_{z} \gets H(SN_{z} || H_{z}^{w}(Rnd_{z}))$} 
		\EndFor
		\EndFor
		\myState{\textbf{return} $CRL_{\Gamma_{CRL}^{i}}$}
		\EndProcedure
	\end{algorithmic}
\end{algorithm}
\afterpage{\global\setlength{\textfloatsep}{\oldtextfloatsep}}


\section{Scheme Analysis and Evaluation}
\label{sec:crl-dis-scheme-analysis-evaluation}

For an analysis on how our scheme satisfies the security and privacy requirements, as well as operational requirements, we refer interested readers to~\cite{khodaei2018VehicleCentric}. In this section, we expand the security and privacy analysis for the $\Delta\textnormal{-}$\ac{CRL} distribution and the \ac{CRL} fingerprint mechanisms. Then, we quantitatively demonstrate the efficiency, scalability, and resiliency of our vehicle-centric scheme, extending the results with respect to~\cite{khodaei2018VehicleCentric}, through an extensive experimental evaluation.

\subsection{Qualitative Analysis}
\label{subsec:crl-dis-qualitative-analysis}

\textbf{\emph{$\Delta\textnormal{-}$\ac{CRL} distribution:}} Upon a new revocation event and $\Delta\textnormal{-}$\ac{CRL} release, an external adversary can try to link the recently revoked pseudonyms backwards, i.e., towards the beginning of the $\Gamma$ interval. However, due to the utilization of a hash chain during the pseudonym issuance process~\cite{khodaei2018VehicleCentric}, it is infeasible to link a revoked pseudonym to the previously non-revoked pseudonyms. Moreover, vehicles can be loosely synchronized with the \ac{VPKI} clock, e.g., through \ac{GPS} used to synchronize \ac{OBU} clocks, in order to be informed about the time of key disclosure (by the \ac{PCA}). Furthermore, each vehicle does not disclose the trip duration. Rather, each vehicle only requests for obtaining \ac{CRL} pieces for the current $\Gamma_{CRL}$, which is a protocol/scheme selectable parameter. More so, the requests are anonymized, i.e., signed by the currently valid pseudonym. For the next $\Gamma_{CRL}$, each vehicle changes its pseudonym and requests to obtain \ac{CRL} pieces for the new interval using a new pseudonymous identifier. The pseudonyms are fully unlinkable, and there is no memory or linkage across such $\Gamma_{CRL}$ intervals. The signed requests do not reveal the actual identity of their owner, thus it does not harm user privacy.

Upon releasing $\Delta\textnormal{-}$\acp{CRL}, vehicle buffer all received pieces with the appended \acp{MAC} in order to validate them when the secret key is disclosed. However, an attacker could aggressively broadcast fake pieces of $\Delta\textnormal{-}$\ac{CRL} to mount a clogging \ac{DoS} attack. The longer the pseudonym lifetimes are, the higher the frequency of broadcast is, and the higher the number of adversaries is, the larger the storage space needed. For example, if there is an adversary broadcasting bogus pieces of $\Delta\textnormal{-}$\ac{CRL} (with $\mathbb{B}=$50 KB/s and $\tau_{P}= $ 5 min), then each neighboring vehicle needs~$\approx$15 MB of memory to store bogus \ac{CRL} pieces. One can apply a rate limiting mechanism by requesting the ``suspicious'' node to piggyback a fingerprint of the $\Delta\textnormal{-}$\ac{CRL} pieces in its successive \acp{CAM}; otherwise, all received packets from the suspicious sender would be dropped. As \acp{CAM} are already signed, any suspicious behavior could be reported to the \ac{VPKI} for further investigation.

\textbf{Reversible pseudonym revocation status:} When a misbehavior detection authority identifies a misbehaving vehicle, it queries the \ac{RA} to initiate a resolution and revocation process. The \ac{RA} queries the corresponding \ac{PCA} to retrieve the ticket, used to obtain that pseudonym. Assume a vehicle has obtained pseudonyms for a period, e.g., a week or a month. The \ac{RA} then \emph{progressively} interacts with the corresponding \ac{H-LTCA} towards obtaining all subsequent tickets issued for that vehicle\footnote{In case of a multi-domain \ac{VC} environment, one more step is required, i.e., the \ac{RA} interacts with all \acp{F-LTCA}~\cite{khodaei2018Secmace}.}. The \ac{RA} queries the \acp{PCA} towards revoking the pseudonyms issued for a given ticket\footnote{Using a ticket, one cannot identify the targeted \ac{PCA}~\cite{khodaei2018Secmace}; thus, the \ac{RA} queries all \acp{PCA} in a domain to revoke pseudonyms issued for that ticket. This strongly protects user privacy: collusion by an \ac{RA} and all \acp{PCA} does not reveal any information to harm user privacy~\cite{khodaei2018Secmace}.}. In case of resolving the stated misbehavior, the \ac{RA} informs the \ac{H-LTCA} to issue more tickets for that vehicle. Moreover, the \ac{RA} stops delivering a ticket (acquired from the \ac{H-LTCA} during pseudonym resolution process~\cite{khodaei2018Secmace}) to the \acp{PCA} towards revoking the pseudonyms issued for that ticket. In this way, a misbehaving entity can be temporarily evicted form the \ac{VC} system; furthermore, upon resolving the issue, the entity can re-enter the system by leveraging the previously obtained pseudonyms without necessarily obtaining a fresh batch of pseudonyms. This flexibility also allows a more efficient and effective \ac{CRL} distribution; at the same time, this prevents from overloading the \ac{VPKI} entities from issuing unnecessary new sets of pseudonyms.

Our vehicle-centric scheme protects user privacy due to separation of duties: no single \ac{VPKI} entity could fully de-anonymize a user by identifying the actual identity of or link successive pseudonyms belonging to a (partially) evicted vehicle. Collusion by an \ac{RA} and a \ac{PCA} does not reveal any information to link the users (long-term identities) with their pseudonyms because the tickets, issued by a \ac{H-LTCA}, are anonymized and they do not reveal their owners' identities. Collusion by an \ac{RA} and the \ac{H-LTCA} does not enable them towards linking the corresponding pseudonyms: time-aligned pseudonyms are issued for all vehicles, thus there is no distinction among pseudonym sets. Upon a misbehavior event, the \ac{H-LTCA} could only infer that a user was evicted for an interval without any other information towards linking, thus tracking, a vehicle. For a detailed information held by each honest-but-curious \ac{VPKI} entity, we refer readers to~\cite{khodaei2018Secmace}.

\textbf{Synchronization with the \ac{VPKI} clock:} Lack of synchronization between the vehicles and the \ac{VPKI} clock could affect integrity of the $\Delta$-\ac{CRL} distribution. More precisely, in case of drifting clocks of the some victim vehicles, an adversary can manipulate a \ac{CRL} piece. Upon disclosing the cryptographic key to compute the \ac{MAC}, the adversary crafts a bogus \ac{CRL} piece, e.g., excluding revoked pseudonym serial numbers from a \ac{CRL} piece or adding valid pseudonyms into a fake \ac{CRL} piece, and broadcast to the victim vehicles whose clock mildly drift from the \ac{VPKI} clock. Due to lack of synchronization, such vehicles would accept the bogus pieces. In order to mitigate such a misbehavior, it suffices to have vehicles periodically synchronizing their clocks with the \ac{VPKI} clock. For example, if the accuracy of an \ac{RTC} is 20 parts-per-million (ppm), i.e., $20 \times 10^{-6}$, and the maximum accepted error in timestamp is 1 sec, then each vehicle should synchronize its clock every 13.8 hours ($\frac{1 \times 10^3 sec}{20 \times 10^{-6} ppm}$), i.e., twice a day, which seems to be practical.

\begin{figure} [!t]
	\vspace{0em}
	\begin{center}
		\centering
		\subfloat[Vehicle-centric scheme]{
			\hspace{-1em} 
			\includegraphics[trim=0cm 0.15cm 0.5cm 1.45cm, clip=true, width=0.265\textwidth,height=0.265\textheight,keepaspectratio]{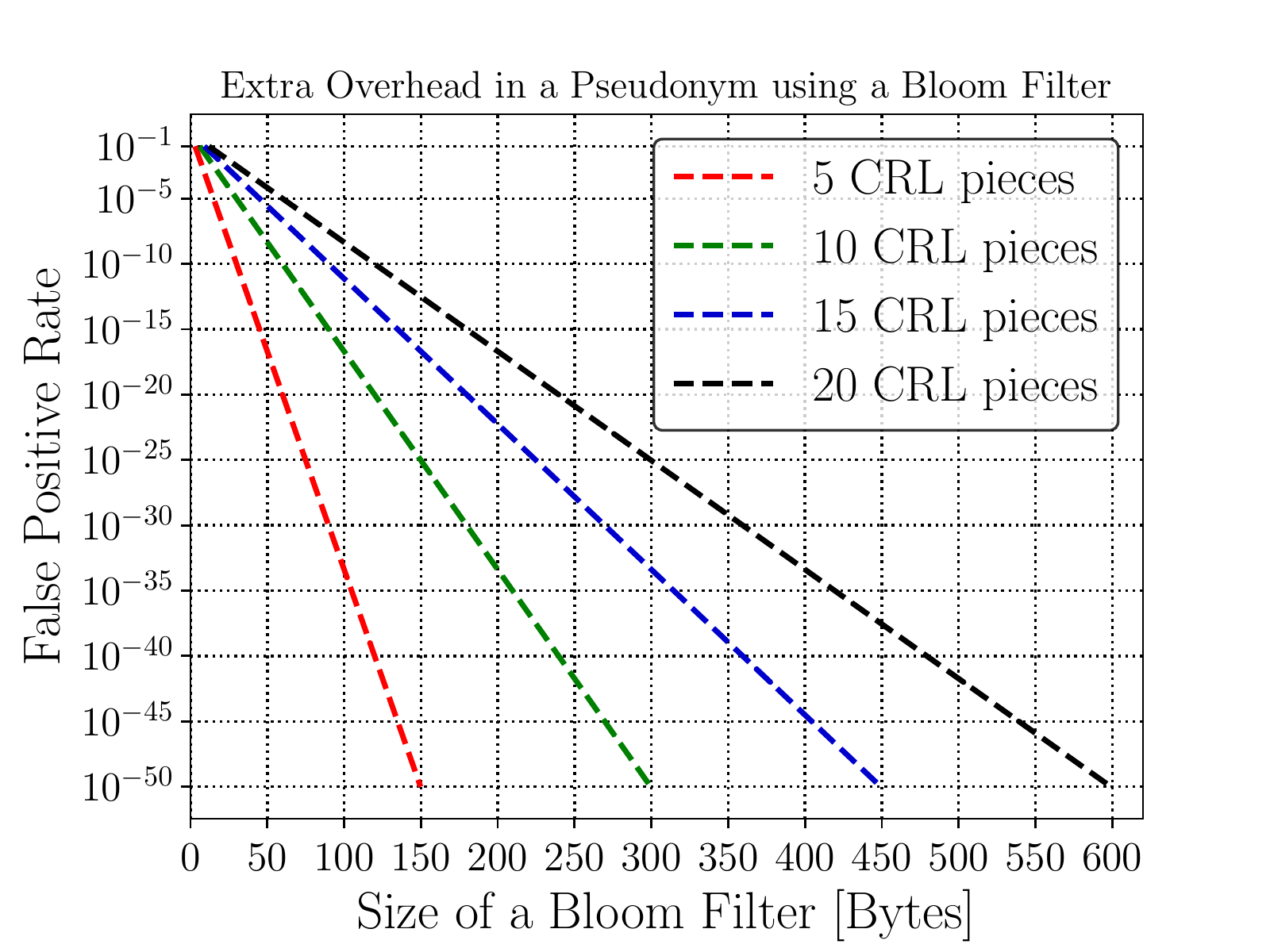}}
		\subfloat[Precode-and-hash scheme~\cite{nguyen2016secure}]{
			\hspace{-1.35em} 
			\includegraphics[trim=0cm 0.15cm 0.5cm 1.45cm, clip=true, width=0.265\textwidth,height=0.265\textheight,keepaspectratio]{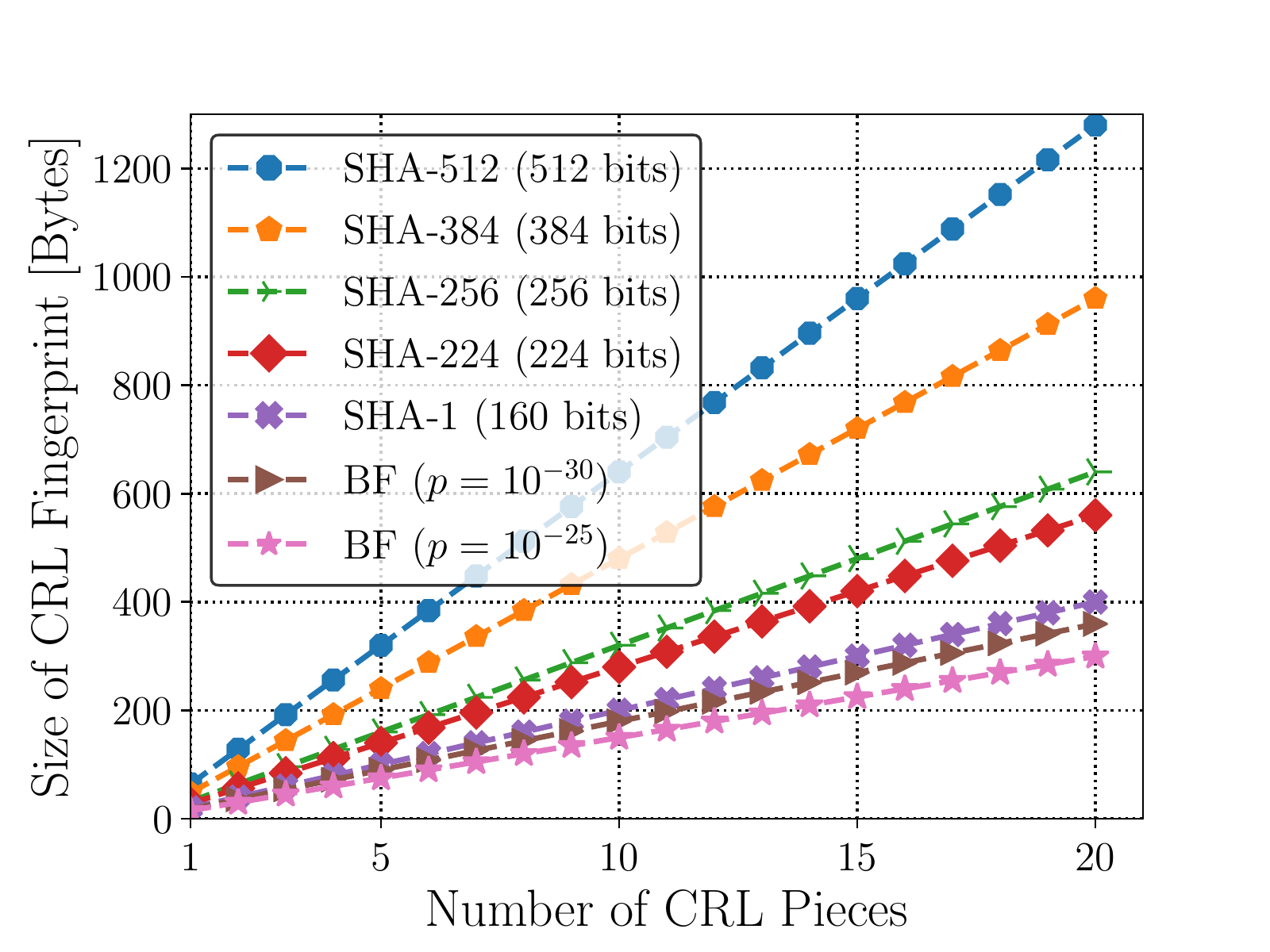}}
		\vspace{-0.25em}
		\caption{Extra overhead for \ac{CRL} fingerprints.}
		\label{fig:crl-dis-extra-overhead-for-crl-fingerprints}
	\end{center}
	\vspace{-1.95em}
\end{figure}

\textbf{Vulnerability window:} Even though our vehicle-centric scheme distributes \acp{CRL} and $\Delta\textnormal{-}$\acp{CRL}, there could be a small vulnerability window. For example, when a new vehicle joins the system within the current $\tau_p$ interval, the \ac{PCA} has already disclosed the cryptographic key corresponding to that interval; thus, that vehicle cannot rely on the received $\Delta\textnormal{-}$\ac{CRL} pieces because the key is already disclosed and the $\Delta\textnormal{-}$\ac{CRL} pieces could have been manipulated. Obviously, the shorter the pseudonym lifetimes are, the narrower the vulnerability window is. In general, there is a trade-off between closing down the vulnerability window and cost, notably communication overhead and deploying uninterrupted connectivity to the \ac{VPKI}, e.g., dense deployment of \acp{RSU} or leveraging cellular communications. Depending on the type of misbehavior and the fraction of recently joined vehicles, the \ac{VPKI} could opt in to geo-cast the signed revocation information at any point in time, even within a pseudonym lifetime. Alternatively, vehicles could request the neighboring \acp{RSU} or vehicles for the signed $\Delta\textnormal{-}$\ac{CRL} pieces. Note that fully closing down the vulnerability window requires an efficient revocation scheme combined with a persistent and reliable connectivity to the \ac{VPKI} entities, e.g., leveraging cellular-based \ac{V2X} communications~\cite{3GPP-TR-36.885-LTE-based-V2X}.

\textbf{\ac{CRL} and fingerprint size comparison:} Representing \ac{CRL} pieces in a space-efficient \ac{BF} trades off communication overhead for a false positive rate ($p$). Fig.~\ref{fig:crl-dis-extra-overhead-for-crl-fingerprints}.a shows that the \ac{BF} size linearly increases as the false positive rate decreases. For example, for 10 \ac{CRL} pieces covering one $\Gamma_{CRL}$ interval, and $p = 10^{-20}$ (with the optimal number of hash functions), the \ac{BF} size and thus the overhead for each pseudonym is 120 bytes. This eliminates the need to sign each \ac{CRL} piece.

The \ac{PCA} can concatenate the hash values for each \ac{CRL} piece~\cite{nguyen2016secure}. Fig.~\ref{fig:crl-dis-extra-overhead-for-crl-fingerprints}.b compares our \ac{BF}-based \ac{CRL} fingerprint size with the five approved hash algorithms~\cite{dang2008recommendation}: SHA-1, SHA-224, SHA-256, SHA-384 and SHA-512, each producing hash digest size of 160, 224, 256, 384 and 512 bits, respectively. For instance, by employing \emph{precode-and-hash}~\cite{nguyen2016secure} with SHA1 (20 bytes output size)~\cite{nguyen2016secure}, the size of a fingerprint for 20 \ac{CRL} pieces becomes 400 bytes; whereas employing our scheme results in an extra overhead of 311 bytes ($p=10^{-25}$) or 371 bytes for the extremely low false positive rate ($p=10^{-30}$). Alternatively, one can utilize truncated hash digests; however, truncated message digest must be carefully used: if the message digest length is too small, computation of pre-image, second pre-image or collisions becomes feasible~\cite{gerbet2015power}. All in all, truncation will not guarantee the expected security strength of a hash digest~\cite{dang2008recommendation}. A detailed \ac{CRL} size comparison with \ac{C$^2$RL} scheme~\cite{raya2006certificaterevocation, raya2007eviction, rigazzi2017optimized} can be found in our earlier work~\cite{khodaei2018VehicleCentric}.

\begin{figure} [!t]
	\vspace{1em}
	\begin{center}
		\centering
		\hspace{-2em}
		\includegraphics[trim=2cm 0cm 0cm 0cm, clip=true, width=0.35\textwidth,height=0.35\textheight,keepaspectratio]{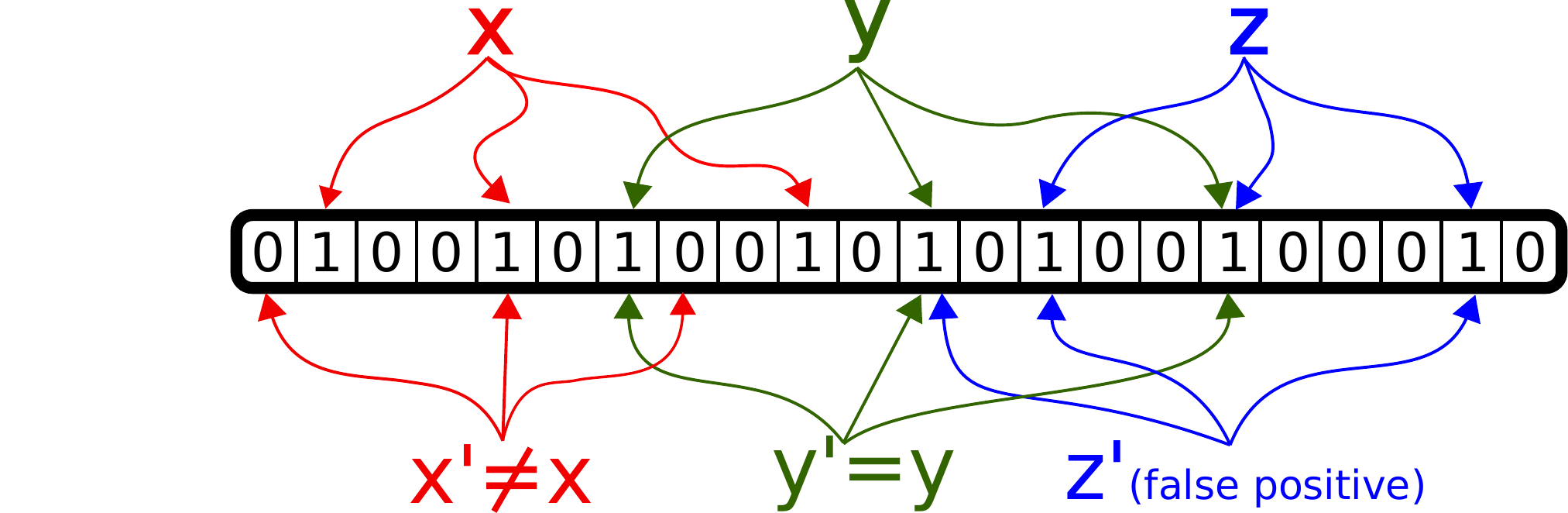}
		\vspace{-0.25em}
		\caption{\ac{BF} insertion and query (m=22 bits, k=3 hash functions, n=3 items).}
		\label{fig:crl-dis-bloom-filter-false-positive-rate-example}
	\end{center}
	\vspace{-1em}
\end{figure}

\textbf{\emph{Compromising the security of the \ac{CRL} fingerprint:}} A \ac{BF} is a space-efficient probabilistic data structure that is used for efficient membership query. It essentially provides condensed authenticators for the inserted items at the cost of false positives. Fig.~\ref{fig:crl-dis-bloom-filter-false-positive-rate-example} illustrates a \ac{BF} with insertion and query operations. In order to insert an item, e.g., $x$, into the \ac{BF}, we feed item $x$ into $k$ different hash functions $\{h_1, \dots, h_k\}$ to identify the corresponding bits of the \ac{BF}, i.e., $I_x=\{h_1(x) \: mod \: m, \dots, h_k(x) \: mod \: m\}$. In this example, the size of the \ac{BF}, $m$, is 22 bits and the number of hash functions, $k$, is 3. Items y, and z are also inserted into the \ac{BF}. In order to query an item to check if it belongs to the \ac{BF}, we check if the item has been inserted into the \ac{BF} by feeding it to the same hash functions. In this example, $x'$ does not belong to the \ac{BF}; $y'$ is equal to $y$ and thus it exists. However, $z'$ appears to be in the set while it has never been added, i.e., a false positive. The false positives arise due to the collision on the condensed hash digests and one can adjust the size of the \ac{BF} to achieve a desired false positive rate~\cite{mitzenmacher2002compressed, broder2004network}.

There are three types of attacks applicable to \acp{BF}~\cite{gerbet2015power}: \emph{chosen-insertion} attack, \emph{query-only} attack, and \emph{deletion} attack. In the chosen-insertion attack, an adversary can either add a new item into the \ac{BF}, or he could make the \ac{BF} constructor, i.e., the \ac{PCA}, insert a new item. In the query-only attack, an adversary targets the false positive rate of a \ac{BF} towards generating a fake \ac{CRL} piece to be accepted as legitimate. Finally, in the deletion attack, an adversary tries to delete an item, or make the \ac{PCA} delete it from the \ac{BF}. Note that this type of attack is for a specific form of \acp{BF} which allows deleting an item, e.g., \emph{counting} \acp{BF}~\cite{broder2004network}.

The chosen-insertion and query-only attacks are not applicable in our vehicle-centric \ac{CRL} distribution scheme exactly because the \ac{BF} is \emph{explicitly} signed by the \ac{PCA}, or it is integrated into a set of recently issued pseudonyms, i.e., \emph{implicitly} signed. However, the query-only attack is applicable: an adversary could try to generate a bogus \ac{CRL} piece in order to exclude revoked pseudonym serial numbers or add valid ones by forging a fake \ac{CRL} piece that passes the \ac{BF} test. Note that this is different from a pollution or a \ac{DDoS} attack: not only would it prevent a legitimate vehicle from obtaining a genuine \ac{CRL} piece, but also disseminate an \emph{authentic-looking} piece that passes the \ac{BF} test. In fact, such an attack relies on sheer computational power; its effectiveness depends on the computational resources allocated. The probability of generating a bogus \ac{CRL} piece, i.e., obtaining a false positive, is~\cite{bloom1970space, mitzenmacher2002compressed}: 

	\begin{equation*}
	P = \Bigg[1 - \bigg(1 - \dfrac{1}{m}\bigg)^{kn}\Bigg]^{k}
	\end{equation*}

\begin{figure} [!t]
	\vspace{1em}
	\begin{center}
		\centering
		\includegraphics[trim=0.1cm 0cm 0.4cm 1.35cm, clip=true, width=0.28\textwidth,height=0.28\textheight,keepaspectratio]{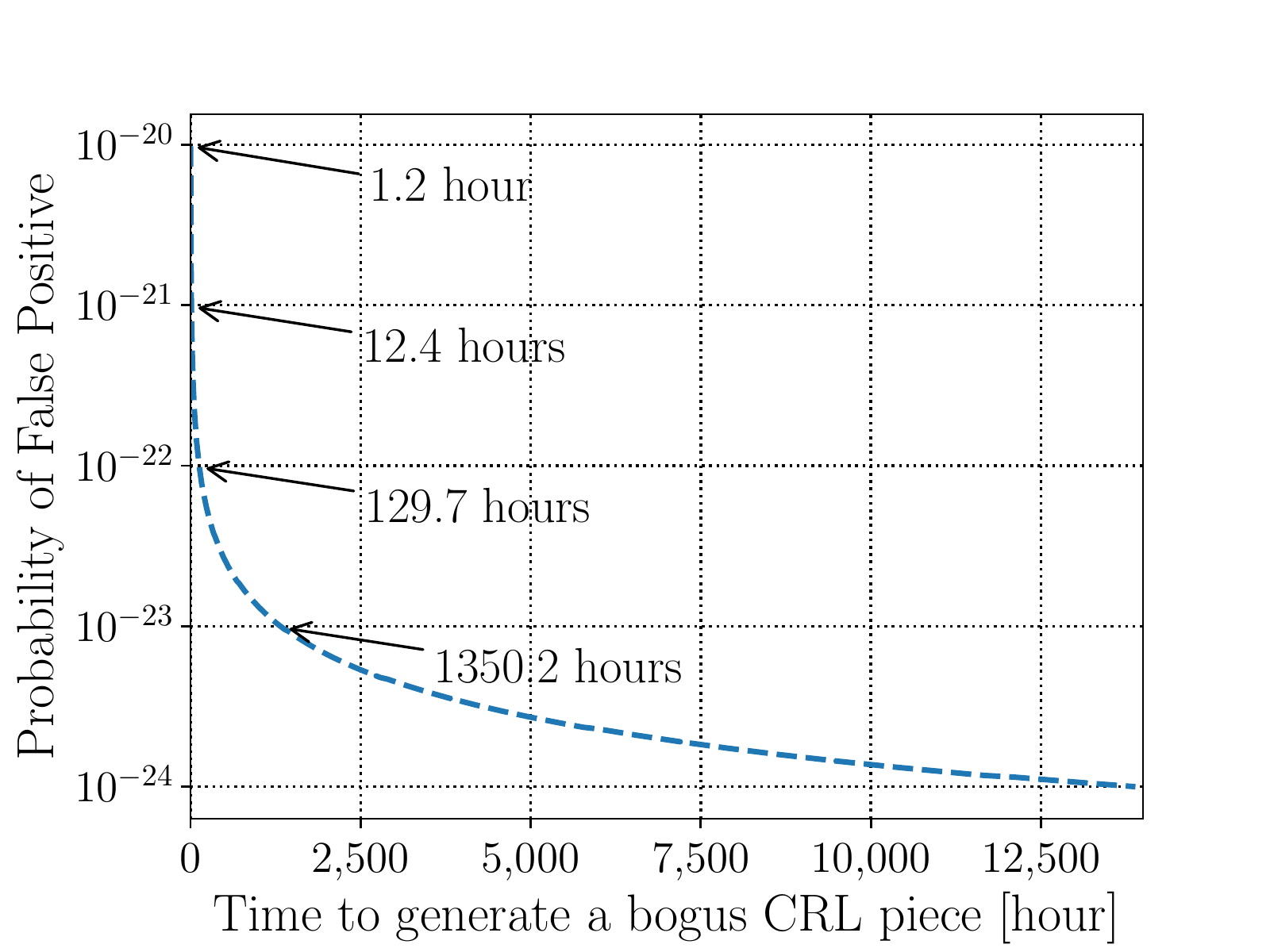}
		\vspace{-0.25em}
		\caption{Query-only attack on the \ac{CRL} fingerprints; adversary's computational power is $1.6 \times 10^{18}TH/sec$.}
		\label{fig:crl-dis-query-only-attack}
	\end{center}
	\vspace{-2em}
\end{figure}

Our scheme resists such attacks that attempt to exclude revoked pseudonym serial numbers or add valid ones by forging a fake \ac{CRL} piece that passes the BF test.\footnote{Generating a fake \ac{BF} (e.g., $p=10^{-20}$) with completely different valid pseudonyms serial number necessitates accessing at least, e.g., $10^{20}$, valid pseudonyms, i.e., a more powerful adversary (\emph{malicious \ac{VPKI} entities}), and is beyond the scope of our adversarial model.} An adversary could buy top-notch bitcoin-mining hardware, Antminer-S9~\cite{antminerS9Review} (14TH/s, \$3,000). If $\Gamma_{CRL}=1$ hour and $p=10^{-20}$, and the optimal number of hash functions, $K=67$, the adversary needs 132,936 Antminer-S9 (\$400M) to generate a bogus piece within a $\Gamma_{CRL}$ interval ({\small $\frac{10^{20}\times67}{14\times10^{12}}$}). Alternatively, he could join AntPool~\cite{antpoolchina}, one of the largest Bitcoin mining pools, ($1,604,608 \: TH/s$) to generate a fake piece. Fig.~\ref{fig:crl-dis-query-only-attack} shows the time to conduct the query-only attack towards generating a bogus \ac{CRL} piece. If $\Gamma_{CRL}=1$ hour and $p=10^{-20}$, and the optimal number of hash functions, $K=67$, the adversary could generate a bogus piece within 70 min, which might seem a practical threat. However, if $p=10^{-22}$ (with $K=73$) or even $p=10^{-23}$ (with $K=76$), the adversary would need 5 or 55 days, respectively ({\small $\frac{10^{22}\times73}{1.6\times10^{18}}=126h$, $\frac{10^{23}\times76}{1.6\times10^{18}}=1,319h$}). With inherently short $\tau_{P}$ (important for unlinkability and thus privacy) and $\Gamma_{CRL}$ interval, proper choice of $p$ makes attacks infeasible; in other words, irrelevant, as forged pieces refer to already expired credentials. Upon receiving conflicting pieces, vehicles report misbehavior to the \ac{VPKI} to take appropriate actions, e.g., adjusting $p$. The results of our experiments in Sec.~\ref{subsec:crl-dis-quantitative-analysis} rely on $p=10^{-30}$ and $K=100$.

\begin{figure} [!t]
	\vspace{-1em}
	\begin{center}
		\centering
		\subfloat[]{
			\hspace{-1.25em} 
			\includegraphics[trim=0.1cm 0.3cm 0.4cm 1.25cm, clip=true, width=0.266\textwidth,height=0.266\textheight,keepaspectratio]{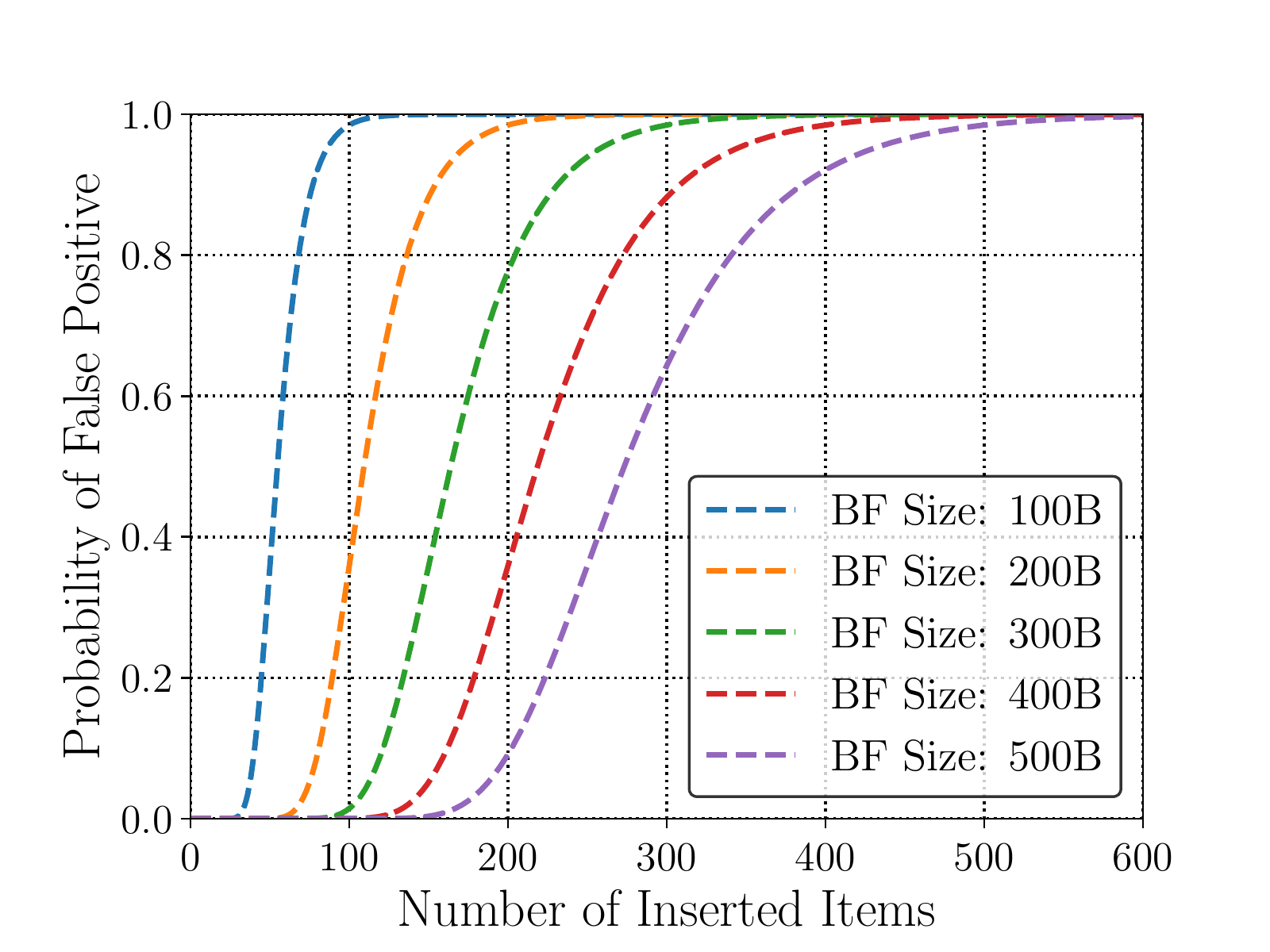}}
		\centering
		\subfloat[]{
			\hspace{-1.25em} 
			\includegraphics[trim=0cm 0.3cm 0.4cm 1.25cm, clip=true, width=0.266\textwidth,height=0.266\textheight,keepaspectratio]{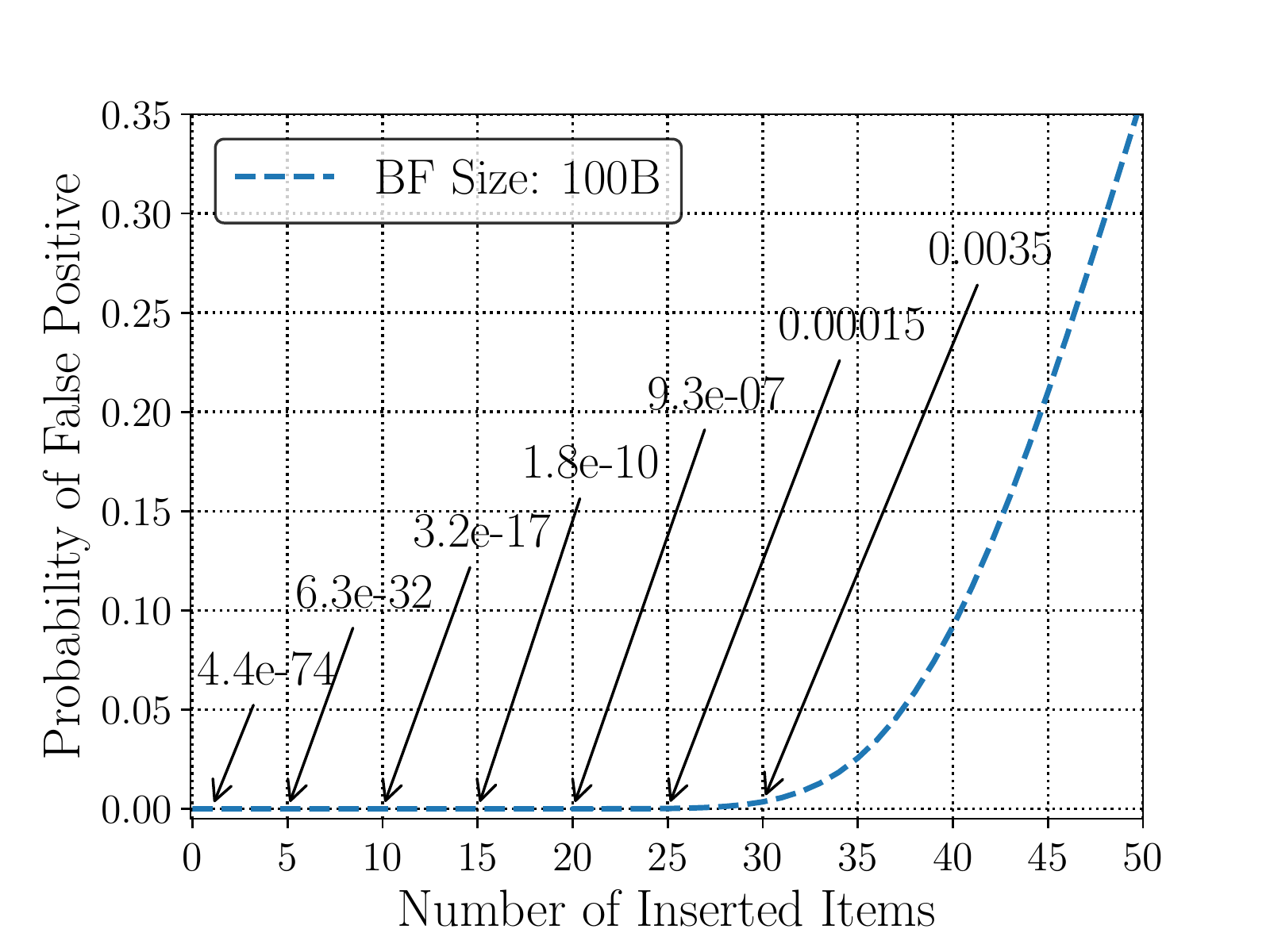}}
		\vspace{-0.85em}
		\caption{Chosen-insertion attack on the \ac{CRL} fingerprint.}
		\label{fig:crl-dis-chosen-insertion-attack}
	\end{center}
	\vspace{-1.35em}
\end{figure}

Under certain circumstances, the \emph{chosen-insertion} attack can also be applicable in our vehicle-centric \ac{CRL} distribution scheme. For example, integrating \ac{CRL} fingerprints into pseudonyms would result in larger pseudonym size, i.e., increasing packet size and overhead, thus higher channel congestion and error rates~\cite{calandriello2011performance, 1609-2016}. In such cases, the size of a \ac{CRL} fingerprint should not exceed a certain amount of bytes for efficiency reasons. Thus, the \ac{PCA} cannot generate a \ac{CRL} fingerprint with the desired size and false positive rate to be integrated into a subset of pseudonyms. This implies that the \ac{PCA} should insert more items (\ac{CRL} pieces) into the \ac{CRL} fingerprint with `constant' size, and thus increase the probability of false positive. Fig.~\ref{fig:crl-dis-chosen-insertion-attack}.a shows that the probability of generating a bogus \ac{CRL} piece grows exponentially when the \ac{BF} size is constant. Fig.~\ref{fig:crl-dis-chosen-insertion-attack}.b shows that the probability of false positive when \ac{BF} size is 100 bytes. The probability of false positive by inserting 5 items in the \ac{BF} is 6.3$\times 10^{-32}$. Even with an extremely computationally powerful adversary, generating a bogus \ac{CRL} piece is practically infeasible. However, by inserting 10 \ac{CRL} pieces into the \ac{BF} (assuming the \ac{BF} size remains constant), the probability of false positive becomes 3.2$\times 10^{-17}$. By considering the computational power of an adversary to be 1.6$\times 10^{18}$~\cite{antpoolchina}, the time to generate a bogus \ac{CRL} piece (with $K=67$) becomes $\approx$14 seconds.

Depending on the type of misbehavior and revocation rate ($\mathbb{R}$), the number of revocation entries within a $\Gamma_{CRL}$ interval could be huge, which yields a larger \ac{CRL} size. If this would result in a high rate of false positive, the \ac{PCA} could opt in enlarging the maximum bandwidth ($\mathbb{B}$) for \ac{CRL} distribution, or decreasing the $\Gamma_{CRL}$ interval (if $\Gamma_{CRL}$ is larger than $\Gamma$). Alternatively, the \ac{PCA} could decrease the percentage of \emph{fingerprint-carrier} pseudonyms; even with 1\% of the vehicles to be fingerprint-carrier nodes, all of the vehicles could obtain the \ac{CRL} fingerprint in time~\cite{khodaei2018VehicleCentric}. In our experiments, even with $\mathbb{R}=$5\%, $\mathbb{B}=$50 KB/s, and probability of false positive $p=10^{-30}$, the number of \ac{CRL} pieces becomes 7, with a 126-byte fingerprint.


\subsection{Quantitative Analysis}
\label{subsec:crl-dis-quantitative-analysis}

\subsubsection{Experimental Setup}
\label{subsubsec:crl-dis-experimental-setup}

We use OMNET++~\cite{omnetpp} and the Veins framework to simulate a large-scale scenario using SUMO~\cite{behrisch2011sumo} with a realistic mobility trace, the \acs{LuST} dataset~\cite{codeca2015lust}. For the cryptographic protocols and primitives (\ac{ECDSA}-256 and SHA-256 as per IEEE 1609.2~\cite{1609-2016} and \acs{ETSI}~\cite{ETSI-102-638}), we use OpenSSL. \ac{V2X} communication is IEEE 802.11p~\cite{IEEE-WAVE-2016} and cryptographic protocols and primitives were executed on a virtual machine (dual-core 2.0 GHz). To evaluate \ac{CRL} pieces construction (with a \ac{BF}) and validate \ac{CRL} pieces (\ac{BF} membership check), we used a Nexcom vehicular \ac{OBU} (Dual-core 1.66 GHz, 1GB memory) from the PRESERVE project~\cite{preserve-url}. For \ac{CRL} fingerprint operations, \ac{BF} insertion and query, we used PYBLOOM~\cite{pybloom}.

\textbf{Effective placement of the \acp{RSU}:} We sorted the intersections with the highest numbers of vehicles passing by~\cite{liang2012optimal}. We then placed the \acp{RSU} based on these \emph{``highly-visited''} intersections with non-overlapping radio ranges.

\textbf{Metrics:} We evaluate the latency to obtain the latest \ac{CRL} pieces, i.e., from the time a vehicle enters the system until it successfully downloads them. We choose a small amount of bandwidth ($\mathbb{B}$) for the distribution, e.g., 5-10 KB/s. Note that request-triggered \ac{CRL} piece broadcasts at 5-100 KB/s (40-800 Kbit/s) are practical because 802.11p supports data-rates up to 24 Mbit/s~\cite{IEEE-WAVE-2016}.

Table~\ref{table:crl-dis-simulation-parameters} shows the simulation parameters; Tables~\ref{table:crl-dis-lust-total-pseudonyms-based-on-lifetime} and~\ref{table:crl-dis-simulation-parameters-with-diff-revocation-rates} show the simulation information for the \acs{LuST} dataset with respect to different pseudonyms lifetimes ($\tau_{P}$), revocation rates ($\mathbb{R}$), and maximum bandwidth for distributing \ac{CRL} pieces ($\mathbb{B}$). We assume that the revocation events are uniformly distributed over a day. For example, if $\tau_{P}=60s$, the total number of pseudonyms for one day is around 1.7M. Assuming 1\% of the pseudonyms are revoked ($\mathbb{R}=$1\%), there will be around 17K revoked pseudonyms in a day. With our \emph{vehicle-centric} approach, each vehicle only needs to obtain pieces of \ac{CRL} for the interval it travels. When $\Gamma_{CRL}=1$ hour, the average number of entries per $\Gamma_{CRL}$ interval is around 710. With $\mathbb{B}=$ 10 KB/s, total number of pieces will be 6. These numbers come from the actual implementation of encoded packets, serialized with the C++ boost library.

\begin{table}[!t]
	\vspace{-0em}
	\centering
	\caption{Simulation parameters (\acs{LuST} dataset).}
	\vspace{-0.25em}
	\label{table:crl-dis-simulation-parameters}
	\resizebox{0.45\textwidth}{!}
	{
		\renewcommand{\arraystretch}{1.2}
		\begin{tabular}{ | c | c ||| c | c | }
			\hline
			\textbf{Parameters} & \textbf{Value} & \textbf{Parameters} & \textbf{Value} \\\hline\hline
			
			\ac{CRL}/Fingerprint TX interval & 0.5s/5s & Pseudonym lifetime & 30s$-$600s \\\hline 
			Carrier frequency & 5.89 GHz & Area size & 15 KM $\times$ 15 KM \\\hline 
			TX power & 20mW & Number of vehicles & 138,259 \\\hline
			Physical layer bit-rate & 18Mbps & Number of trips & 287,939 \\\hline
			Sensitivity & -89dBm & Average trip duration & 692.81s \\\hline
			Thermal noise & -110dBm & Duration of simulation & 4 hour (7$-$9, 17$-$19) \\\hline 
			\ac{CRL} dist. Bandwidth ($\mathbb{B}$) & 5$-$100 KB/s & $\Gamma$ & 1$-$60 min \\\hline 
			Number of \acp{RSU} & 100 & $\Gamma_{CRL}$ & 60 min \\\hline 
		\end{tabular}
	}
	\vspace{-0.5em}
\end{table}

\begin{table}[!t]
	\vspace{-0.25em}
	\caption{Vehicle-centric revocation information for {\footnotesize \acs{LuST} dataset} \protect\linebreak {\footnotesize ($\mathbb{R}=1\%$, $\mathbb{B}=10 KB/s$).}}
	\vspace{-0.25em}
	\label{table:crl-dis-lust-total-pseudonyms-based-on-lifetime}
	\resizebox{0.45\textwidth}{!}
	{
		\renewcommand{\arraystretch}{1.5}
		\begin{tabular}{| c ||| *{4}{c} |}
			\hline
			\shortstack{\\ {} \textbf{Pseudonym} \\ \textbf{Lifetime}} & \shortstack{\textbf{Number of} \\ \textbf{Psnyms}} & \shortstack{\textbf{Number of} \\ \textbf{Revoked Psnyms}} & \shortstack{\textbf{Average} \\ \textbf{Number per $\Gamma_{CRL}$}} & \shortstack{\textbf{Number of} \\ \textbf{Pieces}} \\\hline\hline 
			\textbf{$\tau_{P}$=30s} & \textbf{3,425,565} & \textbf{34,256} & \textbf{1,428} & \textbf{13} \\\hline 
			\textbf{$\tau_{P}$=60s} & \textbf{1,712,782} & \textbf{17,128} & \textbf{714} & \textbf{7} \\\hline 
			\textbf{$\tau_{P}$=300s} & \textbf{342,556} & \textbf{3,426} & \textbf{143} & \textbf{2} \\\hline 
			\textbf{$\tau_{P}$=600s} & \textbf{171,278} & \textbf{1,713} & \textbf{72} & \textbf{1} \\\hline 
		\end{tabular}
	}
	\vspace{-0.5em}
\end{table}

\begin{table}[!t]
	\centering
	\vspace{-0.25em}
	\caption{Simulation parameters for \acs{LuST} dataset ($\tau_{P}=60s$).}
	\vspace{-0.25em}
	\label{table:crl-dis-simulation-parameters-with-diff-revocation-rates}
	\resizebox{0.49\textwidth}{!}
	{
		\renewcommand{\arraystretch}{0.99}
		\hspace{-1em}
		\begin{tabular}{|c|c|c|c|c|c|c|c|c|c|c|c|c|c|}
			\hline
			\multirow{3}{*}{\LARGE {\textbf{\shortstack{\\ {} \\ {} Revocation \\ Rate ($\mathbb{R}$)}}}} & \multicolumn{5}{c|}{{\LARGE \textbf{\shortstack{{} \\ Baseline Scheme}}}} & \multicolumn{5}{c|}{{\LARGE \textbf{\shortstack{{} \\ Vehicle-Centric Scheme}}}} \\ 
			\cline{2-14}
			& \multirow{2}{*}{{\LARGE \textbf{\shortstack{\ac{CRL} \\ Entries}}}} & \multicolumn{1}{c|}{{\LARGE \textbf{\shortstack{{} \\ 10 KB/s}}}} & \multicolumn{1}{c|}{{\LARGE \textbf{\shortstack{{} \\ 25 KB/s}}}} & \multicolumn{1}{c|}{{\LARGE \textbf{\shortstack{{} \\ 50 KB/s}}}} & \multicolumn{1}{c|}{{\LARGE \textbf{\shortstack{{} \\ 100 KB/s}}}} & \multirow{2}{*}{{\LARGE \textbf{\shortstack{\ac{CRL} \\ Entries}}}} & \multicolumn{1}{c|}{{\LARGE \textbf{\shortstack{{} \\ 10 KB/s}}}} & \multicolumn{1}{c|}{{\LARGE \textbf{\shortstack{{} \\ 25 KB/s}}}} & \multicolumn{1}{c|}{{\LARGE \textbf{\shortstack{{} \\ 50 KB/s}}}}& \multicolumn{1}{c|}{{\LARGE \textbf{\shortstack{{} \\ 100 KB/s}}}} \\
			\cline{3-6}\cline{8-11}
			& & {\LARGE \textbf{\shortstack{{} \\ Pieces}}} & {\LARGE \textbf{\shortstack{{} \\ Pieces}}} & {\LARGE \textbf{\shortstack{{} \\ Pieces}}} & {\LARGE \textbf{\shortstack{{} \\ Pieces}}} & & {\LARGE \textbf{\shortstack{{} \\ Pieces}}} & {\LARGE \textbf{\shortstack{{} \\ Pieces}}} & {\LARGE \textbf{\shortstack{{} \\ Pieces}}} & {\LARGE \textbf{\shortstack{{} \\ Pieces}}} \\
			\hline
			{\LARGE \textbf{\shortstack{{} \\ 0.5\%}}} & {\LARGE \shortstack{{} \\ 8,564}} & {\LARGE \shortstack{{} \\ 61}} & {\LARGE \shortstack{{} \\ 24}} & {\LARGE \shortstack{{} \\ 12}} & {\LARGE \shortstack{{} \\ 6}} & {\LARGE \shortstack{{} \\ 357}} & {\LARGE \shortstack{{} \\ 4}} & {\LARGE \shortstack{{} \\ 2}} & {\LARGE \shortstack{{} \\ 2}} & {\LARGE 1} \\
			\hline
			{\LARGE \textbf{\shortstack{{} \\ 1\%}}} & {\LARGE 17,128} & {\LARGE 121} & {\LARGE 48} & {\LARGE 24} & {\LARGE 12} & {\LARGE 714} & {\LARGE 7} & {\LARGE 3} & {\LARGE 2} & {\LARGE 1} \\
			\hline
			{\LARGE \textbf{\shortstack{{} \\ 2\%}}} & {\LARGE 34,256} & {\LARGE 242} & {\LARGE 95} & {\LARGE 48} & {\LARGE 24} & {\LARGE 1,428} & {\LARGE 13} & {\LARGE 6} & {\LARGE 3} & {\LARGE 2} \\
			\hline
			{\LARGE \textbf{\shortstack{{} \\ 3\%}}} & {\LARGE 51,384} & {\LARGE 362} & {\LARGE 142} & {\LARGE 71} & {\LARGE 36} & {\LARGE 2,141} & {\LARGE 20} & {\LARGE 8} & {\LARGE 4} & {\LARGE 2} \\
			\hline
			{\LARGE \textbf{\shortstack{{} \\ 4\%}}} & {\LARGE 68,512} & {\LARGE 483} & {\LARGE 190} & {\LARGE 95} & {\LARGE 47} & {\LARGE 2,855} & {\LARGE 26} & {\LARGE 11} & {\LARGE 6} & {\LARGE 3} \\
			\hline
			{\LARGE \textbf{\shortstack{{} \\ 5\%}}} & {\LARGE 85,640} & {\LARGE 604} & {\LARGE 237} & {\LARGE 118} & {\LARGE 59} & {\LARGE 3,569} & {\LARGE 32} & {\LARGE 13} & {\LARGE 7} & {\LARGE 4} \\
			\hline
		\end{tabular}
	}
	\vspace{-0.5em}
\end{table}

\subsubsection{Summary of Results}
\label{subsubsec:crl-dis-summary-of-results}

The results here complement (and are comparable with) the presented results in~\cite{khodaei2018VehicleCentric} and they share the same configuration and system set up. Our vehicle-centric scheme converges more than 40 times faster than the state-of-the-art~\cite{haas2011efficient, laberteaux2008security, haas2009design}, termed here the \emph{baseline} scheme, with a similar experimental set up (Fig.~\ref{fig:crl-dis-comparison-number-of-cognizant-nodes-in-the-system}.b). Moreover, with the baseline scheme, the number of vehicles that successfully obtained the latest \ac{CRL}, referred to as \emph{cognizant vehicles}, is highly dependent on the revocation rate and it significantly drops when the revocation rate increases from 0.5\% to 5\%. However, the performance of our scheme is not affected by the revocation rate: the number of cognizant nodes remains almost intact even if the revocation rate increases up to 5\%~\cite{khodaei2018VehicleCentric}. Furthermore, our vehicle-centric scheme is more resilient to selfish\footnote{Such nodes do not perform any ``active'' attack, e.g., a clogging \ac{DoS} attack, rather they become silent and they never respond to a \ac{CRL}/$\Delta\textnormal{-}${CRL} piece request.}, pollution, and \ac{DoS} attacks: with 25\% of vehicles in the baseline system compromised, one could prevent almost all legitimate vehicles from obtaining the \acp{CRL}; however, with our scheme, the percentage of informed vehicles remains almost intact even if 50\% of the vehicles are compromised (Fig.~\ref{fig:crl-dis-resilience-against-selfish-nodes},~\ref{fig:crl-dis-e2e-delay-resilience-against-pollution-dos-attacks},~\ref{fig:crl-dis-crl_download-failure-ratio-under-ddos-attacks},~\ref{fig:crl-dis-histogram-of-received-crl-pieces-under-ddos-attack}).

Furthermore, our experiments show that the distribution of $\Delta\textnormal{-}${CRL} pieces as well as the validation keys are efficient and resilient against \ac{DoS} attacks; more specifically, 95\% of the vehicles received the $\Delta\textnormal{-}${CRL} pieces in less than 52s: $F_x(t=52 \: ms)=0.95$, and 95\% of them obtained the validation keys within less than 31s: $F_x(t=31 \: ms)=0.95$ (Fig.~\ref{fig:crl-dis-delta-crl-pieces-validation-keys-distribution}). Moreover, \ac{BF} construction and membership checks are much more efficient than the baseline scheme: the average latency to verify a \ac{CRL} piece for the baseline scheme (i.e., verifying an \ac{ECDSA}-256 bits signature) is 2.346 ms. However, for the vehicle-centric scheme, the latency to validate a \ac{CRL} piece using a \ac{BF} (with probability of false positive rate $p=10^{-25}$ and $K=67$, as the optimal number of hash functions) is 0.352 ms, i.e., 6.6 times faster than the baseline scheme (Table~\ref{table:crl-dis-delay-to-check-bloomfilter-membership-item} and Table~\ref{table:crl-dis-delay-to-insert-items-in-bloomfilter}). Finally, our experiments confirm that our scheme outperforms the baseline scheme in terms of communication overhead (notably security overhead)~\cite{khodaei2018VehicleCentric}.

\begin{figure} [!t]
	\centering
	\begin{center}
		\centering
		\hspace{-1.45em} 
		\subfloat[Vehicle-centric scheme]{
			\includegraphics[trim=0.3cm 0.25cm 0.75cm 1.28cm, clip=true, width=0.25\textwidth,height=0.25\textheight,keepaspectratio]{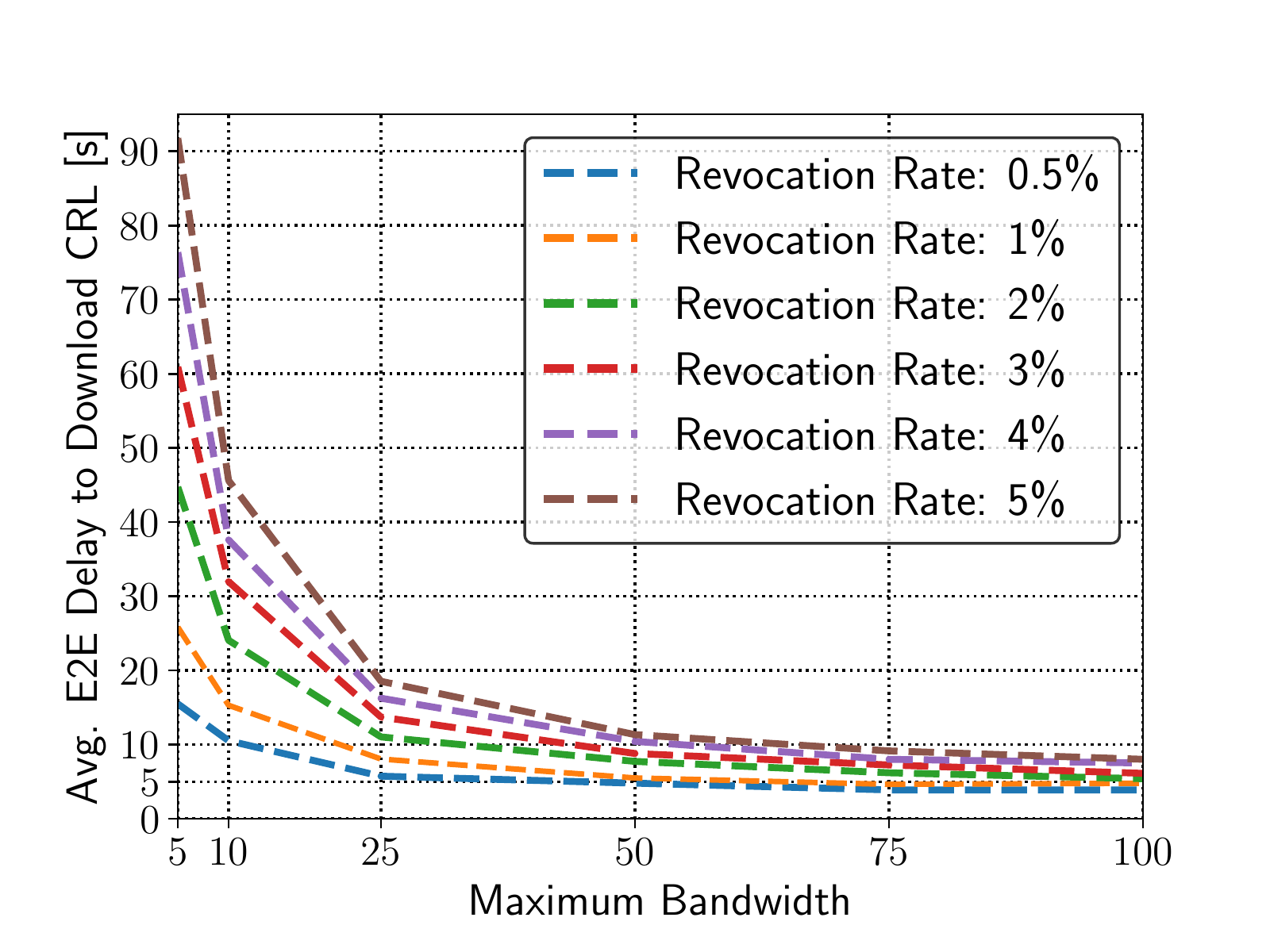}}
		\hspace{-0.45em} 
		\subfloat[Vehicle-centric scheme]{
			\includegraphics[trim=0.3cm 0.01cm 0.75cm 1.35cm, clip=true, width=0.25\textwidth,height=0.25\textheight,keepaspectratio]{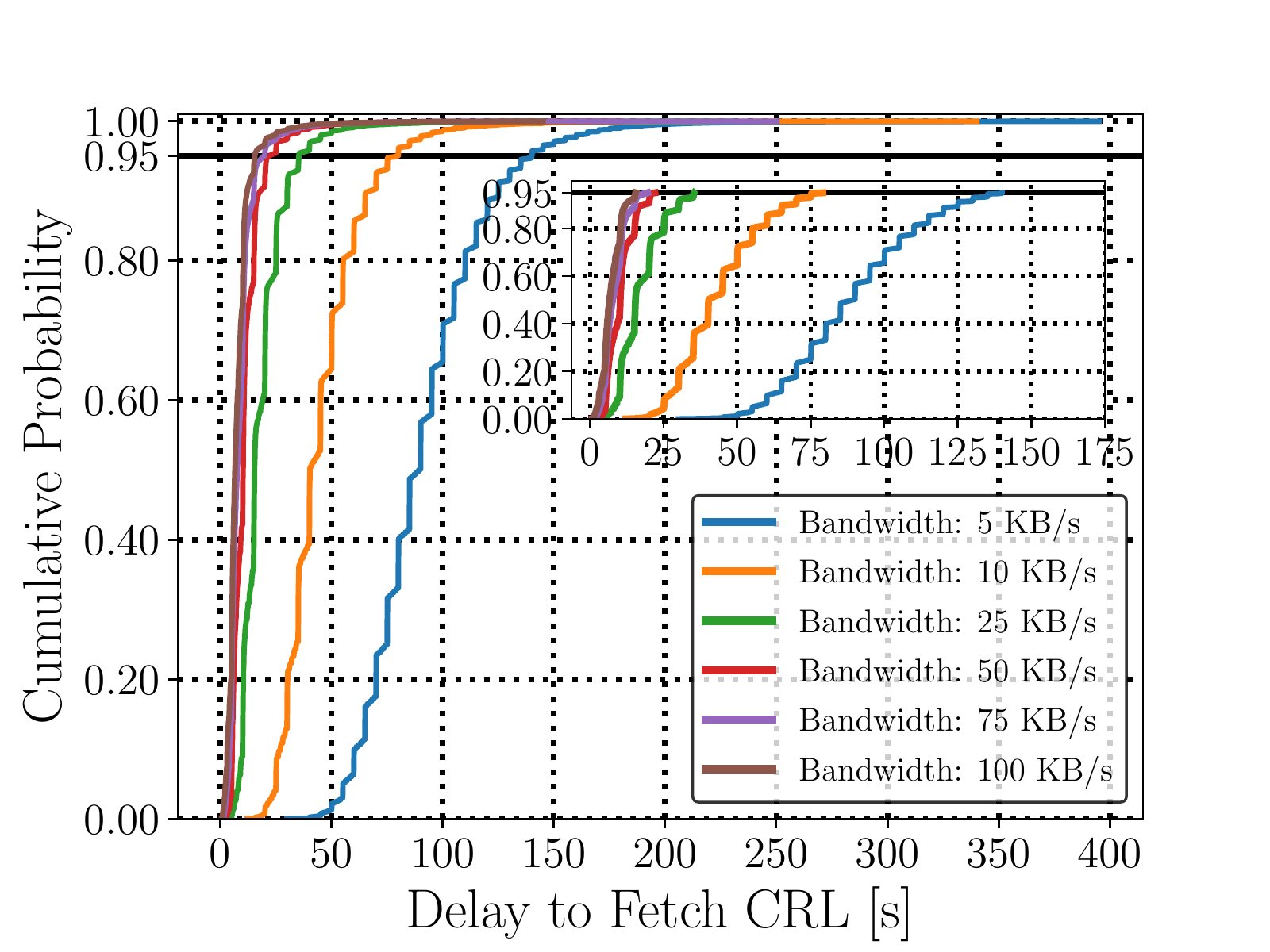}}
		\vspace{-0.25em}
		\caption{(a) Bandwidth-delay trade off ($\tau_{P}=60s$). (b) \acs{CDF} of end-to-end delay with different bandwidth ($\tau_{P}=30s$, $\mathbb{R}=5\%$).}
		\label{fig:crl-dis-avgerage_crl_delay_bandwidth-delay-tradeoff}
	\end{center}
	\vspace{-1.8em}
\end{figure}

\subsubsection{Vehicle-Centric \ac{CRL} Distribution Evaluation}
\label{subsubsec:crl-dis-vehicle-centric-performance-evaluation}

Fig.~\ref{fig:crl-dis-avgerage_crl_delay_bandwidth-delay-tradeoff}.a shows the average end-to-end latency to obtain the \acp{CRL} as a function of maximum bandwidth for the vehicle-centric scheme. The delays were averaged over vehicles operating during rush hours. The total number of pseudonyms is 1.7M ($\tau_{P}=60s$) and the maximum bandwidth ranges from 5 to 100 KB/s. In general, the smaller amount of bandwidth for \ac{CRL} distribution and the higher the revocation rate are, the higher the latency for all vehicles to obtain the \ac{CRL}. For example, the average latency, with $\mathbb{R}=5\%$, decreases from 45.68s to 18.48s as the $\mathbb{B}$ increases from 10 to 25 KB/s. It is imperative to allocate as low bandwidth as possible without compromising the timely distribution of \acp{CRL}. On the one hand, allocating lower bandwidth for the \ac{CRL} distribution diminishes interference with the safety-critical operations and mitigates pollution and \ac{DoS} attacks; but, it degrades the timely \acp{CRL} distribution. On the other hand, allocating large bandwidth would enhance timely \acp{CRL} distribution at the cost of interference with safety operations and/or enabling an attacker to broadcast a fake \ac{CRL} piece at a high rate. Depending on the revocation rates (i.e., events that lead to revocation), the optimal bandwidth for \ac{CRL} distribution can be properly determined to achieve a certain level of quality of service without compromising the operation of time-critical messages. For example, with $\mathbb{R}=5\%$, one can increase the bandwidth from 25 KB/s to 100 KB/s to reduce the delay from 18.48s to 8s, i.e., 2.3 faster \ac{CRL} distribution.

Fig.~\ref{fig:crl-dis-avgerage_crl_delay_bandwidth-delay-tradeoff}.b shows the \ac{CDF} of end-to-end latencies to obtain the needed \acp{CRL} with different bandwidths for the vehicle-centric scheme. In general, the larger the allocated bandwidth for \ac{CRL} distribution, the less the average latency to obtain \acp{CRL}. For example, with $\mathbb{R}=5\%$ and $\mathbb{B}=50KB/s$, 95\% of the vehicles received the needed pieces of the \ac{CRL} in less than 24s: $F_x(t=24 \: ms)=0.95$, i.e., $Pr\{t\leq24 \: ms\}=0.95$.

\subsubsection{Vehicle-Centric $\Delta\textnormal{-}$\ac{CRL} Distribution Evaluation}
\label{subsubsec:crl-dis-vehicle-centric-delta-crl-performance-evaluation}

\begin{figure} [!t]
	\centering
	\begin{center}
		\centering
		\hspace{-1.55em} 
		\subfloat[7:05-7:10 am ($\mathbb{B}=$10 KB/s)]{
			\includegraphics[trim=0cm 0.25cm 1.1cm 1.28cm, clip=true, width=0.25\textwidth,height=0.25\textheight,keepaspectratio]{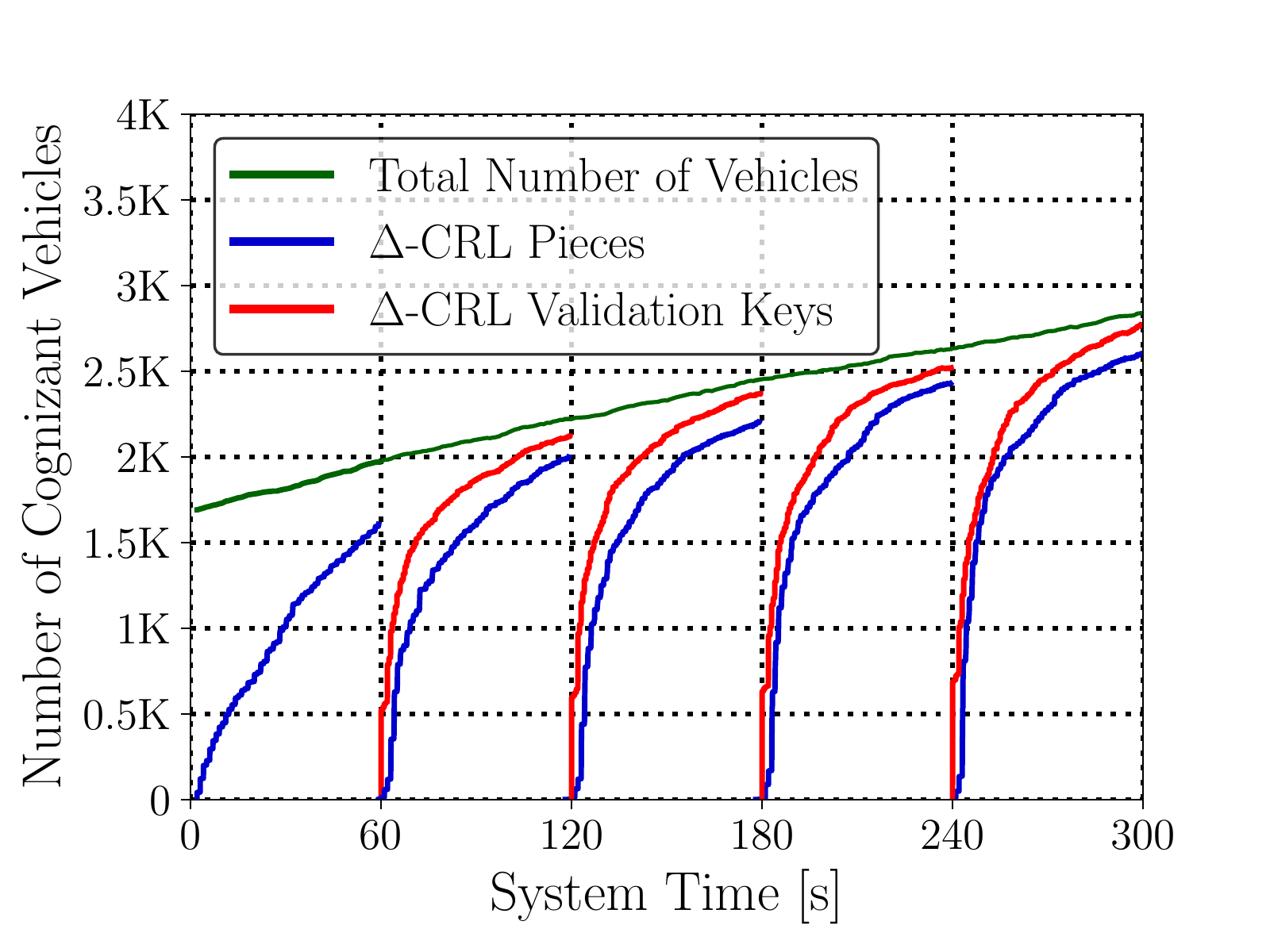}}
		\hspace{-0.3em} 
		\subfloat[7:05-7:10 am ($\mathbb{B}=$10 KB/s)]{
			\includegraphics[trim=0.3cm 0.01cm 0.5cm 1.35cm, clip=true, width=0.25\textwidth,height=0.25\textheight,keepaspectratio]{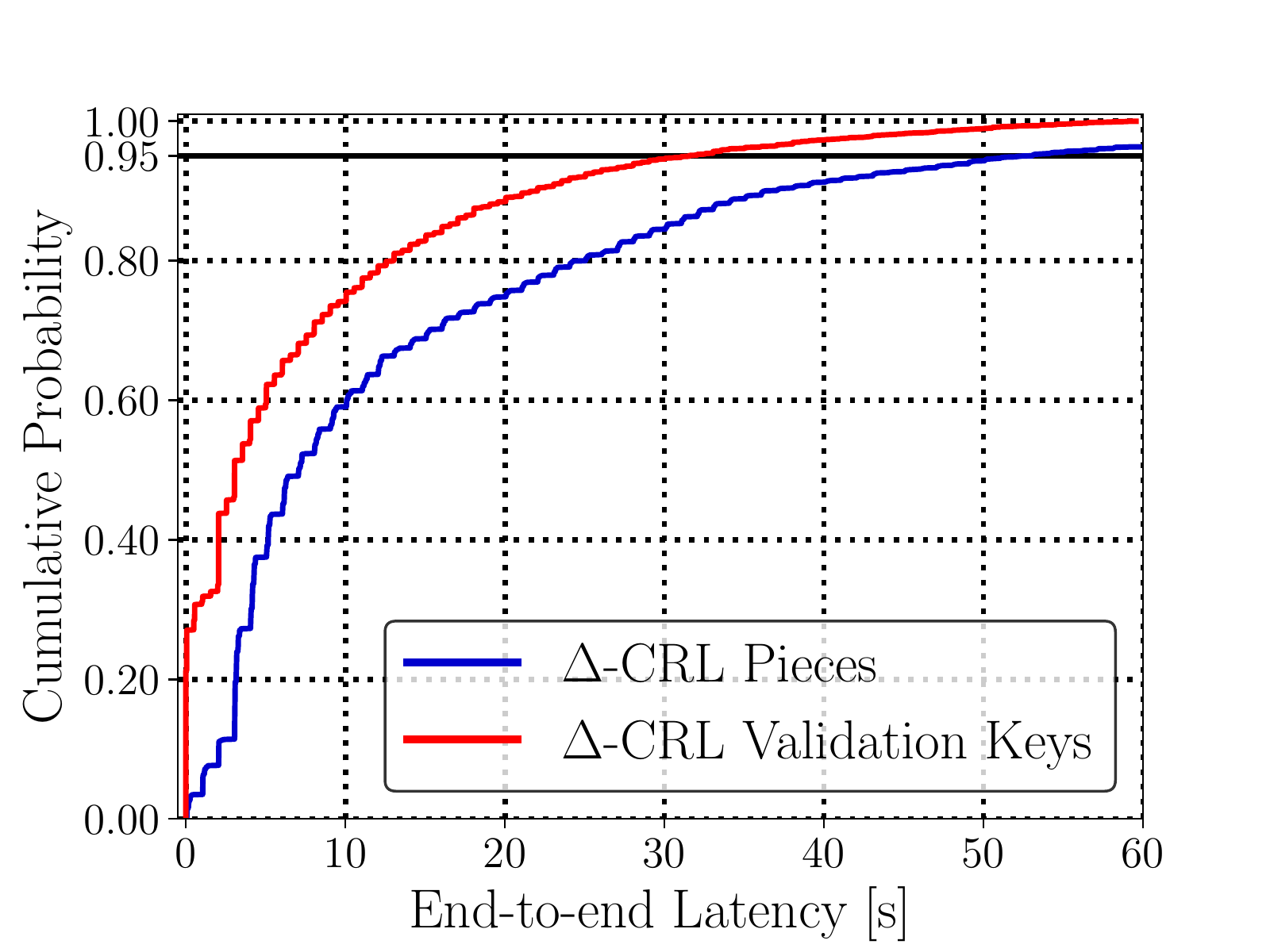}}
		\vspace{-0.25em}
		\caption{End-to-end delay to fetch $\Delta\textnormal{-}$\ac{CRL} pieces and validation keys for vehicle-centric scheme ($\tau_{P}=60$ sec., $\mathbb{R}=5\%$, $\gamma_{key}=0.5$, $\gamma_{piece}=2$).}
		\label{fig:crl-dis-delta-crl-pieces-validation-keys-distribution}
	\end{center}
	\vspace{-1.75em}
\end{figure}

Upon a new revocation event in a $\Gamma_{CRL}$ interval, the \ac{PCA} constructs $\Delta\textnormal{-}$\ac{CRL} pieces by including the recently revoked pseudonyms (not included in the base-\ac{CRL}). We emulate revocation events, e.g., due to malfunctioning of sensors, in every pseudonym lifetime with $\mathbb{R}=$ 5\%, i.e., 5\% of the pseudonyms within each $\tau_{P}$ should be revoked and included in the $\Delta\textnormal{-}$\ac{CRL} pieces. The total number of pseudonyms is 1.7M ($\tau_{P}=60s$) and the maximum bandwidth to distribute $\Delta\textnormal{-}$\ac{CRL} pieces is up to 10 KB/s. In this experiment, $\Delta\textnormal{-}$\ac{CRL} pieces are broadcasted with frequency $\gamma_{piece}=2$ (one piece every 2 sec.) and validation keys are broadcasted with frequency $\gamma_{key}=0.5$ (2 times per second). We evaluate the latency to obtain $\Delta\textnormal{-}$\ac{CRL} pieces and validation keys. The longer the pseudonyms lifetimes combined with higher frequency of broadcasts (and larger coverage of the area by \acp{RSU}), the faster the convergence time is, thus the narrower the revocation vulnerability window becomes.

Fig.~\ref{fig:crl-dis-delta-crl-pieces-validation-keys-distribution} shows the performance of the vehicle-centric scheme for the $\Delta\textnormal{-}$\acp{CRL} distribution. A new revocation event happens at the beginning of a pseudonym lifetime and 5\% of pseudonyms should be revoked. The \ac{PCA} constructs $\Delta\textnormal{-}$\ac{CRL} pieces and the \acp{RSU} broadcast them. As illustrated in Fig.~\ref{fig:crl-dis-delta-crl-pieces-validation-keys-distribution}.a, the majority of vehicles could obtain $\Delta\textnormal{-}$\ac{CRL} pieces until the end of each pseudonym lifetime. Upon a pseudonym transition, the validation key for the $\Delta\textnormal{-}$\ac{CRL} pieces is released. However, there is a time lag until practically all vehicles are informed about the validation key disclosure. To ensure that every vehicle has already obtained the validation keys at the beginning of each $\tau_{P}$, the \ac{PCA} could distribute the $\Delta\textnormal{-}$\ac{CRL} validation keys in the middle of preceding $\tau_{P}$. In other words, in order to optimize the distribution of $\Delta\textnormal{-}$\ac{CRL} pieces to be used during $\tau^{i}_{P}$, the \ac{PCA} would start distributing the $\Delta\textnormal{-}$\ac{CRL} pieces from the middle of $\tau^{i-2}_{P}$ until the middle of $\tau^{i-1}_{P}$; accordingly, the distribution of validation key, corresponding to $\tau^{i}_{P}$, would be started from the middle of $\tau^{i-1}_{P}$.

\begin{figure} [!t]
	\vspace{-0em}
	\begin{center}
		\centering
		\subfloat[7:00-7:10 am ($\mathbb{B}=$25 KB/s)]{
			\hspace{-1.45em} \includegraphics[trim=0.15cm 0.25cm 0.25cm 0.3cm, clip=true, width=0.25\textwidth,height=0.25\textheight,keepaspectratio]{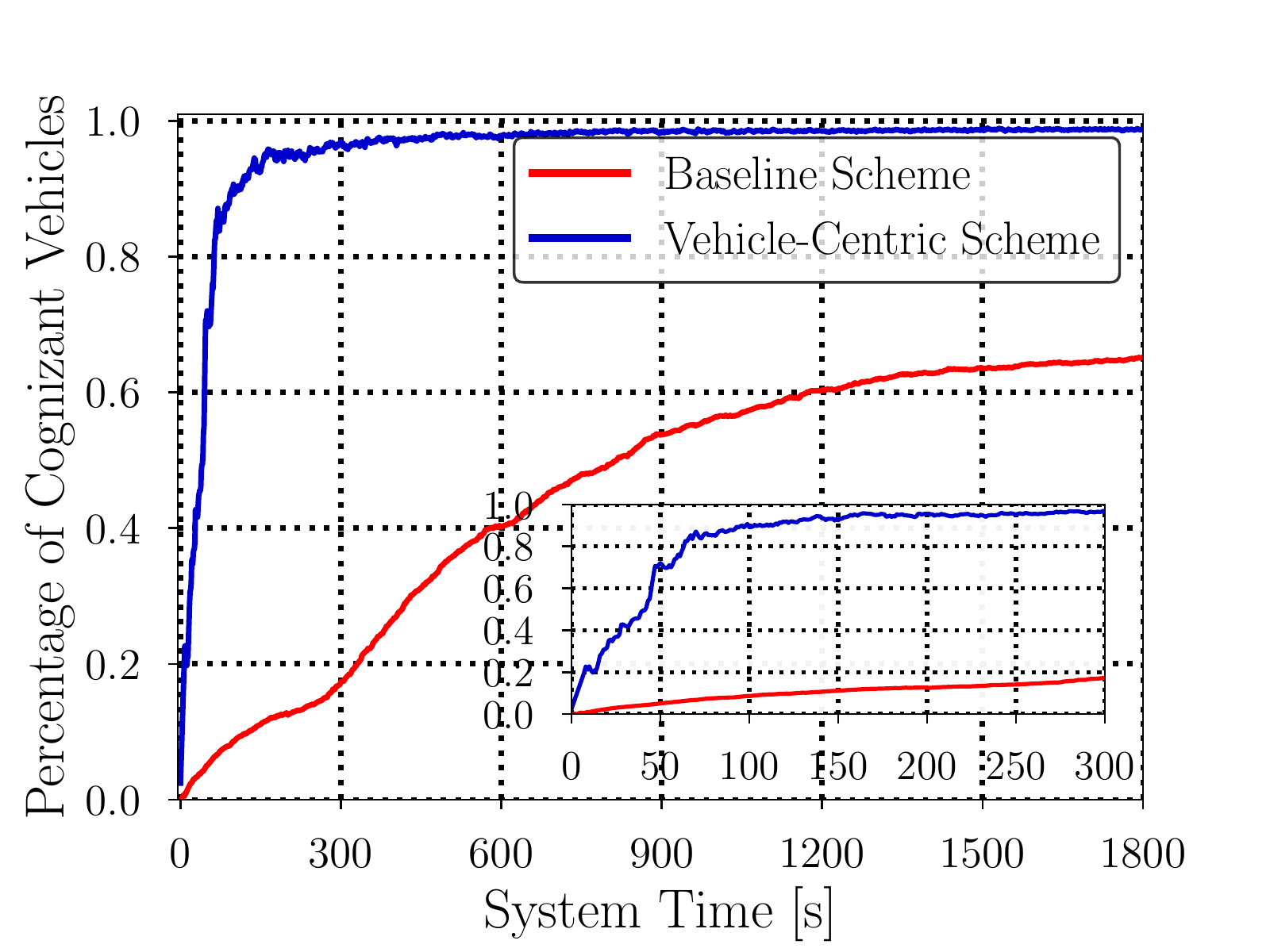}}
		\subfloat[7-9 am, 5-7 pm ($\mathbb{B}=$25 KB/s)]{
			\hspace{-1.15em} \includegraphics[trim=0.15cm 0.25cm 0.5cm 1.3cm, clip=true, width=0.25\textwidth,height=0.25\textheight,keepaspectratio]{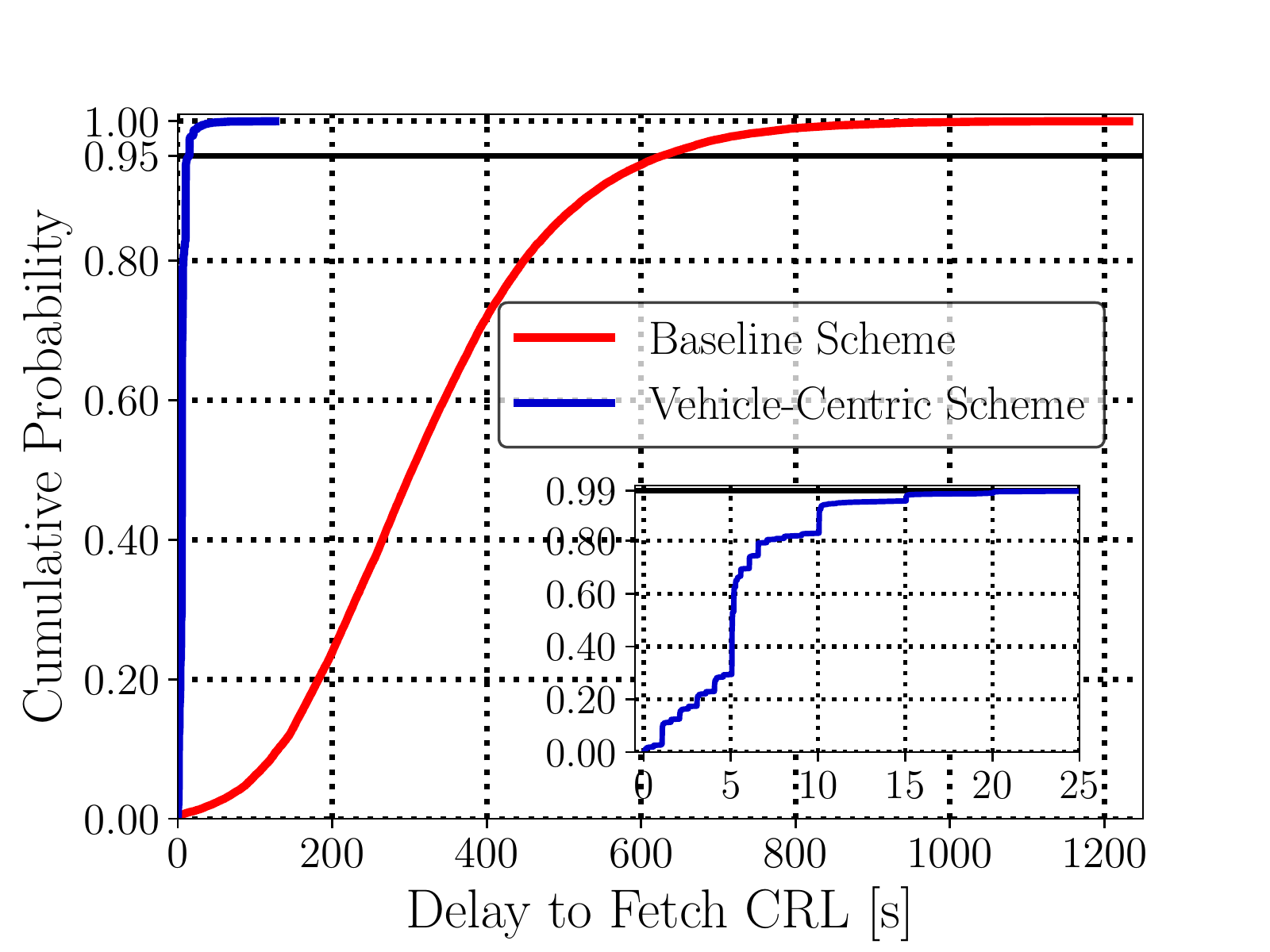}}
		\vspace{-0.25em}
		\caption{End-to-end delay to fetch \acp{CRL} {\small ($\tau_{P}=60s$, $\mathbb{R}=1\%$)}.}
		\label{fig:crl-dis-comparison-number-of-cognizant-nodes-in-the-system}
	\end{center}
	\vspace{-1.65em}
\end{figure}

Fig.~\ref{fig:crl-dis-delta-crl-pieces-validation-keys-distribution}.b shows the \ac{CDF} of delays for distributing $\Delta\textnormal{-}$\ac{CRL} pieces and the validation keys. For the $\Delta\textnormal{-}$\ac{CRL} pieces distribution, $F_x(t=52 \: ms)=0.95$ while for the $\Delta\textnormal{-}$\ac{CRL} validation key distribution, $F_x(t=31 \: ms)=0.95$. Faster convergence of $\Delta\textnormal{-}$\ac{CRL} validation keys, in comparison with $\Delta\textnormal{-}$\ac{CRL} pieces, stems from potentially multiplicity of $\Delta\textnormal{-}$\ac{CRL} pieces and the frequency of distribution ($\gamma$). In order to mitigate a memory-exhaustion \ac{DoS} attack on $\Delta\textnormal{-}$\ac{CRL} pieces distribution, we mandate a rate limiting mechanism. This ensures that a compromised insider cannot `abuse' the allocated bandwidth towards performing a memory-exhaustion \ac{DoS} attack on the distribution of $\Delta\textnormal{-}$\ac{CRL} pieces. Note that the distribution of $\Delta\textnormal{-}$\ac{CRL} validation keys is not vulnerable to a \ac{DoS} attack~\cite{perrig2002tesla}.

\subsubsection{Performance Comparison}
\label{subsubsec:crl-dis-performance-comparison}

We compare our scheme with the \emph{baseline} scheme~\cite{haas2011efficient, laberteaux2008security, haas2009design} that uses \acp{RSU} and car-to-car epidemic distribution, with the same assumptions, configuration, and system parameters. For the baseline scheme, the \acs{CA} signs each \ac{CRL} piece and can specify a \emph{``time interval''} so that each vehicle receives $\mathbb{D}$ pseudonyms during the pseudonym acquisition process. As a result, for each batch of revoked pseudonyms, a single key $s_{i}$ (256 bit) is disclosed. Similarly, the \ac{PCA} in our scheme can be configured to issue $\mathbb{D}$ pseudonyms per $\Gamma$, i.e., {\footnotesize $\mathbb{D} = \dfrac{\Gamma}{\tau_{P}}$}. To revoke a batch of $\mathbb{D}$ pseudonyms, the serial number of the first revoked pseudonym in the hash chain and a random number, each 256 bits long, are disclosed. For both schemes, we assume a fully-unlinkable pseudonym provisioning policy~\cite{khodaei2018Secmace}, i.e., $\Gamma$ = $\tau_{P} = 1 min$.

We assume that vehicles are provided with enough pseudonyms corresponding to their actual trips for a day. Upon a revocation event, information on all revoked pseudonyms for the day is disseminated for the baseline scheme. In contrast, with our scheme, the \ac{CRL} entries are distributed in a time prioritized manner, i.e., revoked pseudonyms whose validity intervals fall within the current $\Gamma_{CRL}$ interval. Moreover, by disseminating signed \ac{BF} in advance, the verification cost is minimal compared to baseline signature verification, i.e., zero delay to verify the \ac{BF} integrated in \emph{fingerprint-carrier} pseudonyms or one signature verification for all \ac{CRL} pieces. The processing delay to perform \ac{BF} membership check is evaluated in Table~\ref{table:crl-dis-delay-to-check-bloomfilter-membership-item}.

Fig.~\ref{fig:crl-dis-comparison-number-of-cognizant-nodes-in-the-system}.a shows the number of cognizant vehicles over time for the two schemes. Vehicle-centric distribution of the \ac{CRL} pieces converges faster: the number of cognizant vehicles is very close to the actual number of vehicles in the system. Fig.~\ref{fig:crl-dis-comparison-number-of-cognizant-nodes-in-the-system}.b shows the \ac{CDF} of delays for the two schemes: for the baseline, $F_x(t=626s)=0.95$, whereas with our scheme, $F_x(t=15s)=0.95$, i.e., converging more than 40 times faster. The principal reasons for such significant improvements are the prioritization of the revocation entries based on their validity intervals, thus a huge reduction in size, as well as the efficient verification of \ac{CRL} pieces.

\begin{figure} [!t]
	\vspace{-1em}
	\centering
	\begin{center}
		\centering
		\hspace{-1.45em} 
		\subfloat[Baseline scheme]{
			\includegraphics[trim=0.1cm 0.25cm 1.3cm 0.5cm, clip=true, width=0.25\textwidth,height=0.25\textheight,keepaspectratio]{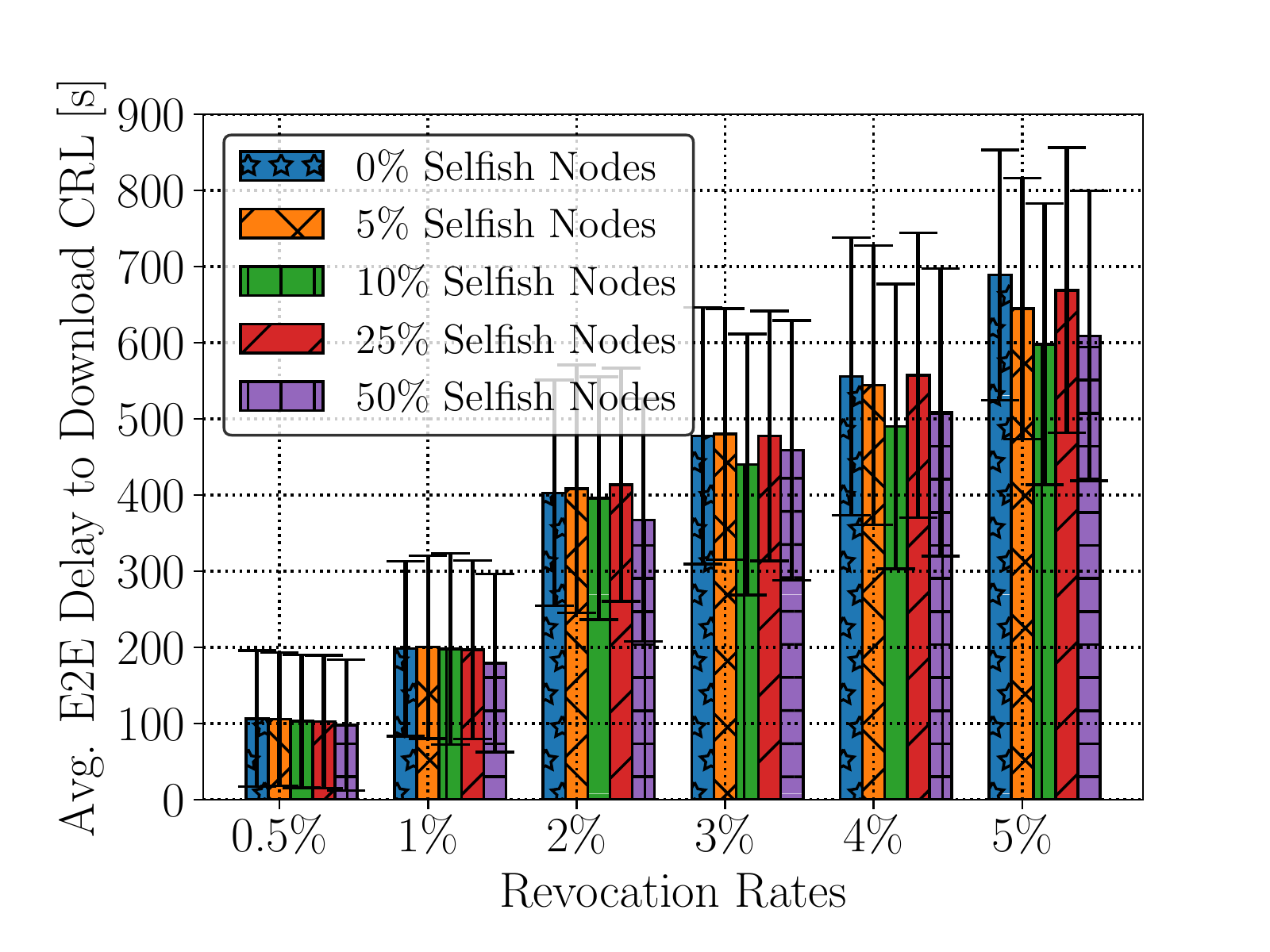}}
		\hspace{-0.45em} 
		\subfloat[Vehicle-centric scheme]{
			\includegraphics[trim=0.3cm 0.01cm 1.1cm 0.5cm, clip=true, width=0.25\textwidth,height=0.25\textheight,keepaspectratio]{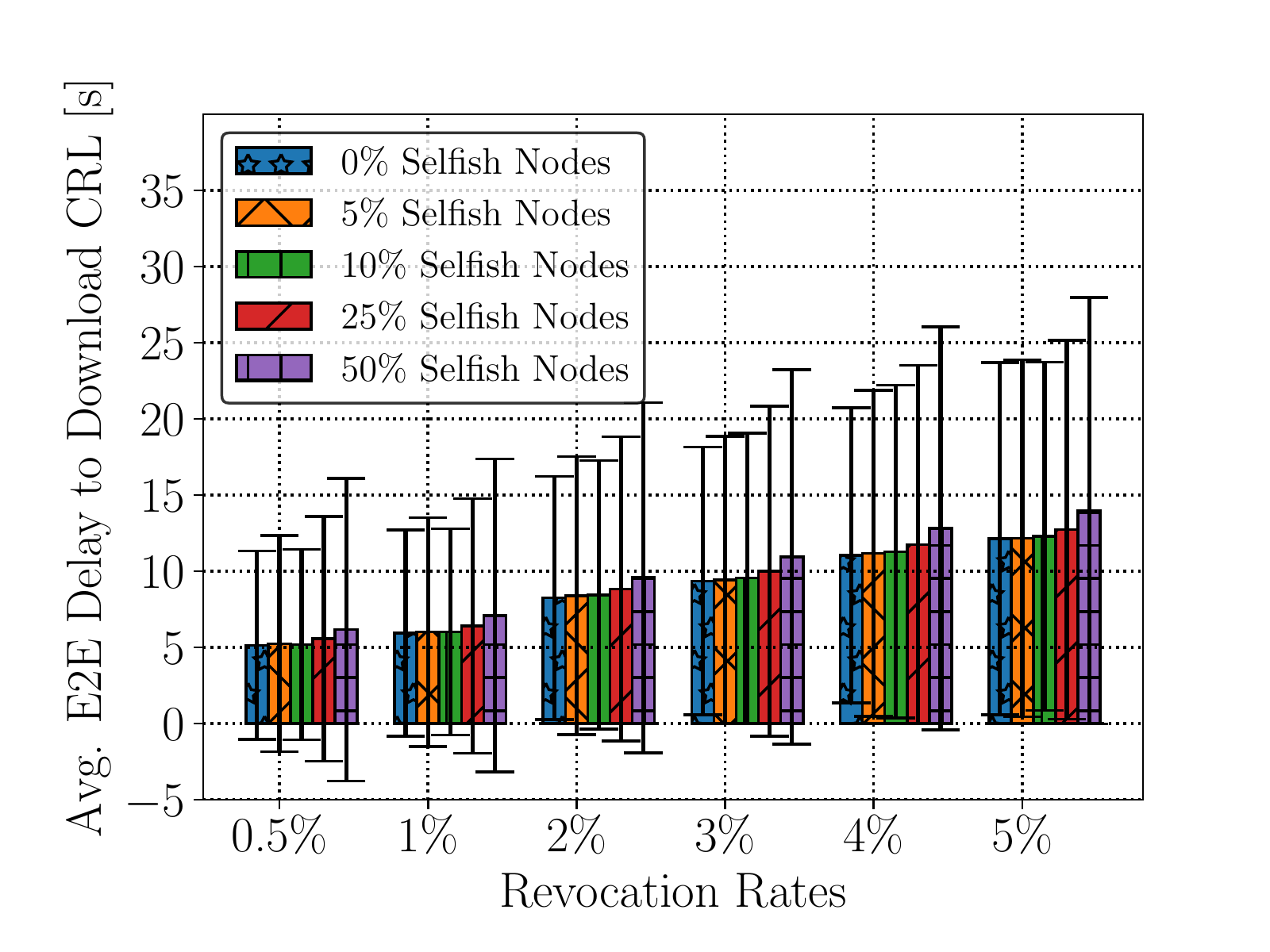}}
		\vspace{-0.25em}
		\caption{Resilience comparison against selfish nodes with different revocation rates (7:00-7:30, $\tau_p=30s$, $\mathbb{B}=50KB/s$).}
		\label{fig:crl-dis-resilience-against-selfish-nodes}
	\end{center}
	\vspace{-2em}
\end{figure}

\begin{figure} [!t]
	\vspace{-1em}
	\centering
	\begin{center}
		\centering
		\hspace{-1.45em} 
		\subfloat[Baseline scheme]{
			\includegraphics[trim=0.3cm 0.25cm 1.3cm 0.5cm, clip=true, width=0.25\textwidth,height=0.25\textheight,keepaspectratio]{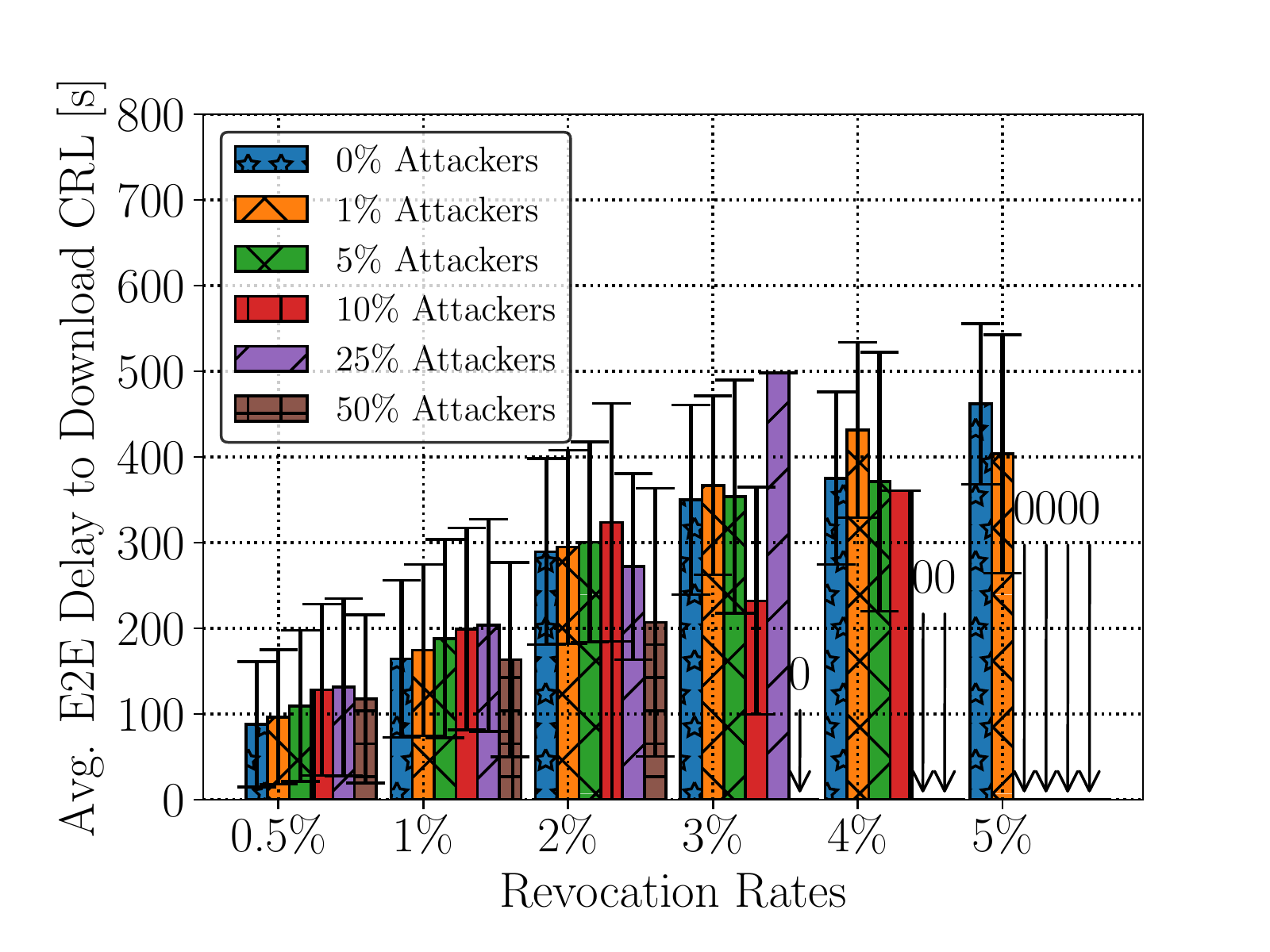}}
		\hspace{-0.45em} 
		\subfloat[Vehicle-centric scheme]{
			\includegraphics[trim=0.3cm 0.01cm 1.1cm 0.5cm, clip=true, width=0.25\textwidth,height=0.25\textheight,keepaspectratio]{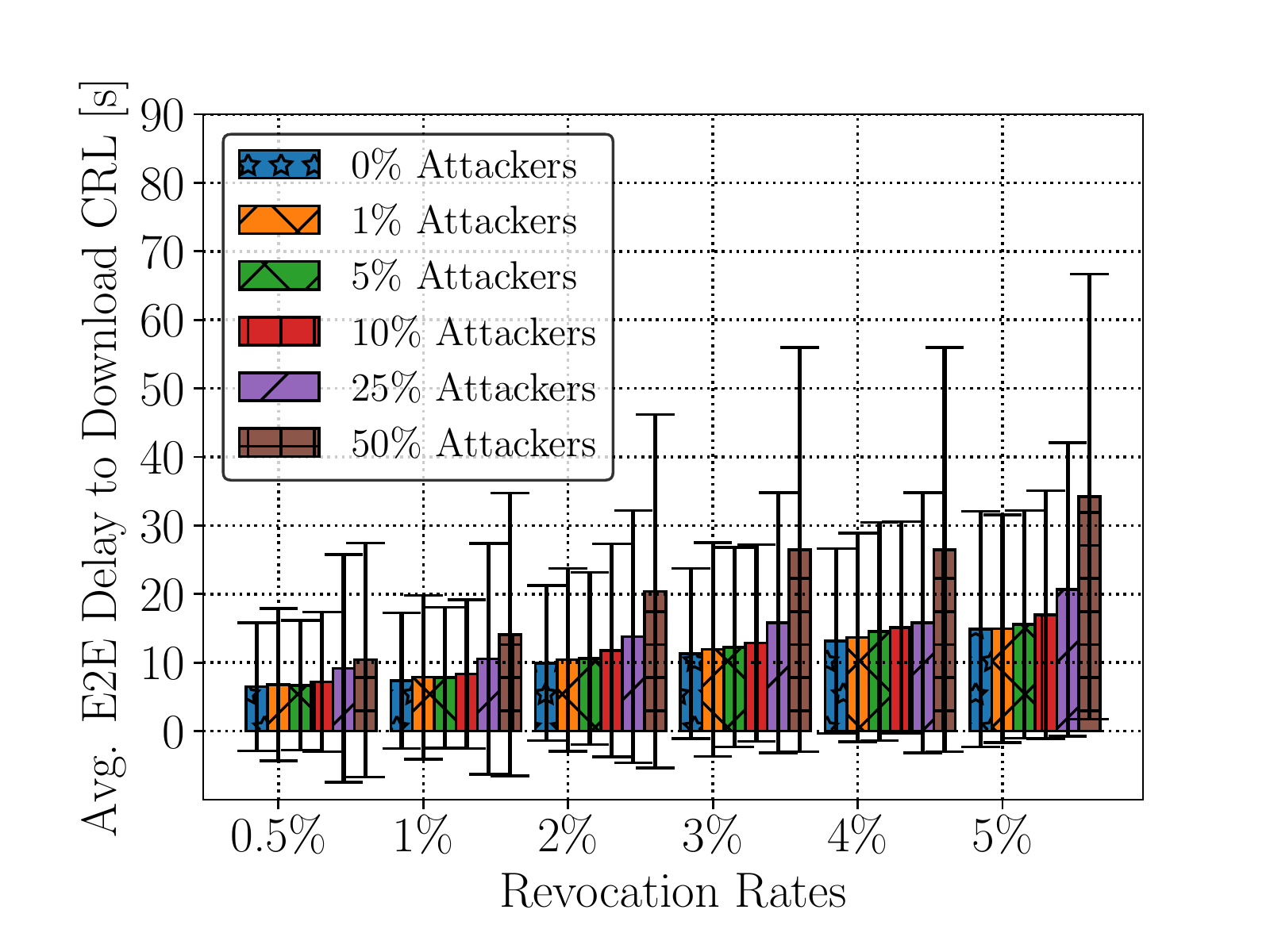}}
		\vspace{-0.25em}
		\caption{Resilience comparison against pollution and \ac{DoS} attacks with different revocation rates (7:00-7:10, $\tau_p=30s$, $\mathbb{B}=50KB/s$).}
		\label{fig:crl-dis-e2e-delay-resilience-against-pollution-dos-attacks}
	\end{center}
	\vspace{-2em}
\end{figure}

\begin{figure} [!t] 
	\vspace{-0em}
	\centering
	\subfloat[{\scriptsize Baseline: no attackers}]{ 
		\hspace{-1.15em}
		\includegraphics[trim=0.5cm 0.3cm 0.75cm 0.75cm, clip=true, totalheight=0.178\textheight, width=0.178\textwidth, angle=0, keepaspectratio] {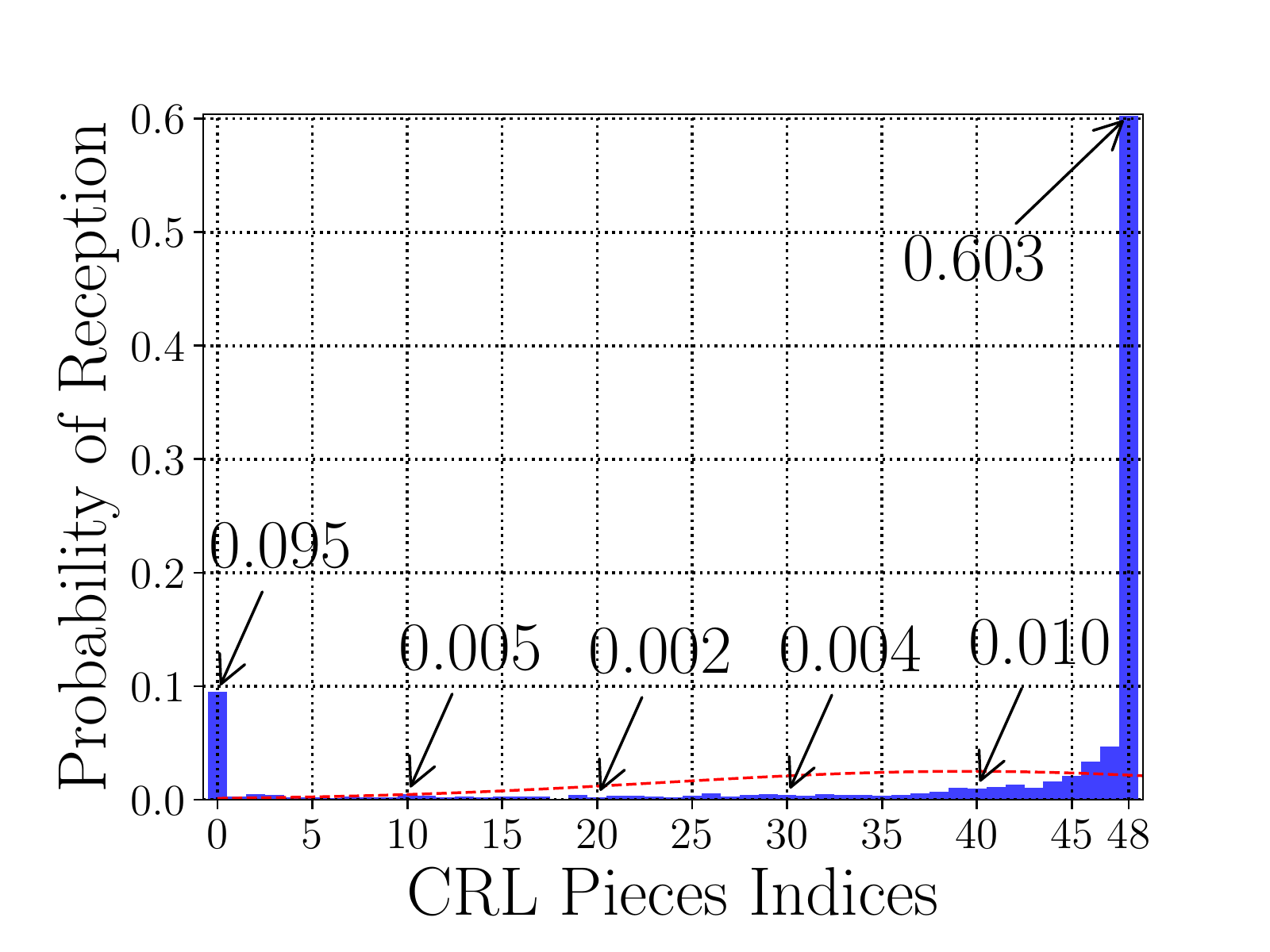}}
	\subfloat[{\scriptsize Baseline: 10\% attackers}]{ 
		\includegraphics[trim=0.5cm 0.3cm 0.75cm 0.75cm, clip=true, totalheight=0.178\textheight, width=0.178\textwidth, angle=0, keepaspectratio] {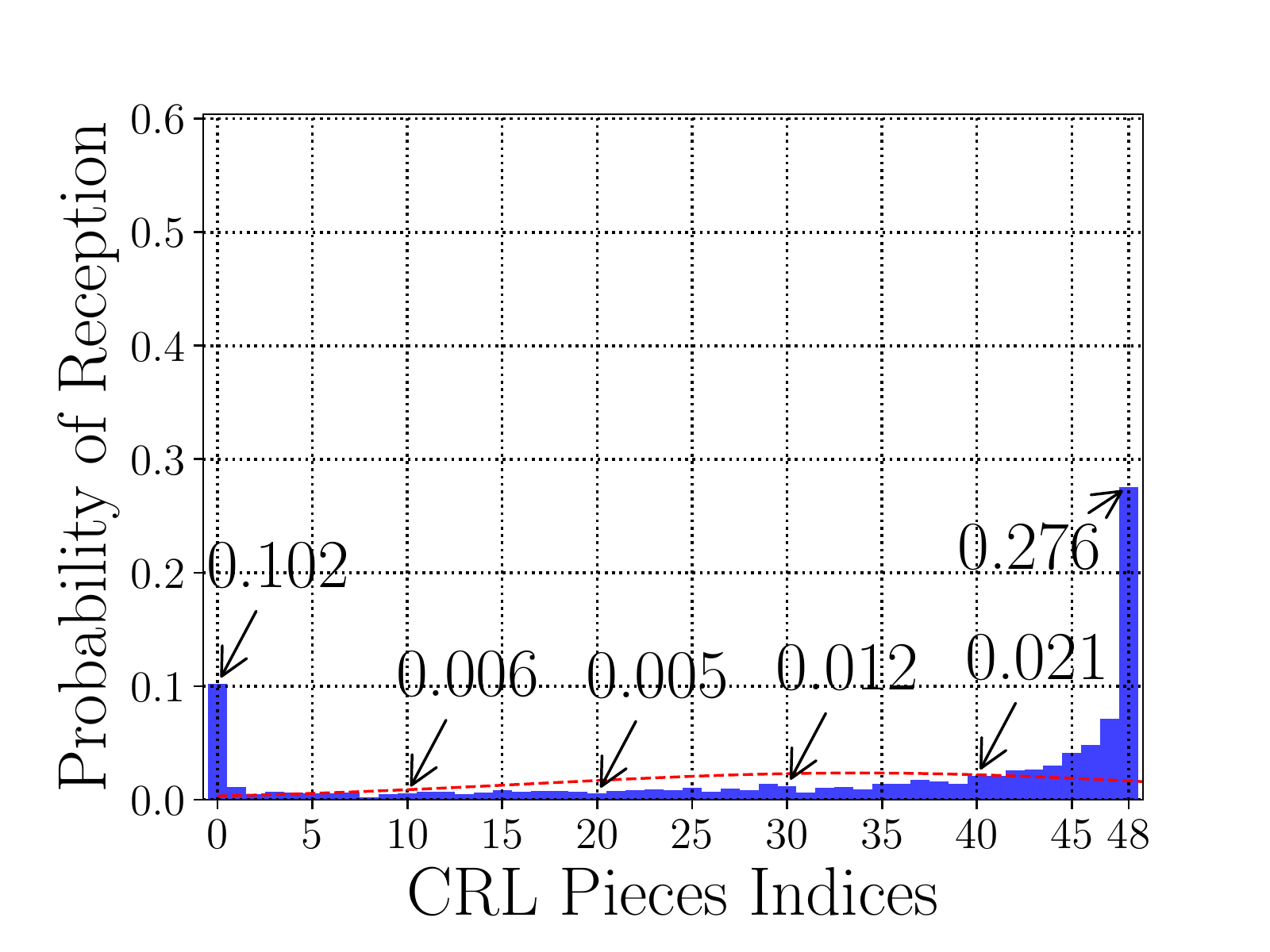}}
	\subfloat[{\scriptsize Baseline: 50\% attackers}]{ 
		\includegraphics[trim=0.5cm 0.3cm 0.75cm 0.75cm, clip=true, totalheight=0.178\textheight, width=0.178\textwidth, angle=0, keepaspectratio] {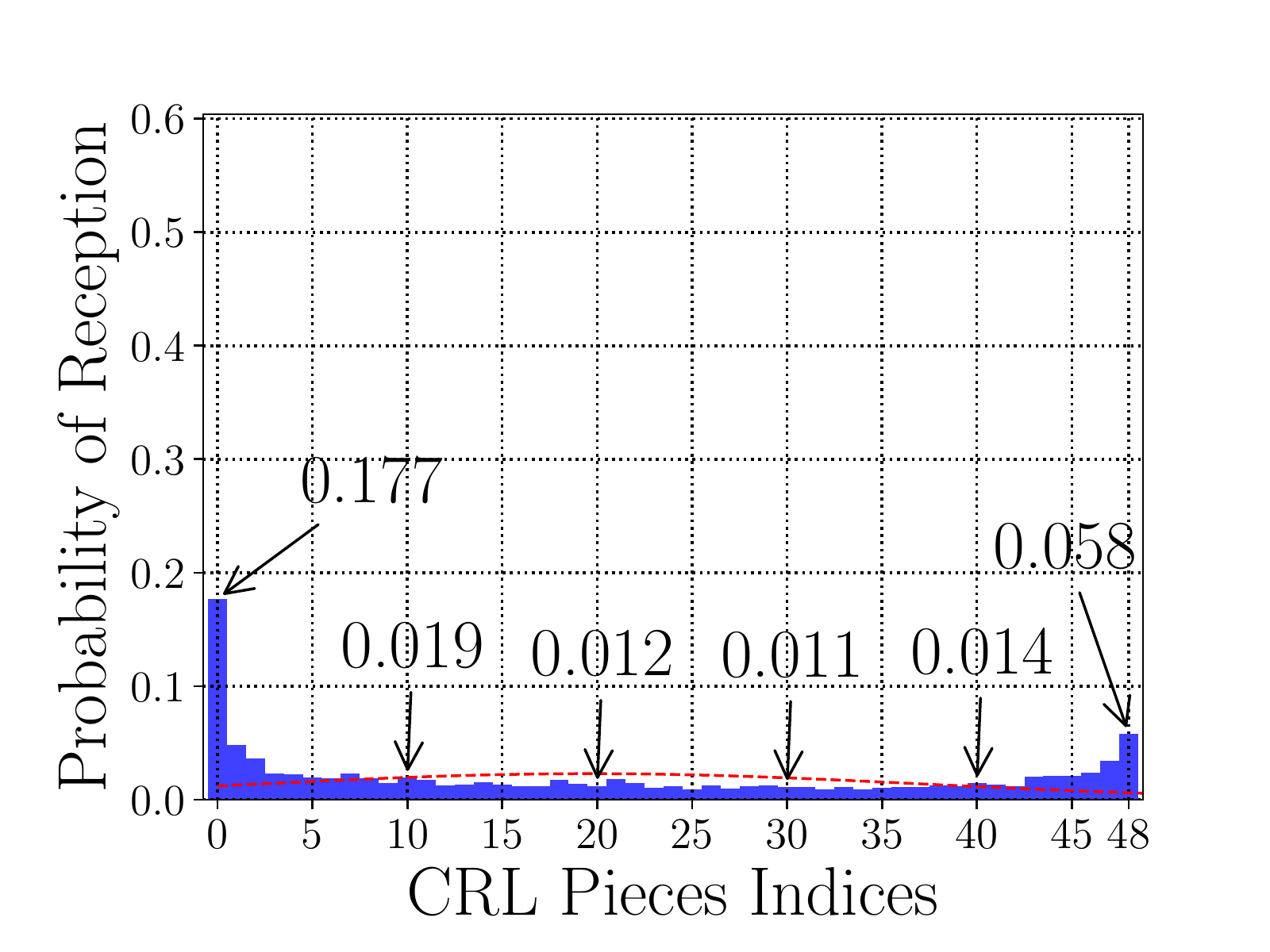}}
	\\
	\vspace{-0.5em}
	\subfloat[{\scriptsize \hspace{-0.75em} Vehicle-centric: no~~~ \protect\linebreak attackers}]{ 
		\hspace{-1.25em}
		\includegraphics[trim=0.5cm 0.3cm 0.75cm 0.75cm, clip=true, totalheight=0.178\textheight, width=0.178\textwidth, angle=0, keepaspectratio] {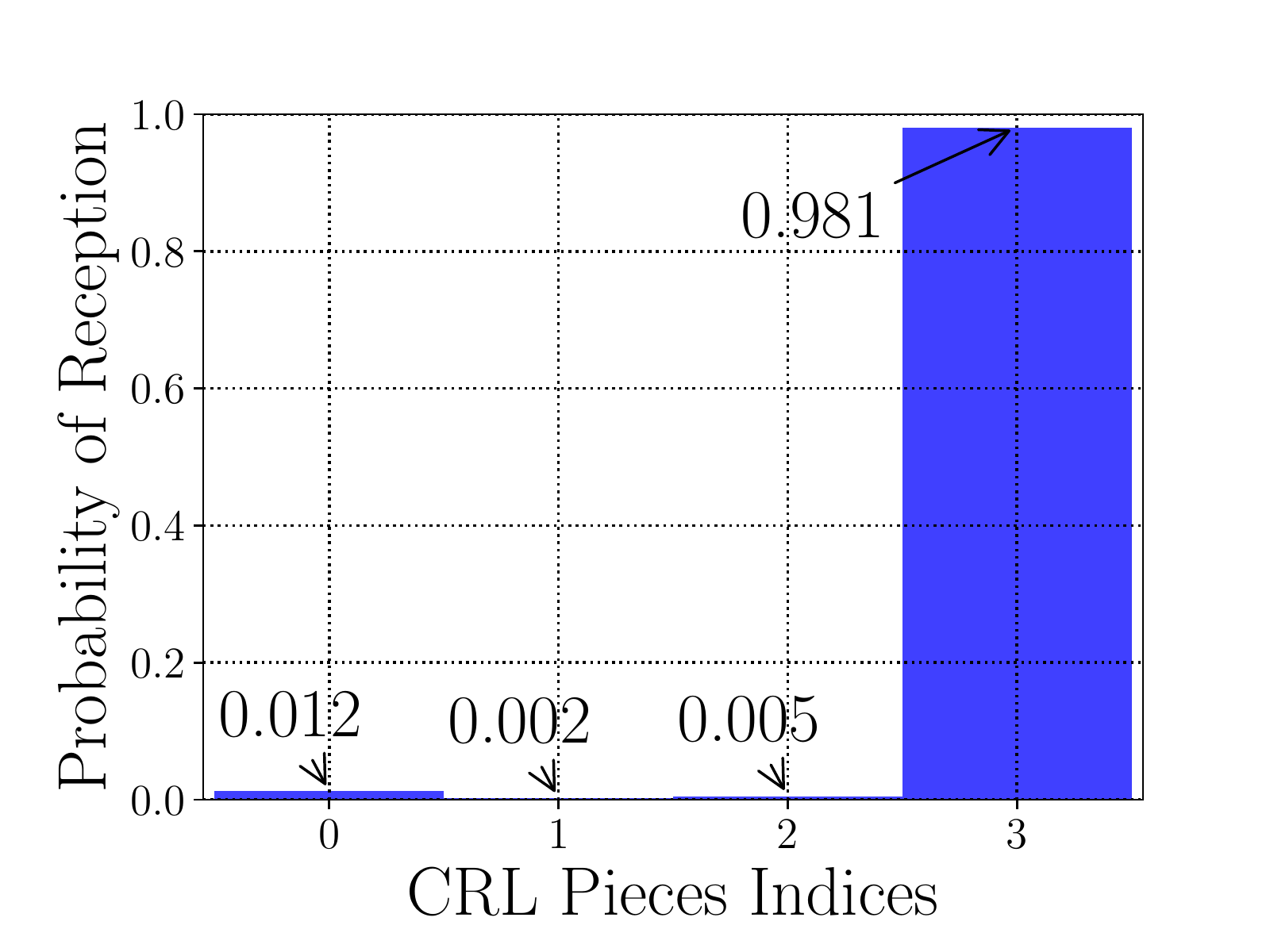}}
	\subfloat[{\scriptsize Vehicle-centric: 10\%~~~~~~~~ \protect\linebreak attackers}]{ 
		\includegraphics[trim=0.5cm 0.3cm 0.75cm 0.75cm, clip=true, totalheight=0.178\textheight, width=0.178\textwidth, angle=0, keepaspectratio] {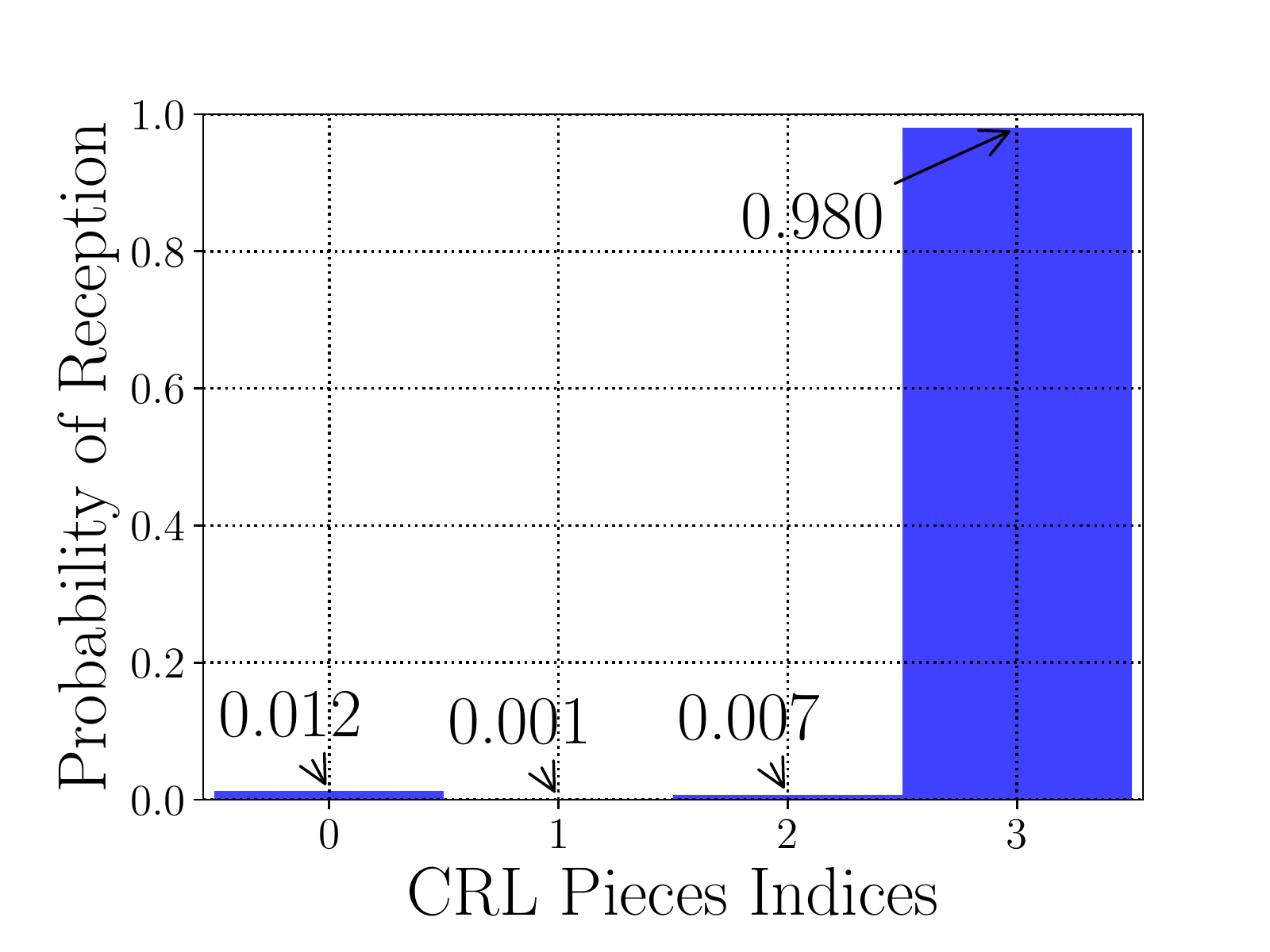}}
	\subfloat[{\scriptsize Vehicle-centric: 50\%~~~~~~~~ \protect\linebreak attackers}]{ 
		\includegraphics[trim=0.5cm 0.3cm 0.75cm 0.75cm, clip=true, totalheight=0.178\textheight, width=0.178\textwidth, angle=0, keepaspectratio] {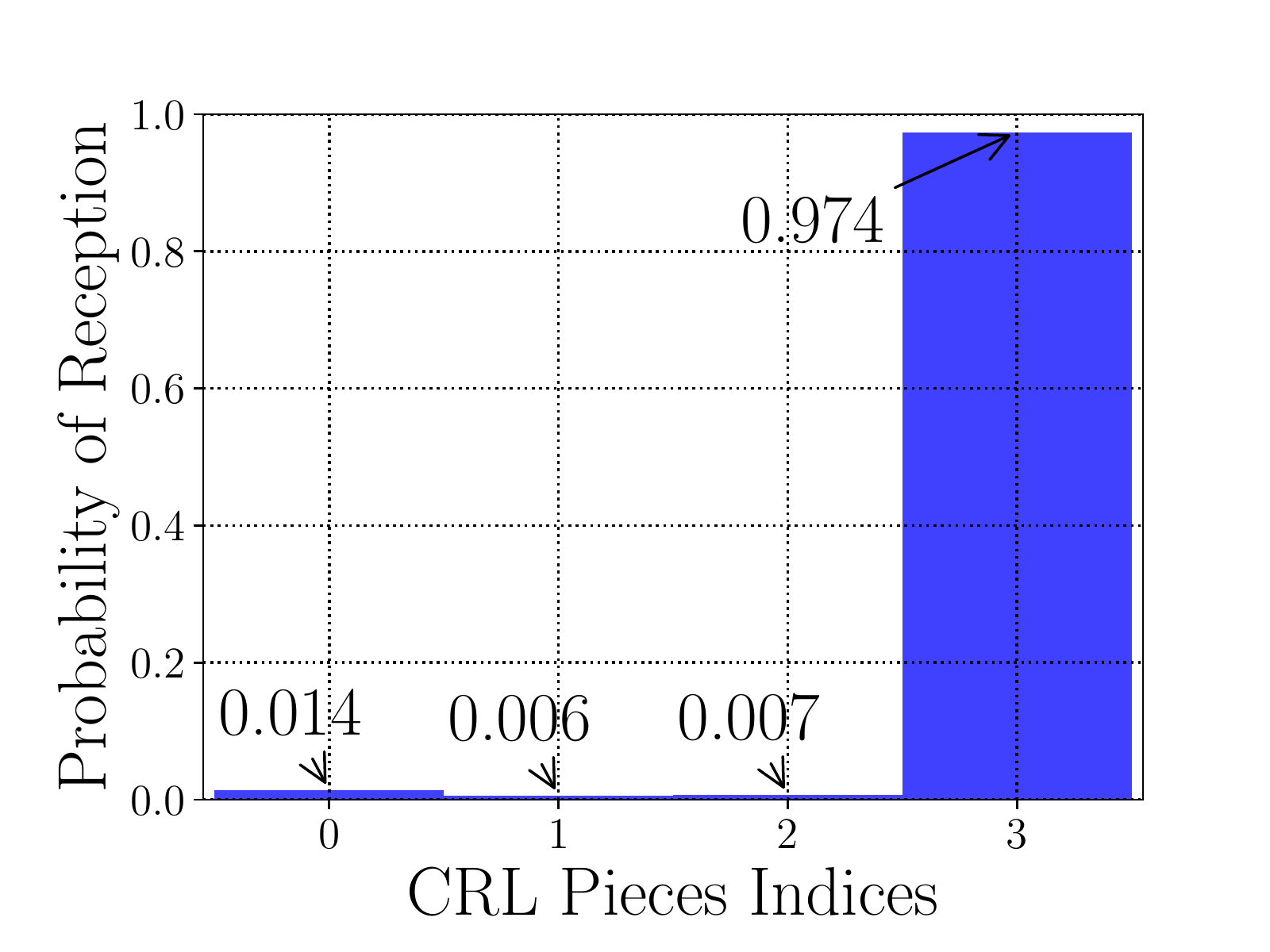}}
	\vspace{-0.25em}
	\caption{Probability of successful \ac{CRL} pieces reception ($\tau_{P}=30s$, $\mathbb{B}=50KB/s$). (a) and (d): no attacks. (b), (c), (e), (f): under a \ac{DoS} attack.}
	\label{fig:crl-dis-histogram-of-received-crl-pieces-under-ddos-attack}
	\vspace{-1.5em}
\end{figure}

Fig.~\ref{fig:crl-dis-resilience-against-selfish-nodes} shows the average end-to-end latency to successfully obtain the entire \acp{CRL} when a fraction of cognizant vehicles are considered to be selfish. Such nodes do not perform any ``active'' attack, e.g., a clogging \ac{DoS} attack; rather, they become silent and they never respond to a \ac{CRL} piece request. Fig.~\ref{fig:crl-dis-resilience-against-selfish-nodes} shows the average end-to-end delay to obtain the entire \ac{CRL} pieces in the presence of selfish nodes. For the baseline scheme, the average end-to-end latency linearly increases when the revocation rate increases; however, the average latency seems to be decreasing when the percentage of selfish nodes increases. 
The reason is that the probability of successfully obtaining the \ac{CRL} decreases, i.e., the higher the revocation rate combined with the higher percentage of the selfish nodes, the less the average number of cognizant vehicles is (this becomes clear in Fig.~\ref{fig:crl-dis-histogram-of-received-crl-pieces-under-ddos-attack}). In contrast, when there are 50\% selfish nodes and $\mathbb{R}=5\%$, the end-to-end latency for our vehicle-centric scheme increases from 12.13s to 14s, i.e.,~$\approx$15\% extra delay. Unlike performing a \ac{DoS} attack that is detectable, i.e., identifying the source of a misbehaving node, detecting such a misbehavior is challenging; detecting and attributing misbehavior to trigger the revocation is beyond the scope of this work.

Fig.~\ref{fig:crl-dis-e2e-delay-resilience-against-pollution-dos-attacks} shows the average end-to-end latency when attackers conduct pollution and \ac{DoS} attacks by periodically broadcasting bogus \ac{CRL} pieces once every 0.5s. The delays were averaged over vehicles successfully obtained \ac{CRL} pieces. Fig.~\ref{fig:crl-dis-e2e-delay-resilience-against-pollution-dos-attacks}.a shows that the baseline scheme is adversely affected when the number of compromised vehicles increases. For example, when $\mathbb{R}=4\%$ and 25\% of the \acp{OBU} are compromised, no vehicle could successfully obtain the entire \ac{CRL} pieces. In contrast, Fig.~\ref{fig:crl-dis-e2e-delay-resilience-against-pollution-dos-attacks}.b shows the performance of our vehicle-centric scheme: with $\mathbb{R}=4\%$ and 25\% of the \acp{OBU} misbehave in this way, the average end-to-end delay reasonably increases (from 13.13s to 15.81s).

\begin{figure} [!t]
	\vspace{-0em}
	\centering
	\begin{center}
		\centering
		\hspace{-1.45em} 
		\subfloat[\acs{CDF} of delays under a \ac{DoS} attack]{
			\includegraphics[trim=0.3cm 0.15cm 0.5cm 1.25cm, clip=true, width=0.252\textwidth,height=0.252\textheight,keepaspectratio]{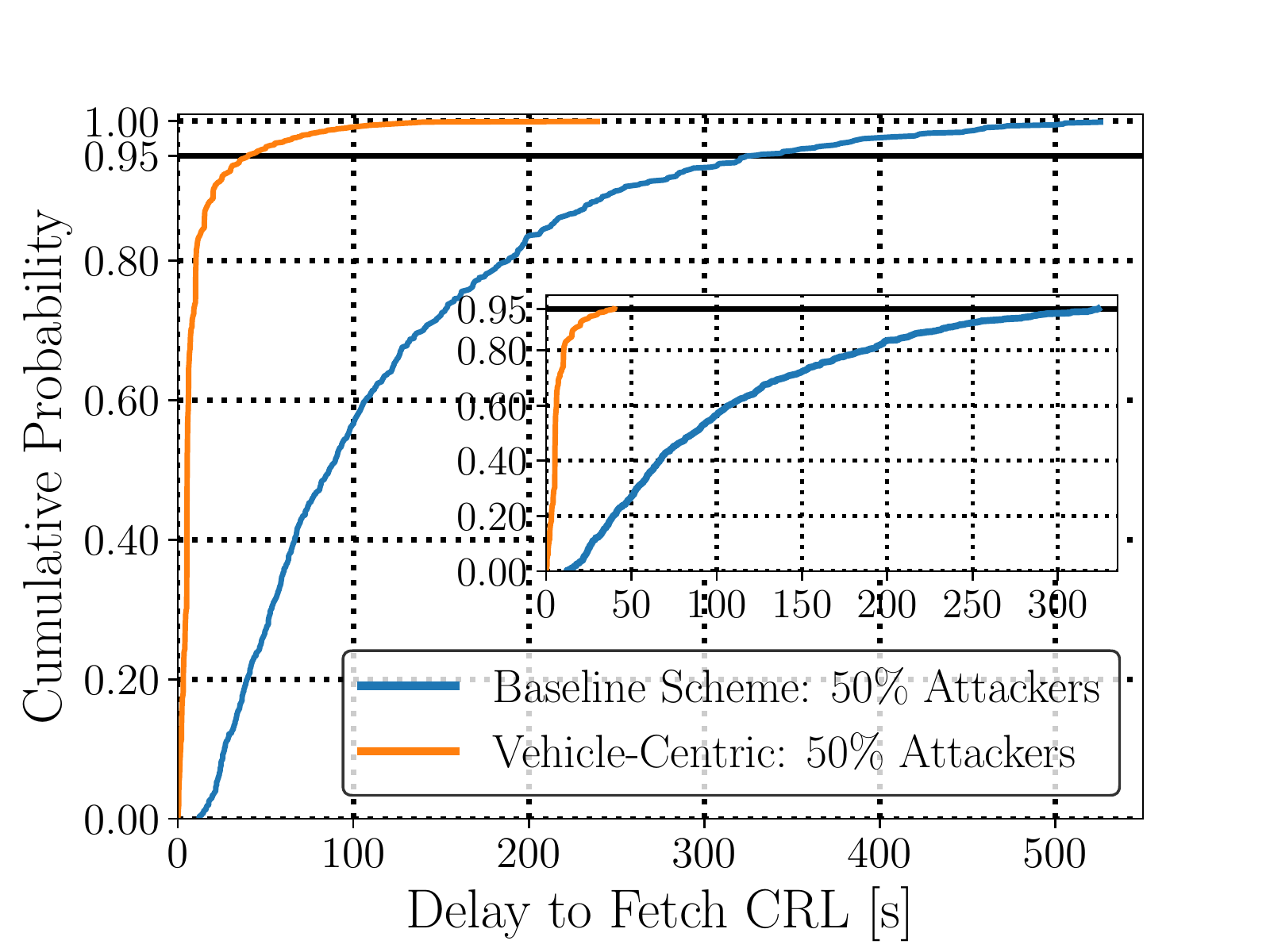}}
		\hspace{-0.45em} 
		\subfloat[Probability of failure]{
			\includegraphics[trim=0.25cm 0.25cm 1cm 1.25cm, clip=true, width=0.251\textwidth,height=0.251\textheight,keepaspectratio]{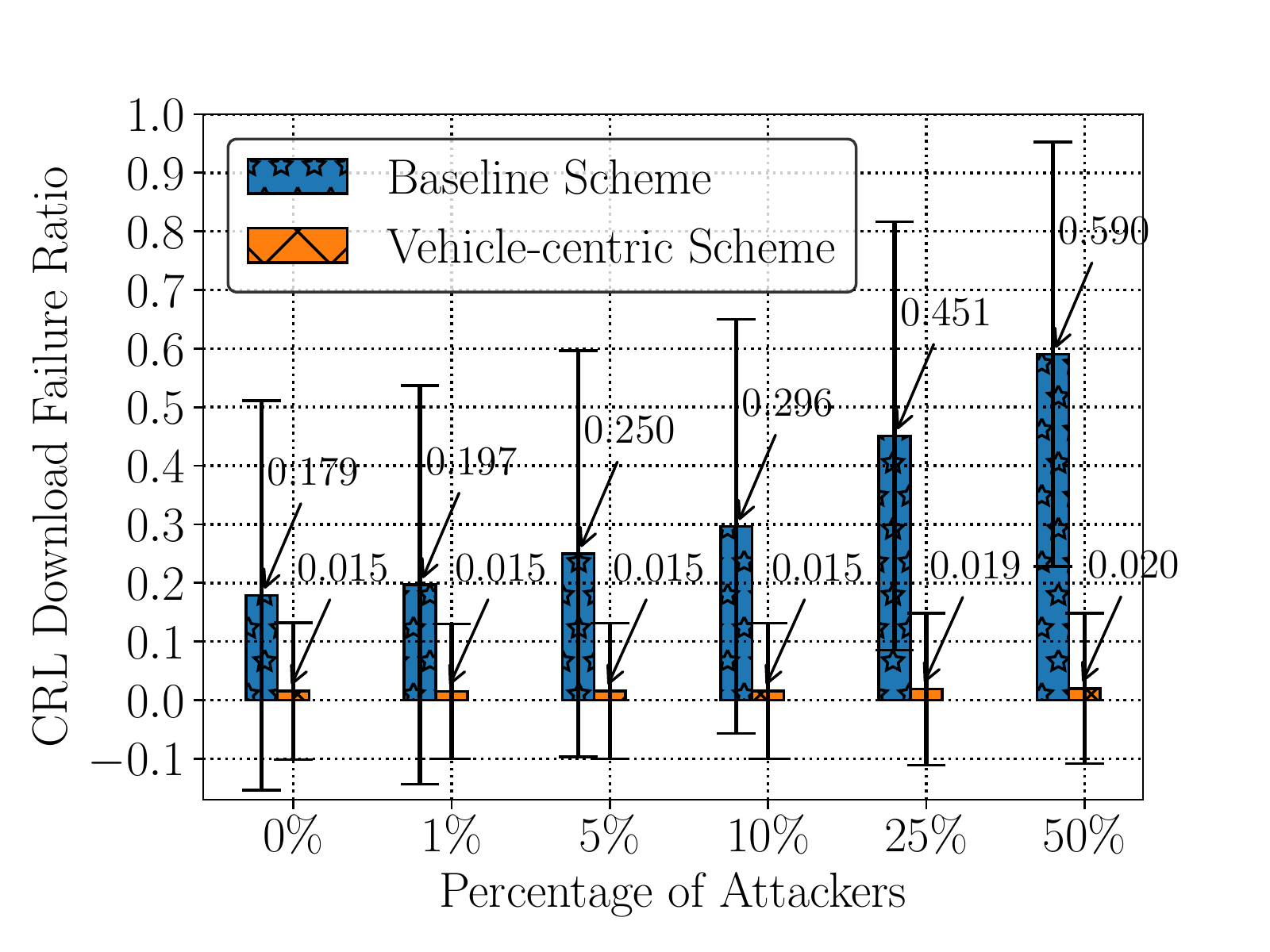}}
		\vspace{-0.25em}
		\caption{(a) \acs{CDF} of latency to successfully obtain \ac{CRL} pieces (50\% attackers). (b) \ac{CRL} download failure ratio as a function of \ac{DoS} attackers ($\tau_{P}=30s$, $\mathbb{B}=50KB/s$).}
		\label{fig:crl-dis-crl_download-failure-ratio-under-ddos-attacks}
	\end{center}
	\vspace{-2em}
\end{figure}

Fig.~\ref{fig:crl-dis-crl_download-failure-ratio-under-ddos-attacks}.a shows the \ac{CDF} of delays when 50\% of the \acp{OBU} are compromised and periodically broadcast bogus \ac{CRL} pieces once every 0.5s. For the baseline scheme, $F_x(t=330s)=0.95$, or $Pr\{t\leq330s\}=0.95$, whereas with vehicle-centric scheme, $F_x(t=40s)=0.95$, or $Pr\{t\leq40s\}=0.95$. Fig.~\ref{fig:crl-dis-crl_download-failure-ratio-under-ddos-attacks}.b shows the probability of failure to obtain the \ac{CRL} pieces for the baseline and vehicle-centric schemes against such a misbehavior. With the baseline scheme, the probability of failure to obtain \ac{CRL} pieces increases from 0.18 to 0.59 when the percentage of \ac{DDoS} attackers increases from zero to 50\%. This implies that the majority of the vehicles would reach their destination without successfully obtaining the \ac{CRL} pieces. In contrast, for the vehicle-centric scheme, even if 50\% of the \acp{OBU} are compromised and conduct a clogging \ac{DoS} attack, the probability of failure is not considerably affected, in fact it is almost negligible, i.e., increasing from 0.015 to 0.020. This shows the resiliency of our vehicle-centric scheme against \ac{DDoS} attacks: the significant improvements that can be achieved by prioritization of the revocation entries based on their validity intervals, i.e., a huge reduction in size, thus significantly increased its resiliency against resource depletion attacks.

Fig.~\ref{fig:crl-dis-histogram-of-received-crl-pieces-under-ddos-attack} shows the histogram of successfully received \ac{CRL} pieces for the baseline scheme and the vehicle-centric scheme. With the baseline scheme, even if there is no attacker in the system, 60\% of the vehicles could successfully obtain the \ac{CRL} pieces within their trip duration. More interesting, when 50\% of the \acp{OBU} are compromised and misbehaving by broadcasting periodically fake \ac{CRL} pieces, only 6\% of the vehicles could successfully obtain the entire \ac{CRL} pieces. In contrast, with the vehicle-centric scheme, even if 50\% of the vehicles misbehave, 97\% of the vehicles would successfully obtain the entire \ac{CRL} pieces within their trip duration. This shows that the operation of our vehicle-centric scheme is not considerably affected even if 50\% of the \acp{OBU} are compromised, and the vehicles could still obtain the needed \ac{CRL} pieces within a reasonable delay.

Table~\ref{table:crl-dis-delay-to-check-bloomfilter-membership-item} shows the latency for validating all \ac{CRL} pieces with different false positive rates. We performed our experiments on the Nexcom \ac{OBU} boxes from the PRESERVE project~\cite{preserve-url}. For example, the latency to validate all \ac{CRL} piece, with $\mathbb{R}=1\%$ and $p=10^{-30}$, is 0.854 ms. Table~\ref{table:crl-dis-delay-to-insert-items-in-bloomfilter} shows the latency for inserting \ac{CRL} pieces in a \ac{BF}\footnote{To directly compare the latencies, we conducted both experiments, i.e., \ac{BF} insertion and membership check, on the Nexcom \ac{OBU} boxes even though a \ac{PCA} would have a stronger computational resource.}. For example, the end-to-end delay to construct the \ac{CRL} fingerprint, with $\mathbb{R}=1\%$ and $p=10^{-30}$, is 1.14 ms.

\begin{table}[!t]
	\centering
	\vspace{-0em}
	\caption{Latency for validation all \ac{CRL} pieces using a \ac{BF}, executed on a \protect\linebreak Nexcom \ac{OBU}, averaged over 10K runs ($\tau_{P}=60s$, $\mathbb{B}=50KB/s$).}
	\vspace{-0.5em}
	\label{table:crl-dis-delay-to-check-bloomfilter-membership-item}
	\resizebox{0.43\textwidth}{!}
	{
		\renewcommand{\arraystretch}{0.99}
		\hspace{-1em}
		\begin{tabular}{|c|c|c|c|c|c|c|c|}
			\hline
			\rowcolor{LightCyan} \multirow{1}{*}{\textbf{\shortstack{Revocation Rate}}} & \multirow{1}{*}{\textbf{\shortstack{\ac{BF} size}}} & \multicolumn{2}{c|}{\textbf{false positive}} & \multicolumn{2}{c|}{\textbf{delay}} & \multicolumn{2}{c|}{\textbf{check/sec.}} \\
			\cline{1-8}
			\hline\hline
			\rowcolor{LightCyan} & 252 bits & \multicolumn{2}{c|}{\shortstack{{} \\ p=$10^{-25}$}} & \multicolumn{2}{c|}{0.709 ms} & \multicolumn{2}{c|}{2,819} \\
			\cline{2-8}
			\rowcolor{LightCyan} & 261 bits & \multicolumn{2}{c|}{\shortstack{{} \\ p=$10^{-26}$}} & \multicolumn{2}{c|}{0.729 ms} & \multicolumn{2}{c|}{2,741} \\
			\cline{2-8}
			\rowcolor{LightCyan} 1\% & 270 bits  & \multicolumn{2}{c|}{\shortstack{{} \\ p=$10^{-27}$}} & \multicolumn{2}{c|}{0.749 ms} & \multicolumn{2}{c|}{2,668} \\
			\cline{2-8}
			\rowcolor{LightCyan} & 282 bits & \multicolumn{2}{c|}{\shortstack{{} \\ p=$10^{-28}$}} & \multicolumn{2}{c|}{0.776 ms} & \multicolumn{2}{c|}{2,577} \\
			\cline{2-8}
			\rowcolor{LightCyan} & 291 bits & \multicolumn{2}{c|}{\shortstack{{} \\ p=$10^{-29}$}} & \multicolumn{2}{c|}{0.839 ms} & \multicolumn{2}{c|}{2,384} \\
			\cline{2-8}
			\rowcolor{LightCyan} & 300 bits & \multicolumn{2}{c|}{\shortstack{{} \\ p=$10^{-30}$}} & \multicolumn{2}{c|}{0.859 ms} & \multicolumn{2}{c|}{2,329} \\
			\cline{2-8}
			\hline\hline
			\rowcolor{LightCyan} & 840 bits & \multicolumn{2}{c|}{\shortstack{{} \\ p=$10^{-25}$}} & \multicolumn{2}{c|}{2.539 ms} & \multicolumn{2}{c|}{2,756} \\
			\cline{2-8}
			\rowcolor{LightCyan} & 957 bits & \multicolumn{2}{c|}{\shortstack{{} \\ p=$10^{-26}$}} & \multicolumn{2}{c|}{2.613 ms} & \multicolumn{2}{c|}{2,678} \\
			\cline{2-8}
			\rowcolor{LightCyan} 5\% & 990 bits  & \multicolumn{2}{c|}{\shortstack{{} \\ p=$10^{-27}$}} & \multicolumn{2}{c|}{2.667 ms} & \multicolumn{2}{c|}{2,624} \\
			\cline{2-8}
			\rowcolor{LightCyan} & 940 bits & \multicolumn{2}{c|}{\shortstack{{} \\ p=$10^{-28}$}} & \multicolumn{2}{c|}{2.774 ms} & \multicolumn{2}{c|}{2,522} \\
			\cline{2-8}
			\rowcolor{LightCyan} & 1067 bits & \multicolumn{2}{c|}{\shortstack{{} \\ p=$10^{-29}$}} & \multicolumn{2}{c|}{2.971 ms} & \multicolumn{2}{c|}{2,355} \\
			\cline{2-8}
			\rowcolor{LightCyan} & 1100 bits & \multicolumn{2}{c|}{\shortstack{{} \\ p=$10^{-30}$}} & \multicolumn{2}{c|}{3.043 ms} & \multicolumn{2}{c|}{2,300} \\
			\cline{2-8}
			\hline
		\end{tabular}
	}
	\vspace{-1em}
\end{table}

\begin{table}[!t]
	\centering
	\vspace{-0em}
	\caption{Latency for inserting \ac{CRL} pieces into a \ac{BF}, executed on a \protect\linebreak Nexcom \ac{OBU}, averaged over 10K runs ($\tau_{P}=60s$, $\mathbb{B}=50KB/s$).}
	\vspace{-0.5em}
	\label{table:crl-dis-delay-to-insert-items-in-bloomfilter}
	\resizebox{0.43\textwidth}{!}
	{
		\renewcommand{\arraystretch}{0.99}
		\hspace{-1em}
		\begin{tabular}{|c|c|c|c|c|c|c|c|}
			\hline
			\rowcolor{LightCyan} \multirow{1}{*}{\textbf{\shortstack{Revocation Rate}}} & \multirow{1}{*}{\textbf{\shortstack{\ac{BF} size}}} & \multicolumn{2}{c|}{\textbf{false positive}} & \multicolumn{2}{c|}{\textbf{delay}} & \multicolumn{2}{c|}{\textbf{check/sec.}} \\
			\cline{1-8}
			\hline\hline
			\rowcolor{LightCyan} & 252 bits & \multicolumn{2}{c|}{\shortstack{{} \\ p=$10^{-25}$}} & \multicolumn{2}{c|}{0.949 ms} & \multicolumn{2}{c|}{2,108} \\
			\cline{2-8}
			\rowcolor{LightCyan} & 261 bits & \multicolumn{2}{c|}{\shortstack{{} \\ p=$10^{-26}$}} & \multicolumn{2}{c|}{0.976 ms} & \multicolumn{2}{c|}{2,049} \\
			\cline{2-8}
			\rowcolor{LightCyan} 1\% & 270 bits  & \multicolumn{2}{c|}{\shortstack{{} \\ p=$10^{-27}$}} & \multicolumn{2}{c|}{1.004 ms} & \multicolumn{2}{c|}{1,991} \\
			\cline{2-8}
			\rowcolor{LightCyan} & 282 bits & \multicolumn{2}{c|}{\shortstack{{} \\ p=$10^{-28}$}} & \multicolumn{2}{c|}{1.041 ms} & \multicolumn{2}{c|}{1,921} \\
			\cline{2-8}
			\rowcolor{LightCyan} & 291 bits & \multicolumn{2}{c|}{\shortstack{{} \\ p=$10^{-29}$}} & \multicolumn{2}{c|}{1.113 ms} & \multicolumn{2}{c|}{1,796} \\
			\cline{2-8}
			\rowcolor{LightCyan} & 300 bits & \multicolumn{2}{c|}{\shortstack{{} \\ p=$10^{-30}$}} & \multicolumn{2}{c|}{1.140 ms} & \multicolumn{2}{c|}{1,754} \\
			\cline{2-8}
			\hline\hline
			\rowcolor{LightCyan} & 840 bits & \multicolumn{2}{c|}{\shortstack{{} \\ p=$10^{-25}$}} & \multicolumn{2}{c|}{3.334 ms} & \multicolumn{2}{c|}{2,099} \\
			\cline{2-8}
			\rowcolor{LightCyan} & 957 bits & \multicolumn{2}{c|}{\shortstack{{} \\ p=$10^{-26}$}} & \multicolumn{2}{c|}{3.432 ms} & \multicolumn{2}{c|}{2,039} \\
			\cline{2-8}
			\rowcolor{LightCyan} 5\% & 990 bits  & \multicolumn{2}{c|}{\shortstack{{} \\ p=$10^{-27}$}} & \multicolumn{2}{c|}{3.545 ms} & \multicolumn{2}{c|}{1,974} \\
			\cline{2-8}
			\rowcolor{LightCyan} & 940 bits & \multicolumn{2}{c|}{\shortstack{{} \\ p=$10^{-28}$}} & \multicolumn{2}{c|}{3.662 ms} & \multicolumn{2}{c|}{1,911} \\
			\cline{2-8}
			\rowcolor{LightCyan} & 1067 bits & \multicolumn{2}{c|}{\shortstack{{} \\ p=$10^{-29}$}} & \multicolumn{2}{c|}{3.898 ms} & \multicolumn{2}{c|}{1,796} \\
			\cline{2-8}
			\rowcolor{LightCyan} & 1100 bits & \multicolumn{2}{c|}{\shortstack{{} \\ p=$10^{-30}$}} & \multicolumn{2}{c|}{3.977 ms} & \multicolumn{2}{c|}{1,760} \\
			\cline{2-8}
			\hline
		\end{tabular}
	}
	\vspace{-1em}
\end{table}


\section{Conclusion}
\label{sec:crl-dis-conclusions}

Paving the way for deploying secure and privacy-preserving \ac{VC} systems, standardization bodies have reached an agreement towards deploying a special-purpose \ac{VPKI} without reaching a consensus on how to efficiently and timely distribute \acp{CRL} in a large-scale environment. The success of secure and privacy-preserving \ac{VC} systems requires effective mechanisms for distributing \acp{CRL}, to guarantee the operations of the systems. We proposed a practical framework to effectively distribute \acp{CRL}: our vehicle-centric scheme distributes necessary \ac{CRL} pieces corresponding to a vehicle's targeted region and actual trip duration, i.e., obtaining only region- and time-relevant revocation information. Through extensive experimental evaluation, we demonstrated that our scheme is highly efficient and scalable, and it is resilient against selfish nodes, as well as pollution and \ac{DoS} attacks. This supports that our scheme is a viable solution towards catalyzing the deployment of the secure and privacy-protecting \ac{VC} systems.


\section*{Acknowledgement}
\label{sec:acknowledgement}

Work supported by the Swedish Foundation for Strategic Research (SSF) SURPRISE project and the KAW Academy Fellowship Trustworthy IoT project.


\bibliographystyle{IEEEtran}
\bibliography{IEEEabrv,references}

\begin{IEEEbiography}[{\includegraphics[width=1in,height=1.25in,clip,keepaspectratio]{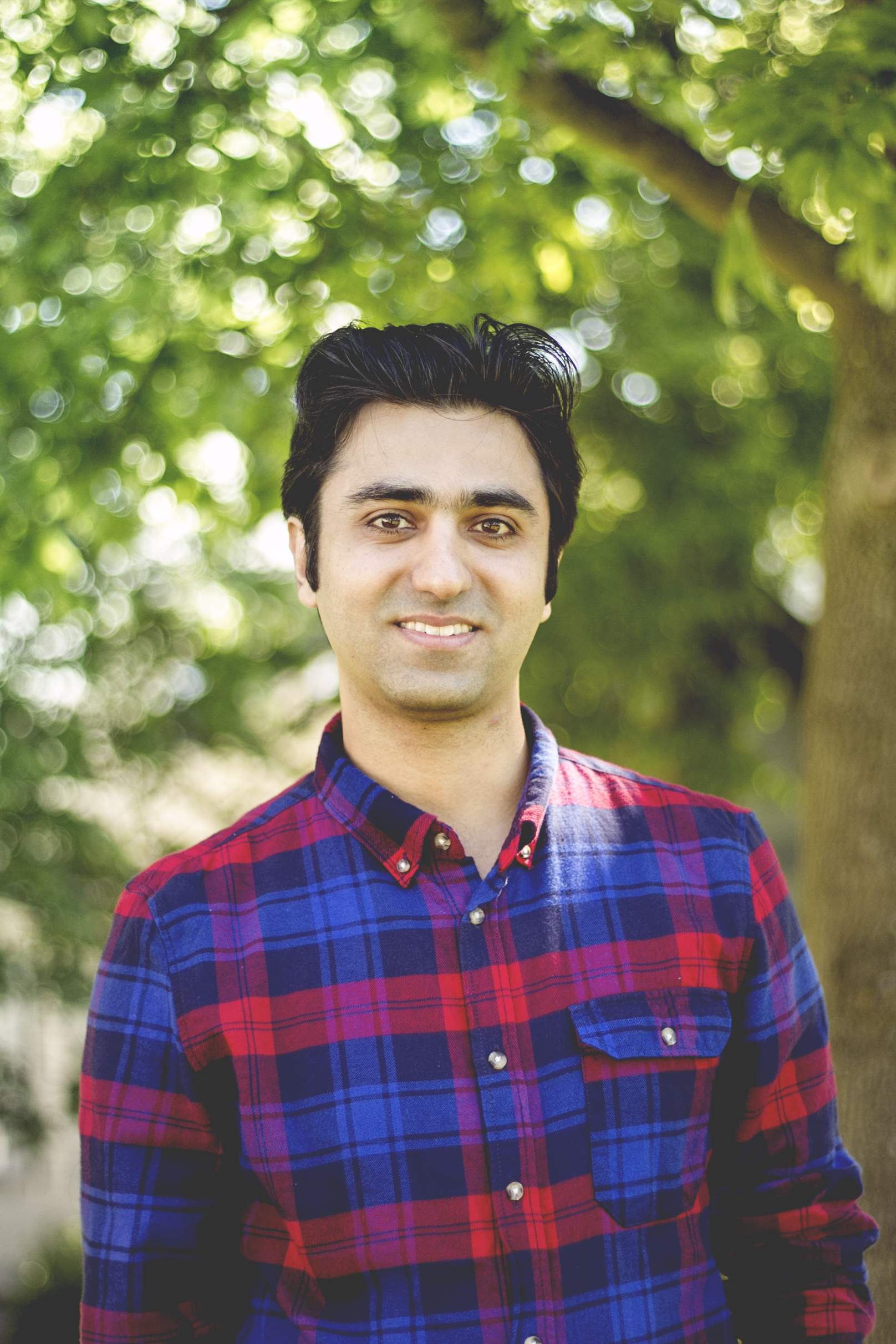}}]{Mohammad Khodaei}
earned his diploma in software engineering from Azad University of Najafabad in Isfahan, Iran, in 2006 and his M.S. degree in information and communication systems security from KTH Royal Institute of Technology, Stockholm, Sweden, in 2012. He is currently pursuing his Ph.D. degree at the Networked Systems Security Group, KTH, under the supervision of Prof. Panos Papadimitratos. His research interests include security and privacy in \acsp{VANET}, smart cities, and the Internet of Things. 
\end{IEEEbiography}

\begin{IEEEbiography}[{\includegraphics[width=1in,height=1.25in,clip,keepaspectratio]{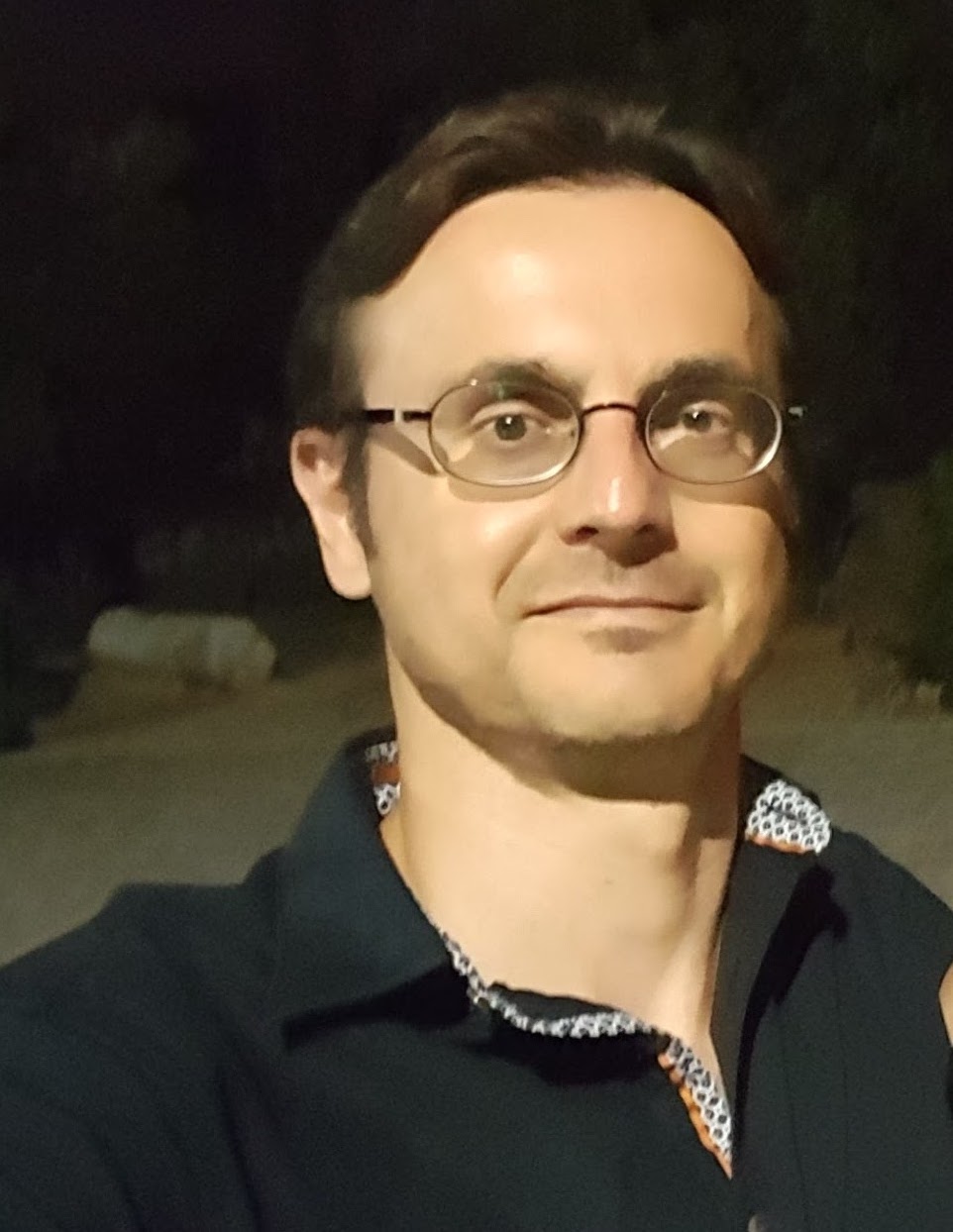}}]{Panos Papadimitratos} earned his Ph.D. degree from Cornell University, Ithaca, NY. At KTH, Stockholm, Sweden, he leads the Networked Systems Security lab, and he is a member of the steering committee of the Security Link center. He has delivered numerous invited talks, keynotes, panel addresses, and tutorials in flagship conferences. He serves or served as: Associate Editor of the IEEE TMC and the ACM/IEEE ToN journals; member of the PETS Editorial and Advisory Boards, and the ACM WiSec and CANS conference steering committees; program chair for the ACM WiSec'16, TRUST'16, CANS'18 conferences; general chair for ACM WISec'18, PETS'19, and IEEE EuroS\&P'19. He is a Fellow of the Young Academy of Europe, a Knut and Alice Wallenberg Academy Fellow, and IEEE Fellow. His group web-page is: \url{https://www.eecs.kth.se/nss}.
\end{IEEEbiography}


\end{document}